\documentclass[12pt]{graphicsclass}

\usepackage[numbers, sort&compress]{natbib}
\usepackage{amsmath}
\usepackage{amssymb}
\usepackage{graphicx}
\usepackage{latexsym}
\usepackage{nicefrac}
\usepackage{multirow}
\usepackage{multicol}
\usepackage{xcolor}
\usepackage{subcaption}
\usepackage{upgreek}
\usepackage{subfiles}
\usepackage{float}

\usepackage{xltabular}
\usepackage[utf8]{inputenc}
\usepackage{palatino}
\usepackage[labelfont=bf]{caption}
\usepackage[margin=1in]{geometry}
\usepackage[hidelinks]{hyperref}
\usepackage{subfiles}
\usepackage[nottoc]{tocbibind}

\newcommand{\abs}[1]{\left\vert#1\right\vert} 
\newcommand{\del}{\partial}
\newcommand{\f}{\frac}
\newcommand{\sub}{\textsubscript}
\newcommand{\super}{\textsuperscript}

\renewcommand{\d}{\mathrm{d}}
\newcommand{\dt}{\mathrm{d}t} 
\newcommand{\ddt}{\f{\mathrm{d}}{\mathrm{d}t}} 

\newcommand{\bbm}{\begin{bmatrix}}
\newcommand{\ebm}{\end{bmatrix}}

\newcommand{\mbf}{\mathbf}

\newcommand{\beq}{\begin{equation}}
\newcommand{\eeq}{\end{equation}}
\newcommand{\beqnn}{\begin{equation*}}
\newcommand{\eeqnn}{\end{equation*}}
\newcommand{\bdis}{\begin{displaymath}}
\newcommand{\edis}{\end{displaymath}}
\newcommand{\beqarr}{\begin{eqnarray}}
\newcommand{\eeqarr}{\end{eqnarray}}
\newcommand{\beqarrnn}{\begin{eqnarray*}}
\newcommand{\eeqarrnn}{\end{eqnarray*}}

\newcommand{\nitrous}{N\sub{2}O}
\newcommand{\nitrogen}{N\sub{2}}
\newcommand{\orckt}{\textit{OpenRocket}}
\newcommand{\engine}{Ma\"{e}lstrom}
\newcommand{\rocket}{Athos}

\newcommand{\Ru}{\mathcal{R}_\text{u}}
\newcommand{\Wa}{\mathcal{W}_\text{a}}
\newcommand{\Wo}{\mathcal{W}_\text{o}}
\newcommand{\gsl}{g_\text{SL}}
\newcommand{\re}{r_\text{E}}
\newcommand{\mprop}{m_\text{prop}}
\newcommand{\cgprop}{\text{CG}_\text{prop}}

\newcommand{\Ai}{A_\text{i}}
\newcommand{\Ci}{C_\text{i}}
\newcommand{\Ni}{N_\text{i}}
\newcommand{\Tt}{T_\text{T}}
\newcommand{\dTank}{d_\text{T}}
\newcommand{\LTank}{L_\text{T}}
\newcommand{\dDip}{d_\text{dt}}
\newcommand{\DDip}{D_\text{dt}}
\newcommand{\LDip}{L_\text{dt}}
\newcommand{\Ttzero}{T_\text{T,0}}
\newcommand{\ptzero}{p_\text{T,0}}
\newcommand{\motot}{m_\text{o,tot}}
\newcommand{\Vt}{V_\text{T}}
\newcommand{\ho}{h_\text{o}}
\newcommand{\hosat}{h_\text{o,sat}}
\newcommand{\mdoto}{\dot{m}_\text{o}}
\newcommand{\nl}{n_\text{l}}
\newcommand{\ndoto}{\dot{n}_\text{o}}
\newcommand{\nv}{n_\text{v}}
\newcommand{\ploss}{p_\text{loss}}
\newcommand{\psat}{p_\text{sat}}
\newcommand{\pt}{p_\text{T}}
\newcommand{\cvv}{c_{v_{\text{v}}}}

\newcommand{\cpv}{c_{p_{\text{v}}}}

\newcommand{\ul}{u_\text{l}}
\newcommand{\ulsat}{u_\text{l,sat}}
\newcommand{\hl}{h_\text{l}}

\newcommand{\hv}{h_\text{v}}

\newcommand{\uv}{u_\text{v}}
\newcommand{\uvsat}{u_\text{v,sat}}
\newcommand{\xt}{\mbf{x}_\text{T}}
\newcommand{\nul}{\nu_\text{l}}
\newcommand{\nulsat}{\nu_\text{l,sat}}
\newcommand{\nuv}{\nu_\text{v}}
\newcommand{\nuvsat}{\nu_\text{v,sat}}
\newcommand{\Vv}{V_\text{v}}
\newcommand{\Vl}{V_\text{l}}
\newcommand{\cgl}{\text{CG}_\text{l}}
\newcommand{\cgv}{\text{CG}_\text{v}}

\newcommand{\Go}{G_\text{o}}
\newcommand{\Lf}{L_\text{f}}
\newcommand{\of}{\text{OF}}
\newcommand{\ofbar}{\bar{\of}}
\newcommand{\Rf}{R_\text{f}}
\newcommand{\Tc}{T_\text{C}}
\newcommand{\Tcmax}{T_\text{C,max}}
\newcommand{\Wc}{\mathcal{W}_\text{C}}
\newcommand{\mf}{m_\text{f}}
\newcommand{\mdotf}{\dot{m}_\text{f}}
\newcommand{\mdotfin}{\dot{m}_\text{f,in}}
\newcommand{\mdotfout}{\dot{m}_\text{f,out}}
\newcommand{\mdotfbar}{\bar{\dot{m}}_\text{f}}
\newcommand{\mo}{m_\text{o}}
\newcommand{\mdotoin}{\dot{m}_\text{o,in}}
\newcommand{\mdotoout}{\dot{m}_\text{o,out}}
\newcommand{\mv}{m_\text{v}}
\newcommand{\ml}{m_\text{l}}
\newcommand{\mc}{m_\text{C}}

\newcommand{\Lv}{L_\text{v}}
\newcommand{\Ll}{L_\text{l}}
\newcommand{\pc}{p_\text{C}}
\newcommand{\rf}{r_\text{f}}
\newcommand{\rfzero}{r_\text{f,0}}
\newcommand{\rdotf}{\dot{r}_\text{f}}
\newcommand{\tburn}{t_\text{burn}}
\newcommand{\xc}{\mbf{x}_\text{C}}
\newcommand{\Vc}{V_\text{C}}
\newcommand{\Vport}{V_\text{port}}
\newcommand{\Vpost}{V_\text{post}}
\newcommand{\Vpre}{V_\text{pre}}
\newcommand{\rhof}{\rho_\text{f}}
\newcommand{\cgf}{\text{CG}_\text{f}}
\newcommand{\cggas}{\text{CG}_\text{gas}}

\newcommand{\Ae}{A_\text{e}}
\newcommand{\At}{A_\text{t}}
\newcommand{\rexit}{r_\text{e}}
\newcommand{\rt}{r_\text{t}}
\newcommand{\Me}{M_\text{e}}
\newcommand{\Mt}{M_\text{t}}
\newcommand{\mftot}{m_\text{f,tot}}
\newcommand{\Te}{T_\text{e}}

\newcommand{\mdotn}{\dot{m}_\text{n}}
\newcommand{\pinf}{p_\infty}
\newcommand{\pe}{p_\text{e}}
\newcommand{\ve}{v_\text{e}}
\newcommand{\isp}{I_\text{sp}}
\newcommand{\itot}{I_\text{total}}
\newcommand{\Fmax}{F_\text{max}}
\newcommand{\Favg}{F_\text{avg}}

\newcommand{\Ar}{A_\text{R}}
\newcommand{\Dr}{D_\text{R}}
\newcommand{\Adrogue}{A_\text{drogue}}
\newcommand{\Cddrogue}{C_{d,\text{drogue}}}
\newcommand{\Amain}{A_\text{main}}
\newcommand{\Cdmain}{C_{d,\text{main}}}
\newcommand{\Dchute}{D_\text{chute}}
\newcommand{\mdry}{m_\text{dry}}
\newcommand{\mr}{m_\text{R}}
\newcommand{\mrzero}{m_\text{R,0}}
\newcommand{\vr}{v_\text{R}}

\newcommand{\xr}{\mbf{x}_\text{R}}
\newcommand{\zr}{z_\text{R}}
\newcommand{\zrzero}{z_\text{R,0}}

\newcommand{\rhoa}{\rho_\text{a}}
\renewcommand{\th}{\theta}

\newcommand{\prun}{p_\text{run}}
\newcommand{\tfeed}{t_\text{feed}}
\newcommand{\tv}{t_\text{v}}
\newcommand{\tpurge}{t_\text{purge}}
\newcommand{\ppurge}{p_\text{purge}}

\title{
{A computational model for the design of a nitrous oxide--paraffin wax hybrid rocket engine}\\~\\~\\
{\large Joel Jean-Philyppe, McGill Rocket Team \\ 
	McGill University, Montreal, Canada \\ 
    January 2023 \\~\\~\\
	{A Technical Report}}\\~\\
}

\author{\textcopyright Joel Jean-Philyppe, January 2023}
\date{}

\begin{document}
\maketitle


\chapter*{Abstract}
\label{sec:abstract}
\addcontentsline{toc}{section}{\nameref{sec:abstract}}

A computational model is developed to assist the design of the first hybrid rocket engine of the McGill Rocket Team, using liquid nitrous oxide as an oxidizer and solid paraffin wax as a fuel. The model is developed in three phases: In the first phase, a steady-state model which neglects transient performance decrease of the hybrid engine is considered. The steady-state model considers: a constant oxidizer mass flow rate; a combustion chamber in chemical equilibrium; a one-dimensional isentropic nozzle; and a one-dimensional constant-thrust rocket ascent. The steady-state model is used to conduct parametric studies on engine performance as a function of design parameters such as: oxidizer mass flow rate, fuel grain dimensions, and nozzle dimensions. The engine design point is selected to optimize specific impulse, given physical and structural constraints of the system. In the second phase, an unsteady model incorporating transient performance decrease of the hybrid engine is considered. The transient model considers: a self-pressurizing oxidizer tank in quasi-steady liquid-vapor equilibrium; a combustion chamber in quasi-steady chemical equilibrium; a one-dimensional non-isentropic nozzle; and a one-dimensional rocket ascent model. The transient performance profile of the engine is established, and the unsteady model is used to predict propellant requirements for the launch vehicle, given a target apogee of 3048~m (10,000~ft). In the third stage, the unsteady model is compared to hot fire testing data of the McGill Rocket Team. Semi-empirical parameters used in the formulation are calibrated against testing data, and the validity of the model is assessed.


\chapter*{Acknowledgements}
\label{sec:ded}
\addcontentsline{toc}{section}{\nameref{sec:ded}} 

The author would like to acknowledge the 2020-2021 and 2021-2022 cohorts of Propulsion Leads, Propulsion Project Leads, and Propulsion Members of the McGill Rocket Team (MRT) for their outstanding work in developing and building the first hybrid engine of the MRT, Maelstr\"{o}m, along with the team's permanent hybrid engine testing facility, in particular, Cyril Mani, Mohammad Ghali, Taj Sobral, Ahmed Raslan, John Issa, Catherine Gaudio, Julian Zhang, Anthony Côté, and Emily Doyle. The author would also like to extend his gratitude to MRT alumni who were pioneers in paving the way to make this possible, in particular, Liem Dam-Quang, Julien Otis-Laperrière, Violet Saathoff, Daniil Lisus, Anthony Ubah, Sandro Papais, Charles Cossette, Kyle Weissman, Jonathan Lesage, Austin L'Écuyer, Thomas Hannaford, and all previous MRT members who contributed to the development of the hybrid engine and hybrid rocket, and the acquisition of the engine testing facility. The author would like to thank: The McGill MacDonald Campus for allowing the MRT to install the testing facility on their premises; The Canadian Student Rocketry community for their input in test site design and safety; And Prof. Andrew J. Higgins for his administrative support of the team. The author would also like to thank all members from the MRT in the 2021-2022 cohort for their remarkable achievement of developing and building the first hybrid engine-powered launch vehicle of the team, Athos, and the 2022-2023 cohort for continuing the development of the hybrid program. 

Finally, the author would like to extend his gratitude to all MRT collaborators and sponsors without whom this project would not have been possible, in particular, McGill University, the Engineering Undergraduate Society of McGill, Swagelok, Altium, FLIR Teledyne, Convergent, the Mechanical Engineering Student Center of McGill, FLO-Tite, ESI Group, Gestion Férique, 3M, Deploy Depot, Texonic, Raybel, ZOOK, Digikey, ASCO, Brown \& Miller Racing Solutions, Airtech, Composites Canada, Maxiil, Electro-Meters, ChemTrend, Huron Alloys Inc., Hawkeye Industries, Rock West Composites, The International Group Inc., Digi-Key, and M1 Composites Technologies.


\tableofcontents


\chapter*{Abbreviations}
\label{sec:abbreviations}
\addcontentsline{toc}{section}{\nameref{sec:abbreviations}}
     
\begingroup
\begin{xltabular}[l]{\linewidth}{l l} 
	 AGL & Above-ground-level \\
	 BC & Body calipers \\
	 CAD & Computer-aided design \\
	 CEA & Chemical Equilibrium with Applications \\
	 CG & Center of gravity \\     
	 COTS & Commercial-off-the-shelf \\
	 CP & Center of pressure \\
	 CV & Control volume \\
	 HF & Hot fire \\
	 IREC & International Rocket Engineering Competition \\
	 LC & Launch Canada Challenge \\
     MRT & McGill Rocket Team \\
	 \nitrogen{} & Nitrogen \\
	 \nitrous{} & Nitrous oxide \\
     NASA & National Aeronautics and Space Administration \\
     ODE & Ordinary differential equation \\
     OF & Oxidizer-to-fuel \\
     PROPEP & Propellant Evaluation Program \\
     RDx & Reaction Dynamics \\
     SAC & Spaceport America Cup \\
     SRAD & Student-researched-and-developed
     
\end{xltabular}
\endgroup


\chapter*{Nomenclature}
\label{sec:nomenclature}
\addcontentsline{toc}{section}{\nameref{sec:nomenclature}}

\textbf{Symbols} 

\begingroup
\begin{xltabular}[l]{\linewidth}{l l} 
	 
	 $\gamma$ & Heat capacity ratio \\
	 $\theta$ & Rocket launch angle \\
	 $\nul$ & Oxidizer liquid molar volume \\
	 $\nulsat$ & Oxidizer liquid saturated molar volume \\
	 $\nuv$ & Oxidizer vapor molar volume \\
     $\nuvsat$ & Oxidizer vapor saturated molar volume \\
     $\rhoa$ & Air density \\
	 $\rhof$ & Fuel density \\
	 $\Adrogue$ & Drogue parachute surface area \\
	 $\Ae$ & Nozzle exit area \\
	 $\Ai$ & Injector hole area (per hole) \\
	 $\Amain$ & Main parachute surface area \\
	 $\Ar$ & Rocket frontal area \\
	 $\At$ & Nozzle throat area \\
	 $a$ & Regression rate scaling constant \\
	 $C_d$ & Rocket drag coefficient \\
	 $\Cddrogue$ & Drogue parachute drag coefficient \\
	 $\Cdmain$ & Main parachute drag coefficient \\
	 $\Ci$ & Injector discharge coefficient \\
	 $\cpv$ & Oxidizer vapor heat capacity at constant pressure \\
	 $\cvv$ & Oxidizer vapor heat capacity at constant volume \\
	 $D$ & Rocket drag force \\
	 $\Dchute$ & Parachute drag force \\
	 $\DDip$ & Dip tube external diameter \\
	 $\Dr$ & Rocket external diameter \\
	 $\dDip$ & Dip tube internal diameter \\	 
	 $\dTank$ & Tank internal diameter \\
	 $F$ & Thrust \\
	 $F_g$ & Gravitational force \\
	 $\Fmax$ & Peak thrust \\
	 $\Favg$ & Average thrust \\
	 $\Go$ & Oxidizer mass flux rate \\ 
	 $g$ & Gravitational constant \\
	 $\gsl$ & Gravitational constant sea level \\
	 $\ho$ & Oxidizer molar enthalpy \\
	 $\hosat$ & Oxidizer saturated molar enthalpy \\
	 $\isp$ & Specific impulse \\
	 $\itot$& Total impulse \\
	 $\LDip$ & Dip tube length \\
	 $\Lf$ & Fuel length \\
	 $\LTank$ & Tank internal length \\
	 $M_1$ & Nozzle first critical Mach number \\
     $M_2$ & Nozzle second critical Mach number, shockwave outlet \\
	 $M_{2x}$ & Nozzle second critical Mach number, shockwave inlet \\
	 $\Me$ & Nozzle exit Mach number \\
	 $\Mt$ & Nozzle throat Mach number \\
	 $\mdry$ & Rocket dry mass \\
	 $\mf$ & Fuel mass storage in combustion chamber \\
	 $\mftot$ & Fuel total mass in propulsion system \\
	 $\mo$ & Oxidizer mass storage in combustion chamber \\
	 $\motot$ & Oxidizer total mass in propulsion system \\
	 $\mr$ & Rocket total mass \\
	 $\mrzero$ & Rocket initial total mass (pad mass) \\
	 $\mdotf$ & Fuel mass flow rate \\
	 $\mdotfin$ & Fuel mass flow rate entering chamber \\
	 $\mdotfout$ & Fuel mass flow rate exiting chamber \\
	 $\mdotfbar$ & Average fuel mass flow rate \\
	 $\mdotn$ & Nozzle total propellant mass flow rate \\
	 $\mdoto$ & Oxidizer mass flow rate \\
  	 $\mdotoin$ & Oxidizer mass flow rate entering chamber \\
	 $\mdotoout$ & Oxidizer mass flow rate exiting chamber \\
 	 $N$ & Augmented regression rate exponent \\
	 $\Ni$ & Injector number of holes \\
	 $m$ & Oxidizer polytropic exponent \\
	 $\mc$ & Chamber gaseous mass storage \\
	 $n$ & Regression rate exponent \\
	 $\nl$ & Tank oxidizer liquid moles \\
	 $\nv$ & Tank oxidizer vapor moles \\
	 $\ndoto$ & Oxidizer molar flow rate \\
	 $p_{0,y}$ & Nozzle stagnation pressure at shockwave outlet\\
	 $p_1$ & Nozzle first critical pressure \\
	 $p_2$ & Nozzle second critical pressure, shockwave outlet \\
	 $p_{2x}$ & Nozzle second critical pressure, shockwave inlet \\
	 $\pinf$ & Ambient pressure \\
	 $\pc$ & Chamber pressure \\
	 $\pe$ & Nozzle gas exit pressure \\
	 $\ploss$ & Feed pressure loss \\
	 $\ppurge$ & Nitrogen purge pressure \\
	 $\prun$ & Run line (pre-injector) pressure \\
	 $\pt$ & Tank pressure \\
	 $\ptzero$ & Tank initial pressure \\
	 $\psat$ & Saturated pressure \\
	 $\Ru$ & Universal gas constant \\
	 $\Rf$ & Fuel external radius \\
	 $\re$ & Earth radius \\
	 $\rexit$ & Nozzle exit radius \\
	 $\rf$ & Fuel internal radius \\
	 $\rfzero$ & Fuel grain initial internal radius \\
	 $\rt$ & Nozzle throat radius \\
	 $\rdotf$ & Fuel grain regression rate \\
	 $\Tc$ & Chamber temperature \\
	 $\Tcmax$ & Chamber peak temperature \\
	 $\Te$ & Nozzle gas exit temperature \\
	 $\Tt$ & Tank temperature \\
	 $\Ttzero$ & Tank initial temperature \\ 
	 $t$ & Time \\
	 $\tburn$ & Burntime \\
	 $\tfeed$ & Timestamp of feed line filling \\
	 $\tpurge$ & Timestamp of nitrogen purge \\
	 $\tv$ & Timestamp of vapor blowdown \\
	 $U$ & Tank ullage factor \\
	 $\ul$ & Oxidizer liquid molar internal energy \\
	 $\ulsat$ & Oxidizer liquid saturated molar internal energy \\
	 $\uv$ & Oxidizer vapor molar internal energy \\
	 $\uvsat$ & Oxidizer vapor saturated molar internal energy \\
	 $\Vc$ & Chamber volume (fuel port internal volume) \\
	 $\Vl$ & Oxidizer liquid volume \\
	 $\Vpost$ & Post-chamber volume \\
	 $\Vpre$ & Pre-chamber volume \\
	 $\Vt$ & Tank internal volume \\
	 $\Vv$ & Oxidizer vapor volume \\
	 $\ve$ & Nozzle gas exit velocity \\
	 $\vr$ & Rocket total velocity \\
	 $\Wc$ & Chamber gas molar weight \\
	 $\Wo$ & Oxidizer molar weight \\
	 $Z$ & Oxidizer vapor compressibility factor \\
	 $\zr$ & Rocket elevation above sea level \\
	 $\zrzero$ & Rocket initial elevation above  sea level \\
	 
\end{xltabular}
\endgroup

\noindent \textbf{Subscripts} 

\begingroup
\begin{xltabular}[l]{\linewidth}{l l} 
	 
	 0 & Initial \\
	 a & Air \\
	 C & Combustion chamber \\
	 dt & Diptube \\
	 E & Earth \\
	 e & Nozzle exit plane \\
	 f & Fuel \\
	 $g$ & Gravitational \\
	 i & Injector \\
	 in & Inlet \\
	 l & Oxidizer liquid \\
	 n & Nozzle \\
	 o & Oxidizer \\
	 out & Outlet \\
	 R & Rocket \\
	 SL & Sea level \\
     sat & Saturated \\
	 T & Tank \\
	 t & Nozzle throat \\
	 v & Oxidizer vapor \\
	 $x$ & Shockwave inlet \\
	 $y$ & Shockwave outlet \\
	 
\end{xltabular}
\endgroup


\listoffigures %


\listoftables


\clearpage 
\pagenumbering{arabic}

\chapter{Introduction}

The development of strong Canadian rocketry expertise is enabled by university-based student-led design teams. These engineering teams have the purpose to build amateur rockets to participate in national and international rocket launch competitions, such as the Launch Canada Challenge (LC) \cite{lc2023} and the Spaceport America Cup (SAC) \cite{sac2023}. In the context of amateur rocketry projects, university students develop expertise in engineering design and project management, along with a knowledge of aerospace and rocketry; these skills translate directly to industry. In fact, an example of a Canadian rocketry company which was founded by university student design team alumni is Reaction Dynamics (RDx) \cite{rdx2023}.

The McGill Rocket Team (MRT) \cite{mrt2022} is the amateur rocket design team associated with McGill University, based in Montreal, Canada. Since its inception in 2014, the team has participated in every iteration of the Intercollegiate Rocket Engineering Competition (IREC), now the SAC \cite{sac2023}. From 2014 to 2021, the MRT designed and flew solid motor-powered launch vehicles, with commercial-off-the-shelf (COTS) motors purchased from industrial suppliers. In its desire to develop expertise in propulsion system design and manufacturing, the MRT began researching and developing a hybrid rocket engine in the 2017-2018 design cycle. The oxidizer chosen was liquid nitrous oxide (\nitrous{}), due to its self-pressurizing capabilities leading to a relatively simple design, and the fuel selected was paraffin wax, due to its ease of accessibility and its high performance as a hybrid engine propellant.

The development of a student-researched-and-developed (SRAD) hybrid rocket engine is a complex process, involving a large number of variables and dynamic systems that interact in a controlled fashion to produce a specific target thrust profile. SRAD engine development is rooted in extensive experimental testing campaigns. However, the number of experimental tests that can be conducted during a design cycle is constrained by time and resources limitations, hence, a necessity to rely on modeling tools to inform experimental design iterations arises. Although commercial software is available to simulate hybrid engine performance, the MRT chose to develop its own in-house model for several reasons. In fact, the MRT has tight budgetary constraints, rendering inaccessible a majority of commercial software, while in-house codes based on MATLAB or Python are accessible freely across several generations of students. But ultimately, the decision to develop an in-house code is primarily rooted in a desire to: Develop a model which is specifically tailored to the needs of the MRT and easily adaptable to future engine iterations; Build modeling expertise within the team; And build a solid theoretical understanding of hybrid rocket engine operation.

The current work presents the hybrid engine computational model which was developed by the author over the course of four years of involvement with the MRT, to guide the design of the first hybrid engine built by the team, \engine{}. The model was developed in parallel with the construction of a permanent rocket engine testing facility. The computational tool relies on simple thermodynamic assumptions, which allows to simplify hybrid engine modeling complexity to a minimum, while maintaining sufficient accuracy to remain a useful design tool. It is assumed that the reader has a basic understanding of: thermodynamic principles involved in rocket propulsion system operation; hybrid propulsion system design and operation; and the thermophysical properties of \nitrous{}. Some excellent resources to acquire such knowledge are Ref.~\cite{genevieve2013, newlands2011a, newlands2011b, newlands2017, newlands2019, sutton2016}.

The current work is structured as follows. In Chapter~2, the first iteration of the hybrid engine model--a simple steady-state formulation--is detailed. This initial design tool is used to conduct design space studies and select target performance parameters. In Chapter~3, the second iteration of the hybrid engine model--an unsteady model--is formulated. The unsteady model is used to assess the transient performance of the engine, and to determine propellant requirements given vehicle characteristics and target apogee. In Chapter~4, the unsteady model is calibrated against experimental engine testing data, and the validity of the model is assessed. Finally, in Chapter~5, possible sources of errors, pathways for model improvements, and concluding remarks are provided.



\chapter{Steady-State Model} \label{chapter:steady}

The initial design of the first hybrid engine of the McGill Rocket Team, \engine{}, is assisted with an ideal steady-state model. The steady-state model neglects transient effects which decrease hybrid engine performance, such as the self-pressurizing blowdown process of the oxidizer tank, or the possible formation of a shockwave in the nozzle. The steady-state model allows to determine overall propulsion system geometry, target design and performance parameters, and an initial estimate of the propellant quantities required to bring the vehicle to target apogee. The current chapter details the formulation and results of the steady-state model.


\section{Model Formulation} \label{sec:steady_model}


\subsection{Model Description}

\begin{figure}
    \centering
    \includegraphics[width=0.8\textwidth]{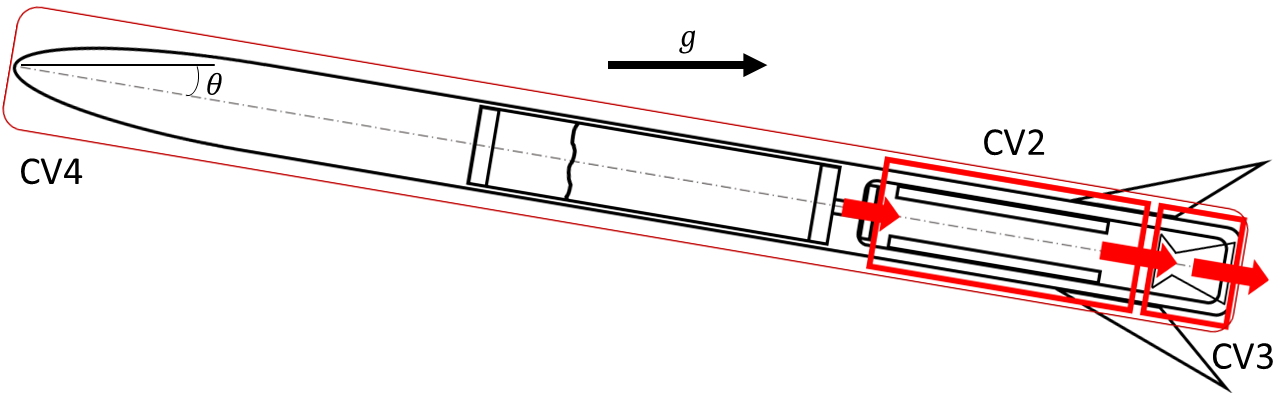}
    \caption{Steady-state hybrid rocket model overview.}
    \label{fig:overviewsteady}
\end{figure}

The steady-state model considers a sequential, non-coupled analysis of the components of the propulsion system, which is segregated into three major control volumes (CV), while a fourth CV encapsulates the rocket, as depicted in Fig.~\ref{fig:overviewsteady}. The steady-state formulation does not explicitly model CV1, the oxidizer tank; instead, the oxidizer is assumed to be delivered to CV2, the combustion chamber, at a constant mass flow rate $\mdoto$. In CV2, the solid fuel grain pyrolosis is analyzed, and the constant oxidizer mass flow rate allows to determine the average oxidizer-to-fuel (OF) mass ratio, $\ofbar$, analytically. The average OF ratio and the design chamber pressure, $\pc$, are input into the Propellant Evaluation Program (PROPEP) chemical equilibrium code \cite{brown1995} to calculate the thermodynamic properties of the combustion gases. These properties are used to determine the dimensions of the ideal, isentropic nozzle, which is modeled in CV3. The resulting nozzle thrust is computed and used to model the one-dimensional ascent of the hybrid rocket, which is encapsulated by CV4.


\subsection{Equilibrium Combustion Chamber}

\begin{figure}[H]
    \centering
    \includegraphics[width=0.4\textwidth]{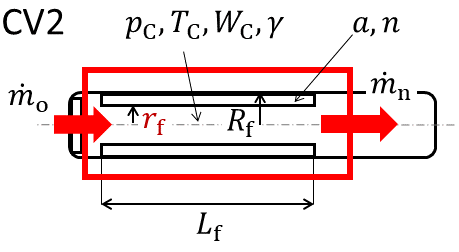}
    \caption{Control volume 2: steady-state combustion chamber.}
    \label{fig:cv2steady}
\end{figure}

A hollow cylindrical fuel grain is considered, and its combustion process in CV2 is depicted in Fig.~\ref{fig:cv2steady}. The semi-empirical hybrid fuel grain regression rate theory is applied to model the process \cite{genevieve2013}. Assumptions used in the formulation include:
\begin{enumerate}
\item The mass flow rate of oxidizer from CV1 to CV2, $\mdoto$, is constant.
\item The combustion chamber pressure, $\pc$, is temporally and axially constant.
\item The fuel grain regression is axially uniform.
\item The combustion gases obey the ideal gas law.
\item The combustion chamber walls are adiabatic.
\end{enumerate}

The fuel grain regression rate semi-empirical expression yields \cite{genevieve2013},
\beq
\f{\d \rf}{\dt} = a \Go^n \equiv a \bigg(\f{\mdoto}{\pi \rf^2}\bigg)^n 
\eeq
where $\rf = \rf(t)$ is the fuel grain port internal radius, $\Go = \Go(t)$ is the oxidizer mass flux rate through the internal port, $a$ is the regression rate scaling constant, and $n$ is the regression rate exponent. Separating variables and integrating in time yields:
\beq \label{eq:rfsteady}
\rf(t) = \bigg[ a N \bigg( \f{\mdoto}{\pi} \bigg)^n t + \rfzero^N \bigg]^{\f{1}{N}}
\eeq
where $ N = 2n+1$ is the augmented regression rate exponent, and $\rfzero$ is the initial fuel port internal radius. Since $\rf(t = \tburn) = \Rf$, where $\tburn$ is the fuel gain burntime, and $\Rf$ is its external radius, Eq.~(\ref{eq:rfsteady}) can be used to show:
\beq
\tburn = \f{1}{aN} \bigg( \f{\pi}{\mdoto} \bigg)^n \big[\Rf^N - \rfzero^N \big].
\eeq
Using $\mf(t) = \pi \rhof \Lf (\Rf^2 - \rf^2)$, where $\mf(t)$ is the remaining fuel grain mass in the chamber over time, $\rhof$ is the fuel grain density, and $\Lf$ is the fuel grain length, the mass flow rate of fuel is:
\beq
\abs{\mdotf(t)} = 2 \pi \rhof \Lf \rf \rdotf.
\eeq
Expanding with the help of Eq.~(\ref{eq:rfsteady}) and its time derivative:
\beq
\abs{\mdotf(t)} = 2 a \pi \rhof \Lf \bigg( \f{\mdoto}{\pi} \bigg)^n \bigg[ a N \bigg(\f{\mdoto}{\pi} \bigg)^n t + \rfzero^N \bigg]^{\f{1-2n}{N}}.
\eeq
The average fuel grain mass flow rate over the burntime is then:
\beqarr
\mdotfbar &=& \f{1}{\tburn} \int_0^{\tburn} \abs{\mdotf(t)} \dt \\
&=& a \pi N \rhof \Lf \bigg[ \f{\Rf^2 - \rfzero^2}{\Rf^N - \rfzero^N} \bigg] \bigg(\f{\mdoto}{\pi} \bigg)^n.
\eeqarr
Hence, the average $\of$ ratio during the combustion is:
\beq \label{eq:ofbarsteady}
\ofbar = \f{\mdoto}{\mdotfbar} \equiv \f{1}{a N \rhof \Lf} \bigg( \f{\mdoto}{\pi} \bigg)^{1-n} \bigg[ \f{\Rf^N - \rfzero^N}{\Rf^2 - \rfzero^2} \bigg].
\eeq
The combustion chamber is assumed to attain chemical equilbrium, and the PROPEP chemical equilibrium code \cite{brown1995} is used to compute the average thermodynamic and physical properties of the combustion gases in the chamber: the temperature, $\Tc$, the molar weight, $\Wc$, and the heat capacity ratio, $\gamma$. These parameters are used as an input to design the ideal isentropic nozzle.


\subsection{Ideal Isentropic Nozzle} \label{sec:steady_nozzle}

\begin{figure}[H]
    \centering
    \includegraphics[width=0.38\textwidth]{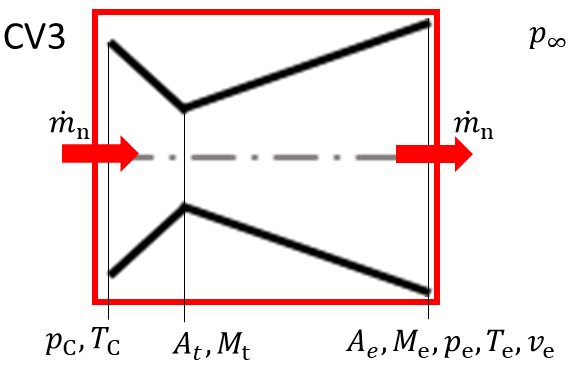}
    \caption{Control volume 3: steady-state isentropic nozzle.}
    \label{fig:cv3steady}
\end{figure}

In CV3, an ideal, isentropic, converging-diverging supersonic nozzle is considered \cite{higgins2021, sutton2016}, as depicted in Fig.~\ref{fig:cv3steady}. Assumptions used in the formulation include:
\begin{enumerate}
\item The nozzle flow is one-dimensional; boundary layer flow effects are neglected.
\item The gaseous mixture is homogeneous.
\item The nozzle gas is in frozen equilibrium, hence its composition does not change across the nozzle axis.
\item The fluid obeys the ideal gas law and is calorically perfect, hence the heat capacity ratio is constant.
\item The nozzle walls are adiabatic.
\item The inlet gas velocity is negligible compared to the exit gas velocity.
\end{enumerate}

Given the chamber thermodynamic properties, the nozzle ideal throat area, $\At$, is determined \cite{higgins2021},
\beq
\At = \f{\mdotn}{\pc \Mt} \sqrt{ \f{\Ru \Tc}{\gamma \Wc} } \bigg( 1 + \f{\gamma-1}{2} \Mt^2 \bigg)^{\f{\gamma+1}{2(\gamma-1)}}
\eeq
where $\mdotn = \mdoto + |\mdotfbar|$ is the total average mass flow rate through the nozzle, $\Mt = 1$ is the gas Mach number at the throat plane, and $\Ru$ is the universal gas constant. Given the ambient pressure, $\pinf$, and $\pe = \pinf$, where $\pe$ is the nozzle gas pressure at the exit plane, the nozzle gas exit Mach number $\Me$, temperature $\Te$, and velocity $\ve$ are \cite{higgins2021}:
\beqarr
\Me &=& \sqrt{ \f{2}{\gamma-1} \bigg[ \bigg( \f{\pc}{\pe} \bigg) ^ \f{\gamma-1}{\gamma} - 1 \bigg] } \\
\Te &=& \Tc \bigg( 1 + \f{\gamma-1}{2}\Me^2 \bigg)^{-1} \\
\ve &=& \Me \sqrt{\f{\gamma \Te \Ru}{\Wc} }.
\eeqarr
Additionally, the nozzle ideal area expansion ratio is \cite{higgins2021},
\beq
\f{\Ae}{\At} = \f{1}{\Me} \bigg[ \bigg( \f{2}{\gamma+1} \bigg) \bigg( 1 + \f{\gamma-1}{2}\Me^2 \bigg) \bigg]^\f{\gamma+1}{2(\gamma-1)}
\eeq
where $\Ae$, the nozzle area at its exit plane, can be determined. The thrust produced by the nozzle, $F$, is then,
\beq
F = \mdotn \ve + (\pe - \pinf) \Ae
\eeq
and the specific impulse of the engine, $\isp$, is,
\beq
\isp = \f{F}{\mdotn \gsl}
\eeq
where $\gsl$ is the gravitational constant at sea level. 


\subsection{Constant-Thrust Rocket Ascent} \label{sec:steady_ascent}

\begin{figure}[H]
    \centering
    \includegraphics[width=0.8\textwidth]{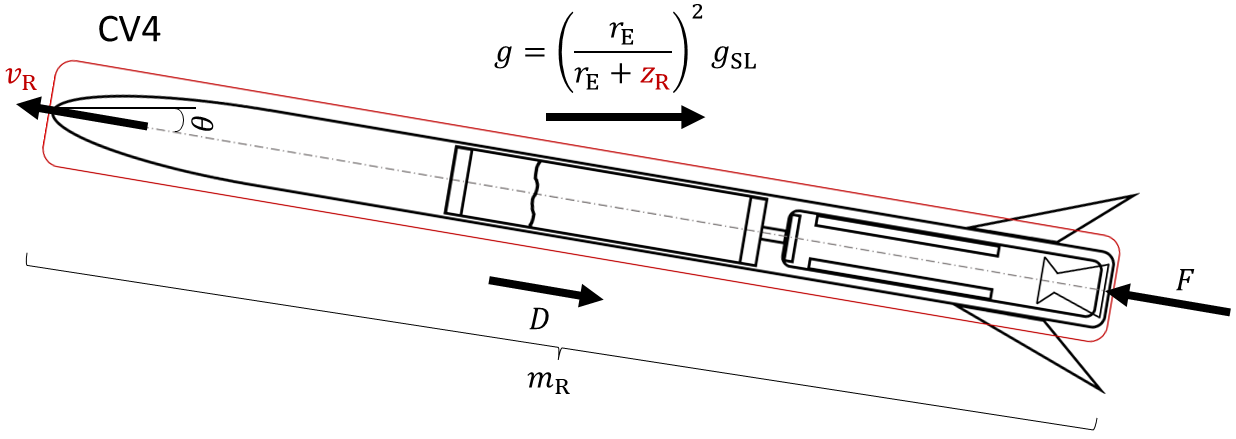}
    \caption{Control volume 4: one-dimensional rocket.}
    \label{fig:cv4steady}
\end{figure}

The rocket is encapsulated by CV4, and its ascent is modeled with a one-dimensional formulation, as depicted in Fig.~\ref{fig:cv4steady}. Assumptions used in the formulation include:
\begin{enumerate}
\item The engine thrust is constant over the burntime.
\item The engine thrust is perfectly aligned with the rocket axis.
\item The flight angle, $\theta$, is temporally constant.
\end{enumerate}
A set of ordinary differential equations are derived and integrated numerically for the rocket total mass, $\mr(t)$, its altitude, $\zr(t)$, and its total velocity, $\vr(t)$.

The conservation of mass yields:
\beq \label{eq:mrsteady}
\f{\d \mr}{\dt} = - \mdotn \Rightarrow \mr(t) = \mrzero - \mdotn t
\eeq
where $\mdotn$ is a constant, $\mrzero = \mdry + \mdotn \tburn$ is the initial total mass of the rocket, and $\mdry$ is the rocket dry mass. The time rate change of the altitude of the rocket is,
\beq \label{eq:zrsteady}
\f{\d \zr}{\dt} = \vr \cos{\th}
\eeq
where $\theta$ is the launch angle with the vertical axis. The conservation of momentum along the axis of the rocket yields \cite{mit2022},
\beq \label{eq:vrsteady}
\f{\d \vr}{\dt} = \f{1}{\mr} [ F - D - F_g \cos{\th} ]
\eeq
where $D$ and $F_g$ are the drag force and the gravitational force acting on the rocket, respectively. The drag force is,
\beq \label{eq:drag}
D = \f{1}{2} C_d \Ar \rhoa \vr^2
\eeq
where $C_d$ is the drag coefficient of the rocket, $\Ar$ is its frontal area, and $\rhoa = \rhoa(\zr)$ is the air density, computed as a function of the rocket altitude with the National Aeronautics and Space Administration (NASA) atmospheric model \cite{hall2021} (Appendix~\ref{a1:thermo}). The gravitational force acting on the rocket is,
\beqarr  \label{eq:gravity}
F_g &=&  \mr g \\
g &=& \bigg( \f{\re}{\re + \zr} \bigg)^2 \gsl
\eeqarr
where $g = g(\zr)$ is the gravitational acceleration as a function of altitude, and $\re$ is the radius of the Earth.

Equations~(\ref{eq:mrsteady})--(\ref{eq:vrsteady}) are valid from $t_0 = 0$ to $\tburn$, at which point the propellants are depleted. Subsequently, $\mdotn = 0$ and $F = 0$, and Eq.~(\ref{eq:zrsteady}) and (\ref{eq:vrsteady}) can be integrated numerically until the apogee of the rocket is attained, hence until $\vr = 0$. The equations are integrated numerically in MATLAB with the ordinary differential equations (ODE) solver \textit{ode45}.


\section{Results: Initial Design of \engine{}}

\subsection{Design Chamber Pressure}

The PROPEP code is used to obtain the performance curve of the propellant mixture as a function of the OF ratio and the chamber pressure. An ideally-expanded isentropic nozzle is considered (Section~\ref{sec:steady_nozzle}). The gases are expanded to ambient pressure at 1402~m (4600~ft) altitude, the ground level at the SAC; this corresponds to an exit gas pressure of 0.846~atm, as computed with the NASA atmospheric model (Appendix~\ref{a1:thermo}). Design chamber pressures between 20.7-34.5~bar (300-500~psia) are considered.

\begin{figure}[h]
    \centering
    \includegraphics[width=0.5\textwidth]{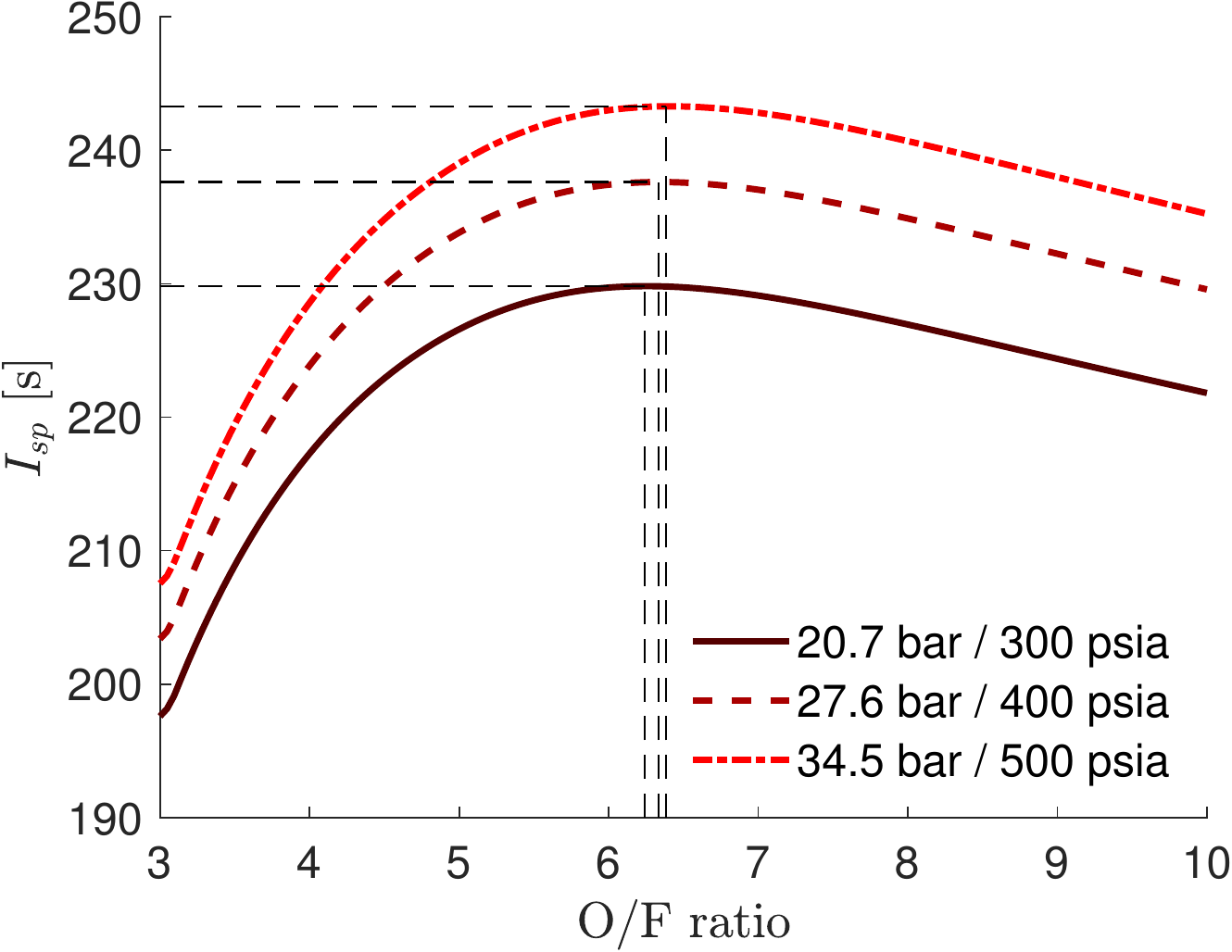}
    \caption{Specific impulse of the \nitrous{}--paraffin wax mixture at different chamber pressures and oxidizer-to-fuel mass ratios. The combustion products are expanded to 0.846~atm (12.4~psia).}
    \label{fig:ispvsof}
\end{figure} 

Figure~\ref{fig:ispvsof} shows at a similar OF ratio, an increase in chamber pressure monotonically results in an increase in engine ideal performance, in accordance with literature. However, structural design considerations limit the operating pressure of the system. Additionally, although the steady-state model does not resolve this phenomenon, the propulsion system of \engine{} is not pressure-fed, but rather relies on the self-pressurizing capabilities of \nitrous{} at ambient conditions to establish chamber pressure. Hence, the transient tank-emptying process results in a decrease of the available head pressure in the tank. The $\Delta p$ across the injector is reduced, which limits mass flow rate, and pressure oscillations in the combustion chamber could result in a reversal of the flow direction if $\Delta p$ across the injector becomes insufficient. With these constraints and trade-offs in mind, a design chamber pressure of 27.6~bar (400~psia) is selected, which is below half the vapor pressure of \nitrous{} at 25~\super{$\circ$}C--57.3~bar (831~psia). This ensures the flow through the injector remains choked. The theoretical maximum specific impulse is $\isp = 238$~s, at $\of = 6.34$.

\subsection{Determination of Drag Coefficient} \label{sec:steady_orckt}

\begin{table}
	\small
	\centering
	\caption{Initial steady-state model parameters for determination of drag coefficient.}
	\label{tab:steady_openrocket}
	\begin{tabular}{c|c|c|c|c}
    & \textbf{Parameter} & \textbf{Symbol} & \textbf{Value} & \textbf{Reference} \\ \hline

	\multirow{4}{*}{\textbf{Constants}} 
	& Launch site altitude & $\zrzero$ & 1402~m (4600~ft) & SAC altitude \\
	& Gravitational constant & $\gsl$ & 9.80665~m/s\super{2} & \cite{wikipedia2022} \\
	& Earth radius & $\re$ & 6378.137~km & \cite{williams2021} \\ 
    & Target apogee (AGL) & - & 3048~m (10,000~ft) & SAC competition \\ \hline
	
	\multirow{4}{*}{\textbf{Rocket}} 
	& Dry mass & $\mdry$ & 52.5~kg (ll6~lbs) & Estimated \\
	& Drag coefficient & $C_d$ & 0.75 & \cite{benson2021} \\ 
	& External diameter & $\Dr$ & 157~mm (6.20~in) & MRT engine casing \\
	& Launch angle & $\theta$ & 6\super{$\circ$} & Past competitions \\ \hline
	
	\multirow{7}{*}{\textbf{Chamber}} 	
	& Oxidizer mass flow rate & $\mdoto$ & 2 kg/s & Estimated \\
	& Chamber pressure & $\pc$ & 27.6~bar (400~psia) & Selected \\
    & Fuel density & $\rhof$ & 900 kg/m\super{3} & \cite{genevieve2013} \\
	& Fuel external radius & $\Rf$ & 39.497~mm (1.555~in) & MRT engine casing \\ 
	& Fuel length & $\Lf$ & 813~mm (32.0~in) & MRT engine casing \\
	& Regression rate constant & $a$ & 1.32 x 10\super{-4} & \cite{genevieve2013} \\
	& Regression rate exponent & $n$ & 0.555 & \cite{genevieve2013} \\	 
	
	\end{tabular}
\end{table}

\begin{figure}
    \centering
    \includegraphics[width=0.5\textwidth]{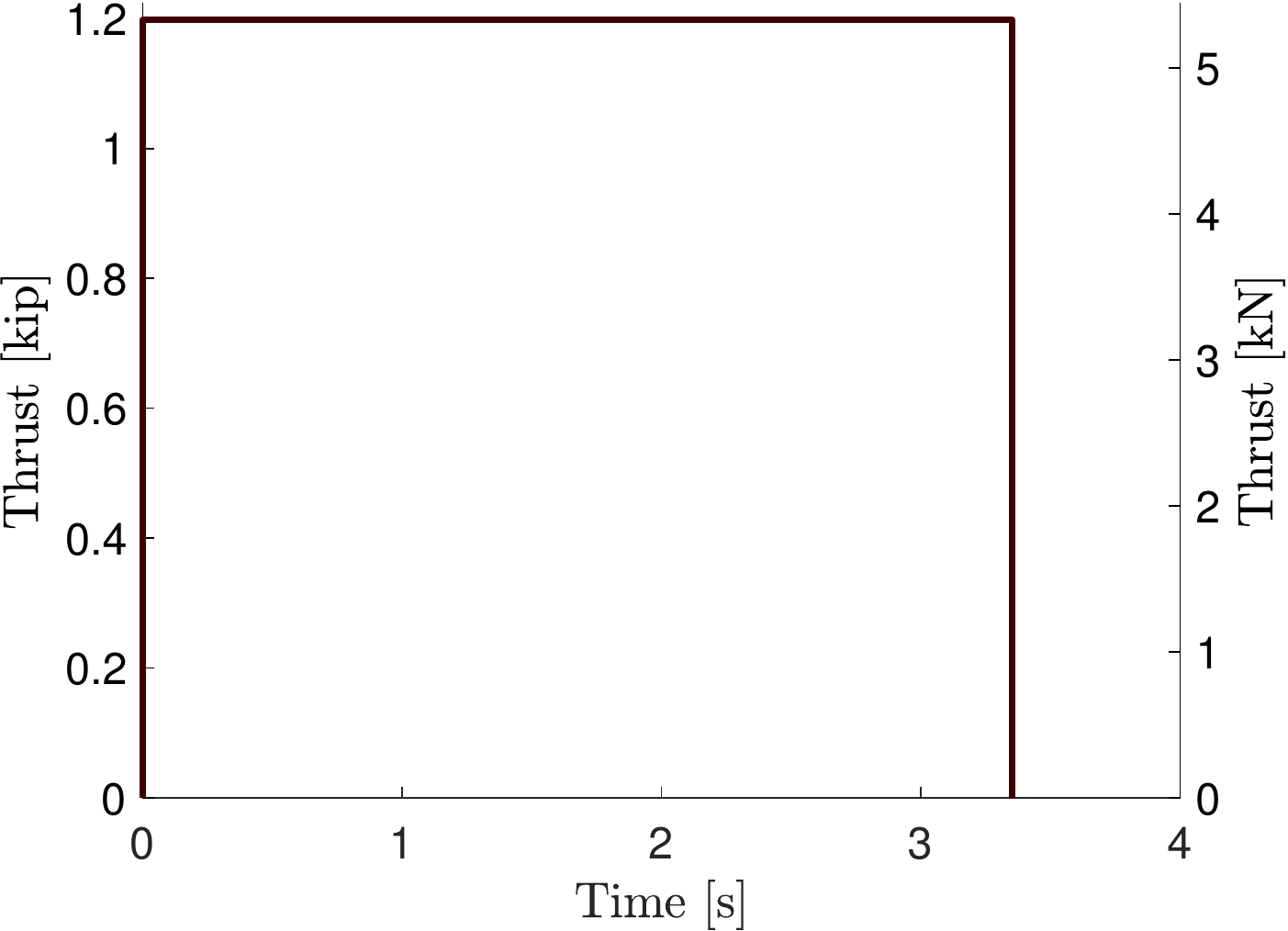}
    \caption{Initial thrust curve based on the steady-state model generated for \orckt{} simulation.}
    \label{fig:orthrust}
\end{figure}

\begin{figure}
    \centering
    \includegraphics[width=\textwidth]{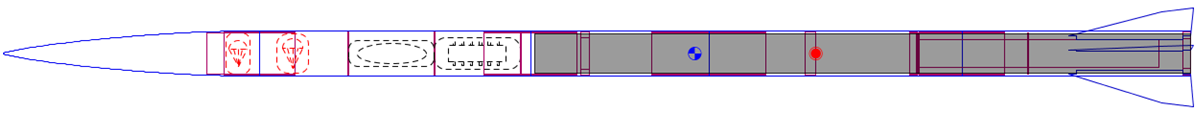}
    \caption{\orckt{} model for \rocket{}. The overall rocket length is 4.24~m (167~in), and the outside diameter is 157~mm (6.20~in).}
    \label{fig:athosorckt}
\end{figure}

A two-way coupling between the steady-state model and the flight simulation software \textit{OpenRocket} \cite{openrocket2021b} is considered to determine the drag coefficient of the vehicle. The mathematical formulation of \textit{OpenRocket} can be found in \cite{niskanen2009}. An initial set of parameters is selected for the steady-state model, as shown in Table~\ref{tab:steady_openrocket}. The fuel mass which allows the rocket to reach the target apogee is resolved iteratively with the steady-state model formulation, keeping all other parameters constant, and is found to be $\mftot = 2313$~g. The resulting engine thrust, burntime, and oxidizer mass are $F = 5.33$~kN, $\tburn = 3.35$~s, and $\motot = \mdoto \tburn = 6707$~g, respectively. These values are used to generate a .eng thrust curve for \orckt{}; the file format is detailed in \cite{thrustcurve2022}, and the thrust curve is shown in Fig.~\ref{fig:orthrust}. Given a model rocket based on an initial design for \rocket{}, shown in Fig.~\ref{fig:athosorckt}, flight simulations are performed in \orckt{}. The drag coefficient as a function of time is exported for 20 simulations--10 with "smooth paint" surface finish, and 10 with "rough paint"--at a wind speed of 16.1 km/h (10 mph). The $C_d$ time-resolved data is then numerically integrated with a trapezoidal method and averaged over the flight duration for the smooth and the regular paint, yielding $C_{d, \text{smooth}}$ and $C_{d, \text{regular}}$, respectively. The value $C_d = 0.75 C_{d, \text{smooth}} + 0.25  C_{d, \text{regular}}$ is then selected to be the average drag coefficient of the vehicle; the smooth paint results are given a larger weight based on previous designs by the MRT. The new value of $C_d$ is input into the steady-state model, and the fuel mass is resolved again. The process to obtain the $C_d$ from \orckt{} is then repeated until the average $C_d$ converges. The value $C_d = 0.529$ is found.

\subsection{Parametric Study and Selection of Geometry}

\begin{figure}[h]
    \centering
    \includegraphics[width=0.9\textwidth]{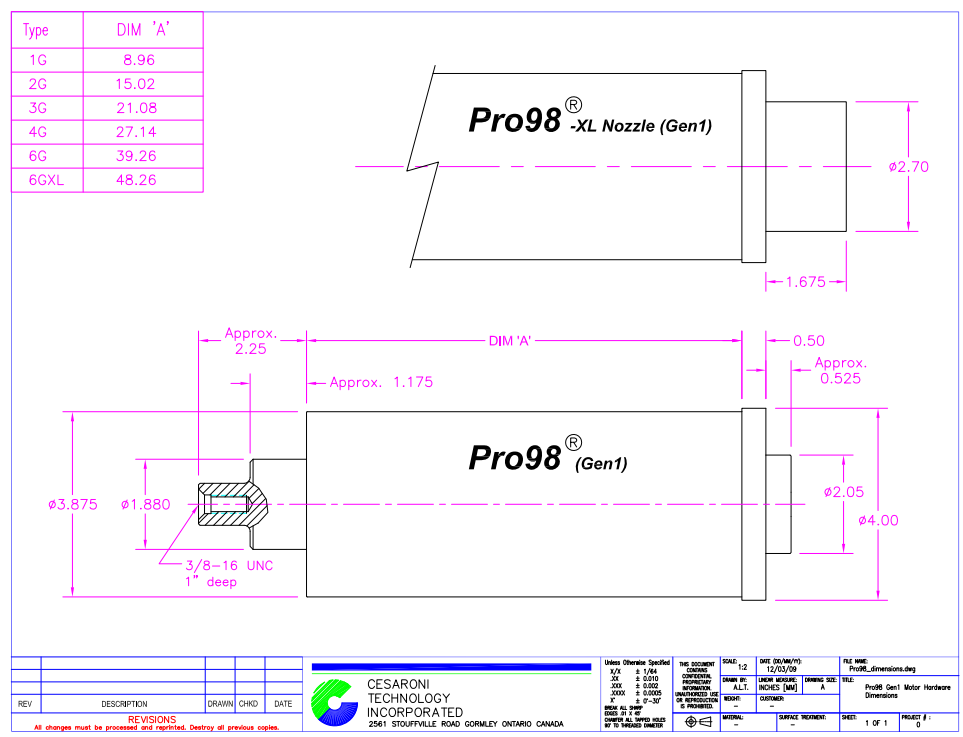}
    \caption{Cesaroni Pro98 6G engine casing specifications. Dimensions provided in inches. From \cite{cesaroni2022}.}
    \label{fig:pro98}
\end{figure}

Design parameters that are to be selected with the steady-state model include: oxidizer mass flow rate, $\mdoto$; fuel length, $\Lf$; and nozzle outlet radius, $\rexit$. The \engine{} engine casing is a Cesaroni Pro98 6G; dimensions of the casing are shown in Fig.~\ref{fig:pro98}. The total length of the casing is 997~mm (39.26~in), and its external diameter is 98.4~mm (3.875~in). Taking into account pre- and post- combustion chamber spacing, engine casing wall thickness, and nozzle wall thickness, the maximum allowable $\Lf$ and $\rexit$--without requiring an external nozzle insert--are $\Lf = 813$~mm (32~in) and $\rexit = 38.1$~mm (1.50~in).

\begin{figure}[h]
	\centering 
	
	\begin{subfigure}{.49\textwidth}
	\centering 
	\includegraphics[width=\textwidth]{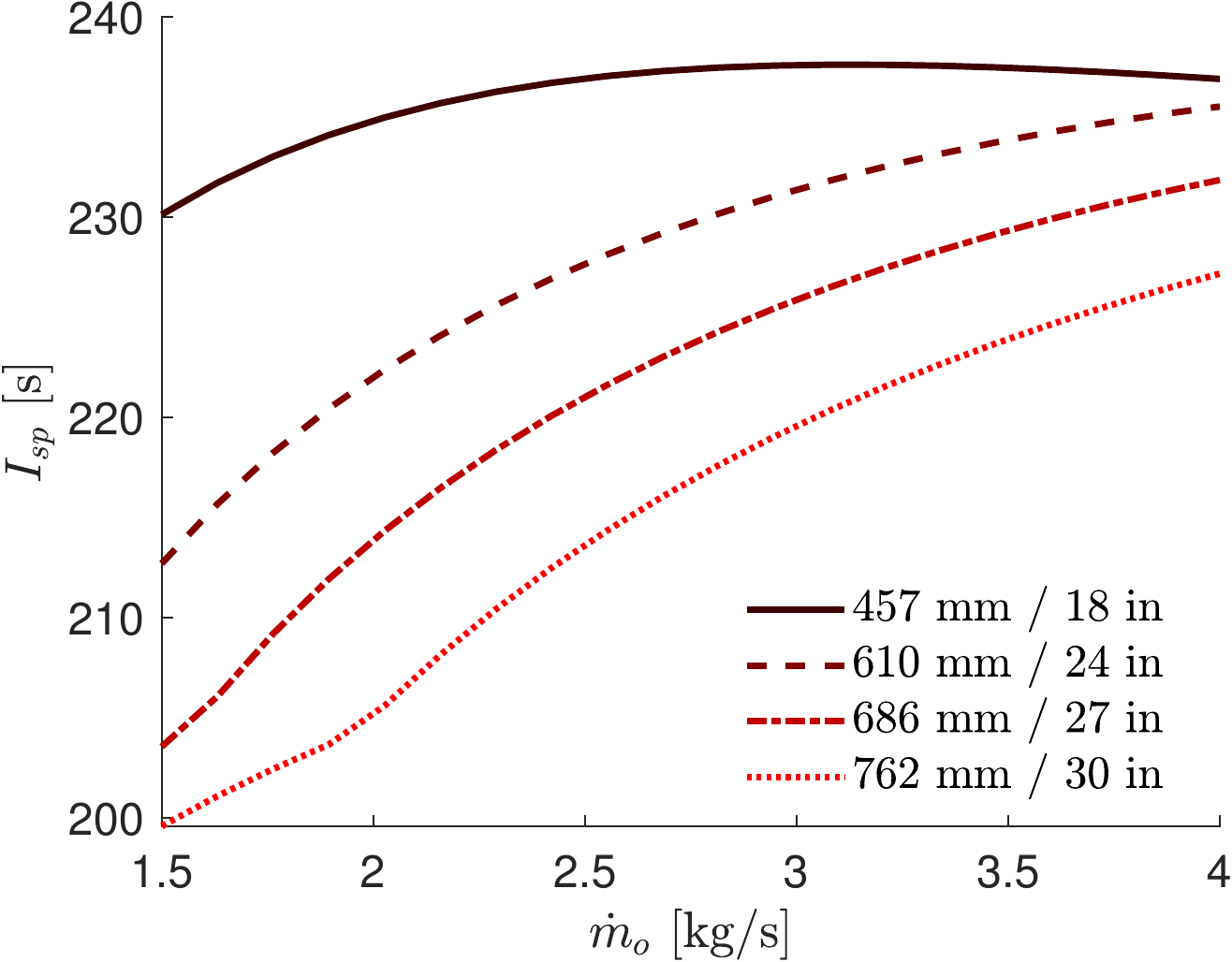}
	\caption{$\isp$ with ideally-expanded nozzle.}
	\label{fig:ispvsmdot_free} 
	\end{subfigure} \hfill
	\begin{subfigure}{.49\textwidth}
	\centering 
	\includegraphics[width=\textwidth]{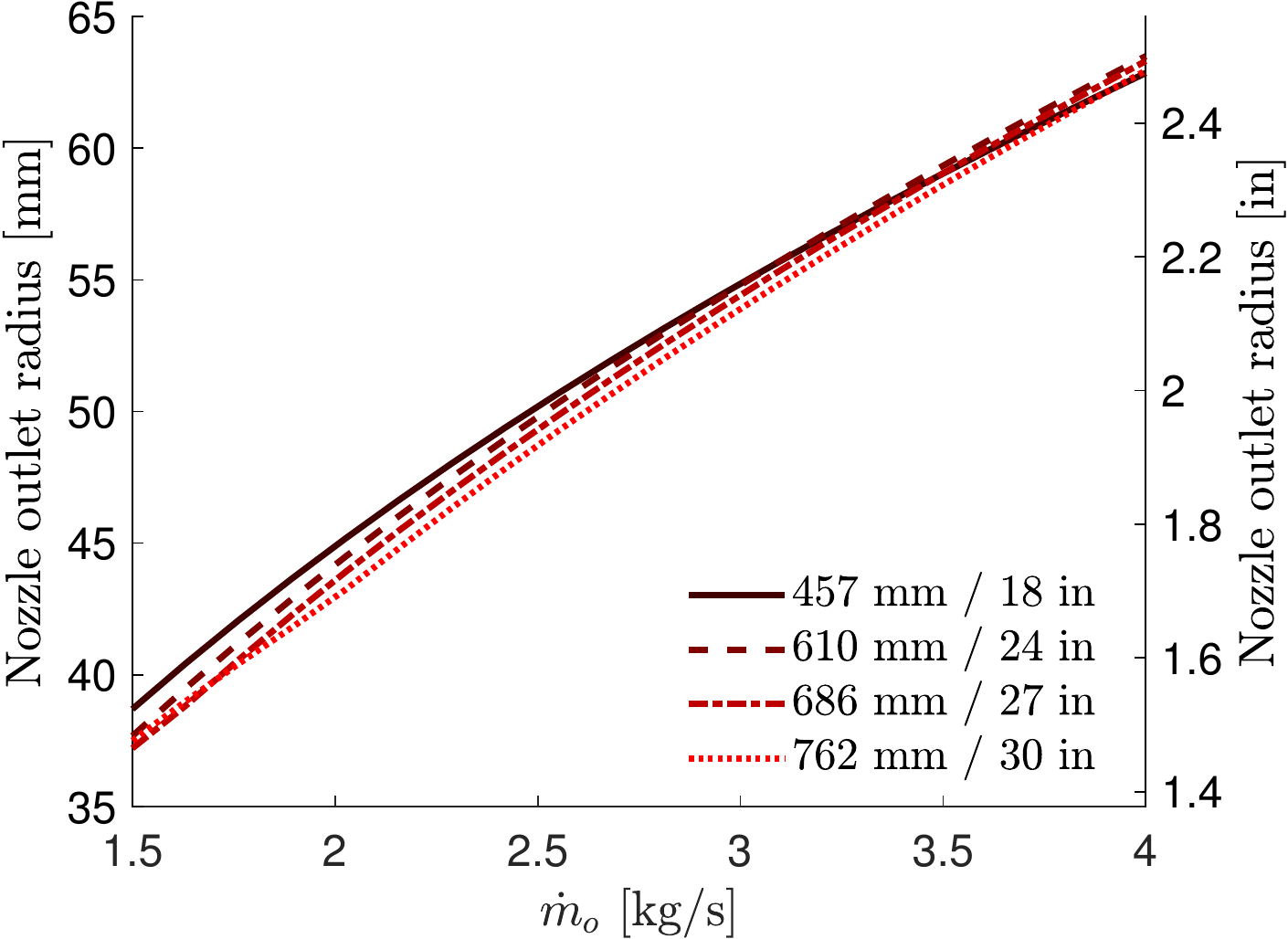}
	\caption{Ideally-expanded nozzle outlet radius.}
	\label{fig:rnozvsmdot_free} 
	\end{subfigure}
	
	\begin{subfigure}{.49\textwidth}
	\centering 
	\includegraphics[width=\textwidth]{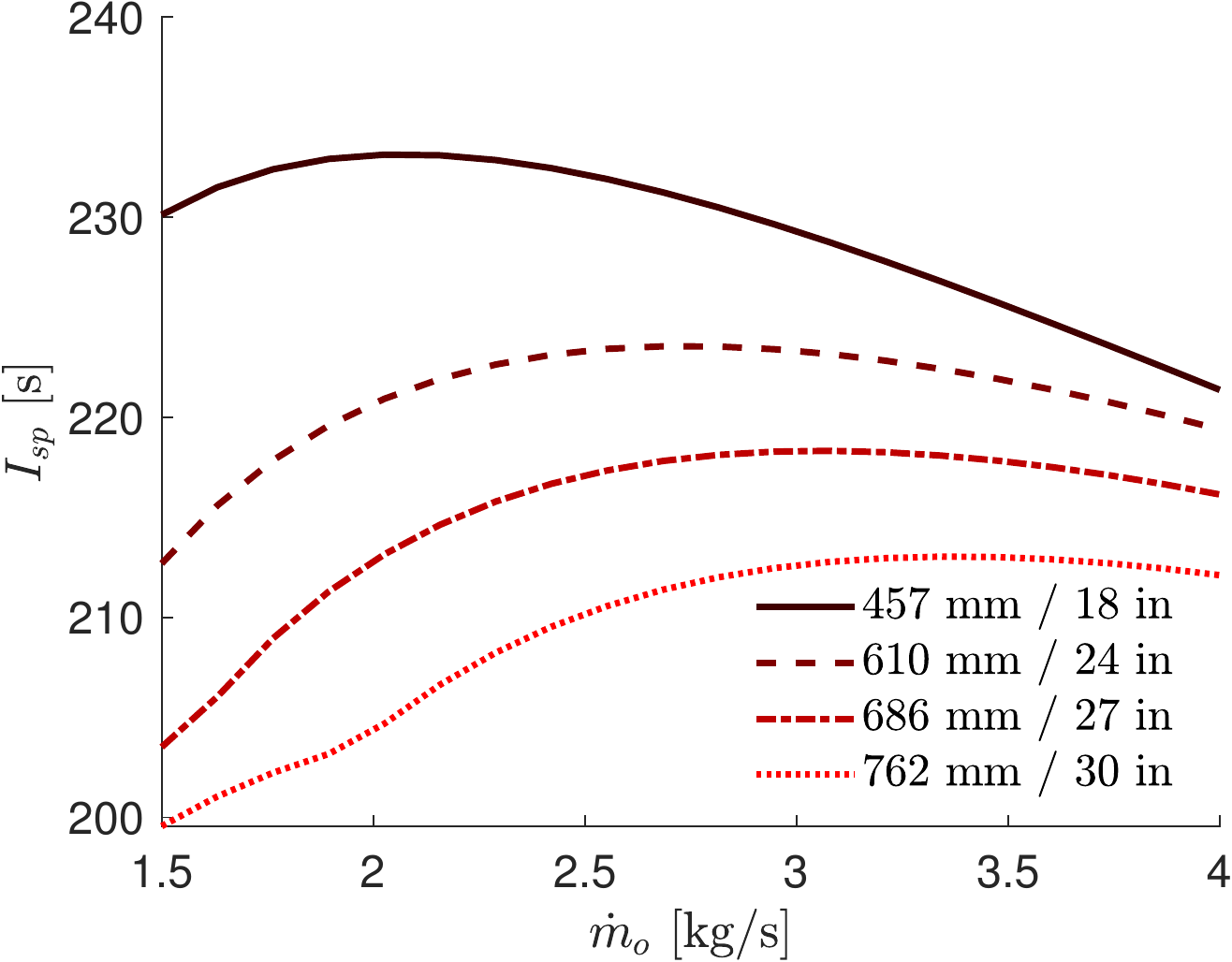}
	\caption{$\isp$ with fixed nozzle outlet radius--38.1~mm (1.50~in).}
	\label{fig:ispvsmdot_const} 
	\end{subfigure} \hfill
	\begin{subfigure}{0.49\textwidth}
	\centering 
	\includegraphics[width=\textwidth]{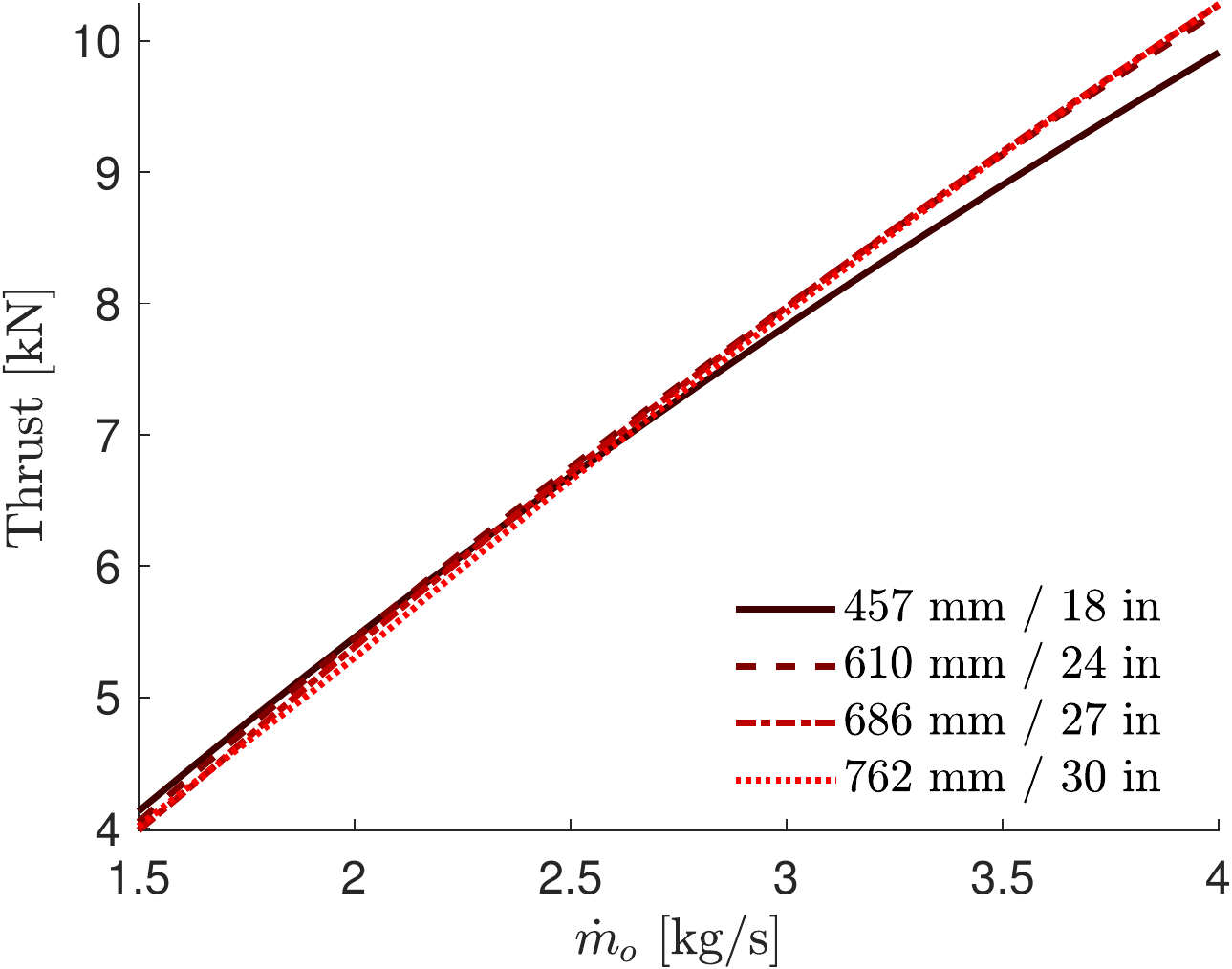}
	\caption{Thrust with fixed nozzle outlet radius--38.1~mm (1.50~in)}
	\label{fig:fvsmdot_const} 
	\end{subfigure}

	\caption{Parameteric study on specific impulse, nozzle outlet radius, and thrust, as a function of oxidizer mass flow rate, at different fuel lengths.}	
	\label{fig:steady_parametric}
\end{figure}

A parametric study on the engine performance as a function of $\mdoto$, $\Lf$, and $\rexit$ is conducted; results are shown in Fig.~\ref{fig:steady_parametric}. For each data point, the fuel grain mass $\mftot$ is computed iteratively such that the rocket always reaches target apogee. Figure~\ref{fig:ispvsmdot_free} shows the $\isp$ of the engine with an ideally-expanded nozzle as a function of oxidizer mass flow rate at different fuel lengths. The $\isp$ is a function of the average OF ratio over the burn, as shown in Fig.~\ref{fig:ispvsof}, and is maximized at $\ofbar = 6.34$ for a 27.6~bar (400~psia) chamber pressure. Equation~\ref{eq:ofbarsteady} shows $\ofbar$ is directly proportional to $(\mdoto)^{1-n}$ and inversely proportional to $\Lf$. In the range of $\mdoto$ considered, a shorter $\Lf$ leads to higher $\isp$, allowing to conclude the curves shown in Fig.~\ref{fig:ispvsmdot_free} are on the fuel-rich side. For $\Lf = 457$~mm (18~in), the curve flattens, which shows the optimal $\ofbar$ is attained and the $\isp$ can be maximized with this shorter fuel grain length. However, it should be noted that the regression rate constants $a$ and $n$ used in Eq.~\ref{eq:ofbarsteady} are taken from literature and are subject to uncertainty. Hence, an excessive shortening of the fuel grain length could lead to an inability to load a sufficient mass of fuel to reach target apogee if $a$ and $n$ are found to be largely inaccurate during engine testing campaigns. A conservative approach should therefore be prioritized. 

Another practical limitation is the maximum allowable $\rexit$ which does not require a nozzle insert. Figure~\ref{fig:rnozvsmdot_free} shows the ideal nozzle outlet radius as a function of $\mdoto$, and it exceeds 38.1~mm (1.50~in) for $\mdoto > 1.75$~kg/s. Hence, a performance gain against design complexity trade-off analysis is conducted. The $\isp$ for a (non-ideal) nozzle expanded to a fixed outlet radius of 38.1~mm (1.50~in) is calculated and plotted in Fig.~\ref{fig:ispvsmdot_const}. The peak performance obtained with $\Lf = 457$~mm (18~in) is reduced from 238~s to 233~s, while the total impulse required by the rocket is around 16,500~Ns. At 7~kN thrust, a reduction of $\isp$ by 5~s leads to an increase in propellant weight by 70~g, which is negligible considering the rocket dry mass is 52.5~kg. Hence, the performance penalty is minimal, and a non-ideally expanded nozzle is therefore preferred, avoiding additional design complexity of a nozzle insert. 

\begin{figure}
	\centering 
	\begin{subfigure}{.49\textwidth}
	\centering 
	\includegraphics[width=\textwidth]{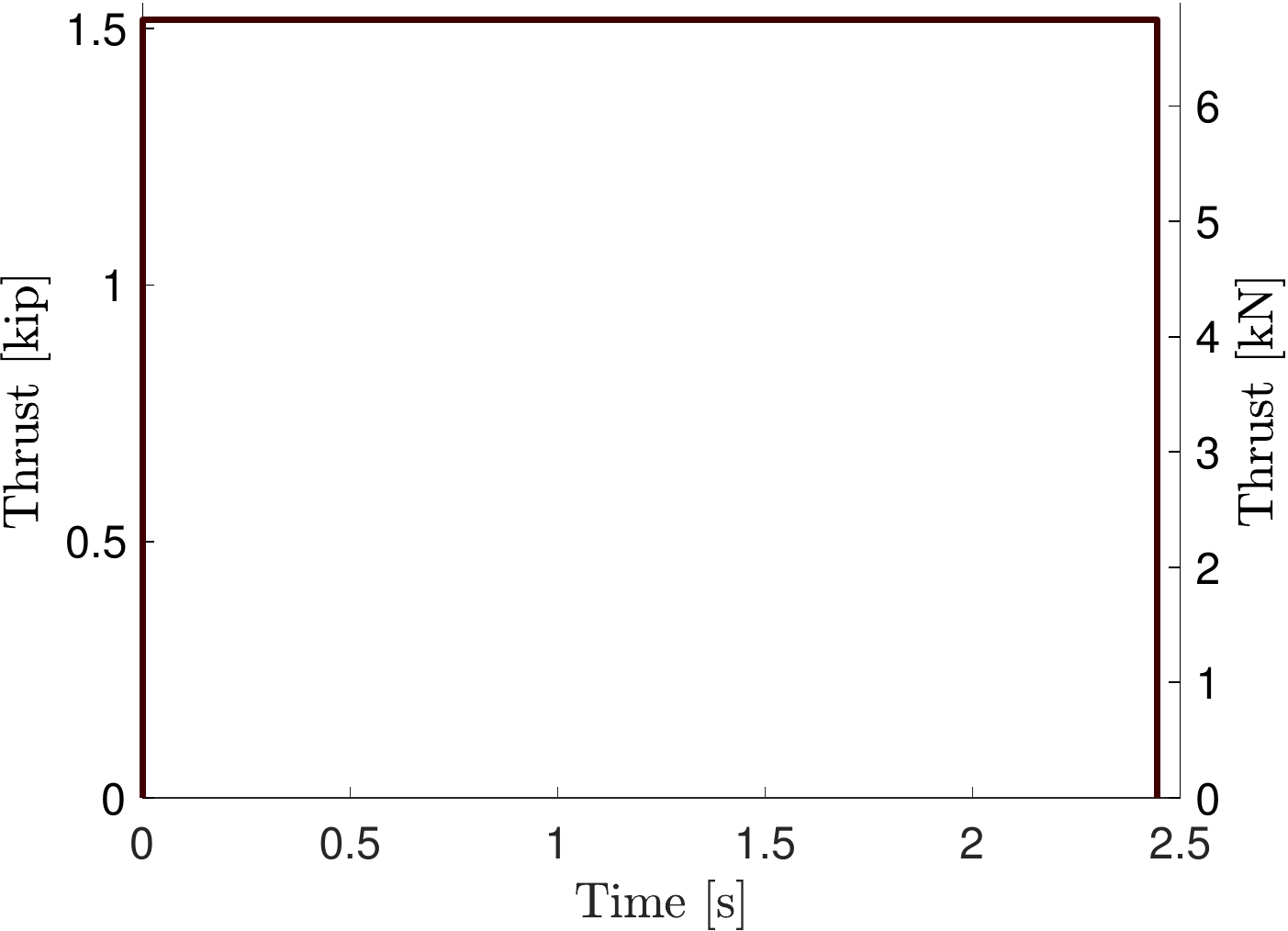}
	\end{subfigure} \hfill
	\begin{subfigure}{.49\textwidth}
	\centering 
	\includegraphics[width=\textwidth]{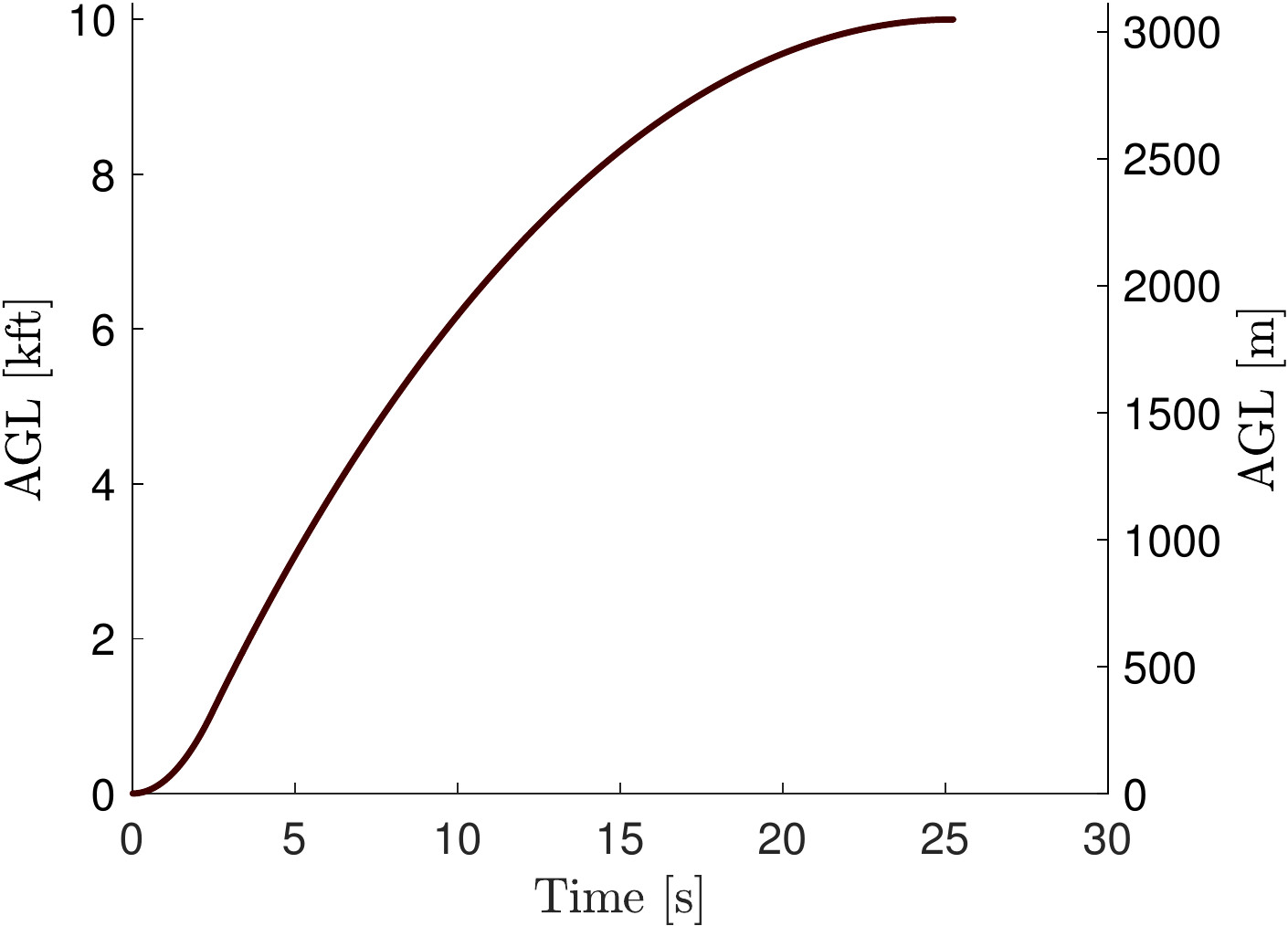}
	\end{subfigure}
	\caption{Steady-state design point thrust curve (left) and ascent path (right).}
	\label{fig:steady_design_point}
\end{figure}

Another observation from Fig.~\ref{fig:ispvsmdot_const} is that for $\Lf = 610$~mm (24~in), a peak $\isp \approx 223$~s is achievable at $\mdoto \approx 2.7$~kg/s. Although the shorter fuel grain theoretically leads to higher performance, due to the uncertainty in $a$ and $n$, a fuel grain of $\Lf = 610$~mm (24~in) is selected. Additionally, Fig.~\ref{fig:fvsmdot_const} shows increasing $\mdoto$ leads to an increase in the thrust of the engine. For structural design considerations of the overall rocket, the thrust of the engine should be limited. The design $\mdoto$ selected is 2.50~kg/s. The resulting rocket parameters and performance metrics are shown in Table~\ref{tab:steady_design}. As well, the thrust curve and the flight ascent path are shown in Fig.~\ref{fig:steady_design_point}.

\begin{table}
	\small
	\centering
	\caption{Steady-state model design point selected and performance metrics.}
	\label{tab:steady_design}
	\begin{tabular}{c|c|c|c}
	& \textbf{Parameter} & \textbf{Symbol} & \textbf{Value} \\ \hline 
	
	\textbf{Rocket} & Drag coefficient & $C_d$ & 0.529 \\ \hline
	
	\multirow{3}{*}{\textbf{Chamber}} 
	& Oxidizer mass flow rate & $\mdoto$ & 2.50~kg/s \\	
	& Chamber pressure & $\pc$ & 27.6~bar (400~psia) \\
    & Fuel length & $\Lf$ & 610~mm (24.0~in) \\ \hline
    
    \multirow{12}{*}{\textbf{Performance}}
    & Burntime & $\tburn$ & 2.44~s \\
	& Oxidizer mass & $\motot$ & 6111~g \\
	& Fuel mass & $\mftot$ & 1424~g \\
	& Pad OF ratio & $\ofbar$ & 4.29 \\
	& Peak chamber temperature & $\Tcmax$ & 2663~K \\ 
	& Peak thrust & $F$ & 6.75~kN \\
	& Pad thrust-to-weight & - & 11.5 \\
	& Nozzle throat radius & $\rt$ & 23.3~mm (0.916~in) \\
	& Nozzle outlet radius & $\rexit$ & 38.1~mm (1.50~in) \\
	& Total impulse & $\itot$ & 16,500~Ns \\
	& Specific impulse & $\isp$ & 223~s \\
	& Apogee predicted & - & 3048~m (10,000~ft) \\
	\end{tabular}
\end{table}

\chapter{Unsteady Model} \label{chapter:unsteady}


The steady-state model described in Chapter~\ref{chapter:steady} allows to define the overall propulsion system geometry, and provides an initial estimate of the propellant quantities required to bring \rocket{} to target apogee. However, transient effects, such as the self-pressurizing blowdown process of the nitrous oxide tank, and the resulting decreasing chamber pressure, are neglected. These effects lead to a decrease of the engine thrust over time. Hence, the steady-state model under-predicts propellant requirements. The geometry determined from the steady-state model is used as an input to the unsteady model, and the transient behavior of the engine is resolved, allowing to determine with greater accuracy propellant requirements. The current chapter details the formulation and results of the unsteady model.

\section{Model Formulation} \label{sec:unsteady}


\subsection{Model Description}

\begin{figure}
    \centering
    \includegraphics[width=0.9\textwidth]{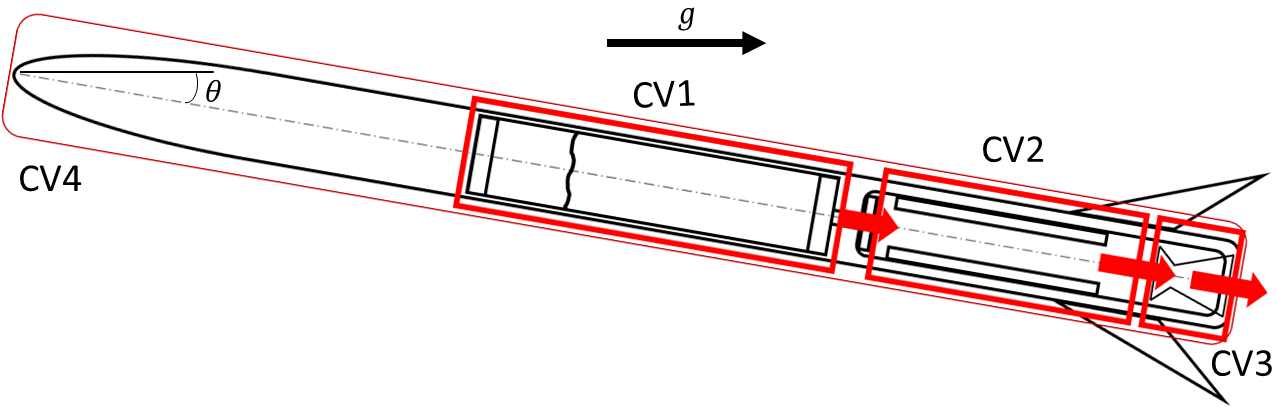}
    \caption{Unsteady hybrid rocket model overview.}
    \label{fig:overviewunsteady}
\end{figure}

The unsteady model considers a coupled analysis of the four CVs depicted in Fig.~\ref{fig:overviewunsteady}. In CV1, the self-pressurizing oxidizer tank is modeled. The \nitrous{} is delivered to CV2, the combustion chamber, through the feed plumbing system and the injector. The pressure differential between the oxidizer tank and the combustion chamber, along with the injector geometry, determine the mass flow rate of \nitrous{} entering the chamber. The solid fuel grain pyrolysis is analysed in CV2, and the combustion process is assumed to be quasi-steady. Hence, the reactants instantly reach chemical equilibrium in CV2 at each time step, and the thermodynamic properties of the combustion products are computed with the PROPEP code \cite{brown1995}. In CV3, the nozzle is analyzed while accounting for the possible formation of a normal standing shock wave. The thrust of the engine is resolved in time, and the one-dimensional ascent of the rocket is modeled over CV4. The state vector $\mbf{x} = \bbm \nv & \nl & \Tt & \rf & \mo & \mf & \pc & \zr & \vr \ebm^\mbf{T}$ tracks the state variables:
\begin{itemize}
\item $\nv$: moles of \nitrous{} in vapor phase in the tank;
\item $\nl$: moles of \nitrous{} in liquid phase in the tank;
\item $\Tt$: tank temperature;
\item $\rf$: fuel grain port internal radius;
\item $\mo$: oxidizer mass storage in the combustion chamber;
\item $\mf$: fuel mass storage in the combustion chamber;
\item $\pc$: combustion chamber pressure;
\item $\zr$: rocket altitude (elevation above sea level);
\item $\vr$: rocket total velocity.
\end{itemize}

A set of differential equations is derived for the state vector $\mbf{x}$ and numerically integrated with the MATLAB ODE solver \textit{ode15s}, until the propellants are depleted. Subsequently, the system is reduced to the state variables $\zr$ and $\vr$, and the flight profile of the vehicle is resolved with the \textit{ode45} solver. 

To lighten the notation of the current chapter, the dependencies on time are not explicitly indicated.


\subsection{Self-Pressurizing Oxidizer Tank}

\begin{figure}[H]
    \centering
    \includegraphics[width=0.4\textwidth]{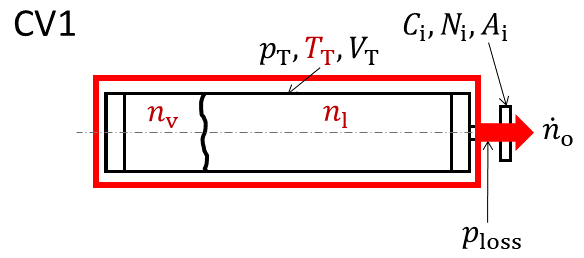}
    \caption{Control volume 1: unsteady oxidizer tank.}
    \label{fig:cv1unsteady}
\end{figure}

The self-pressurizing blowdown process of the \nitrous{} is depicted in Fig.~\ref{fig:cv1unsteady}. Assumptions used in the formulation include:
\begin{enumerate}
\item The \nitrous{} vapor obeys the real gas law.
\item \label{item:quasisteady1} The liquid and vapor are in quasi-steady thermal equilibrium; there is no thermal stratification within the tank.
\item \label{item:quasisteady2} The tank walls are adiabatic. 
\item \label{item:raoult} The liquid-vapor mixture obeys Raoult's law, hence the mixture remains saturated.
\item Evaporation at the liquid-vapor interface is not influenced by boiling phenomena.
\item The potential and kinetic energy of the propellant are neglected.
\end{enumerate}
Using the mass and energy conservation laws, a set of differential equations are derived for the tank state variables $\xt = \bbm \nv & \nl & \Tt \ebm ^\mbf{T}$. 

\subsubsection{Liquid Nitrous Oxide Blowdown}
The conservation of mass in CV1, written in molar form, yields,
\beq \label{eq:cv1mass}
\f{\d \nv}{\dt} + \f{\d \nl}{\dt} = -\ndoto
\eeq
where $\ndoto$ is the discharge flow rate from the oxidizer tank through the feed system and injector \cite{genevieve2013}:
\beq \label{eq:ndotox}
\ndoto = \Ci \Ni \Ai \sqrt{\f{2(\pt - \ploss - \pc)}{\Wo \nul}}.
\eeq
In Eq.~(\ref{eq:ndotox}), $\Ci$ is the dimensionless injector discharge coefficient, $\Ni$ is the number of injector discharge holes, $\Ai$ is their area, $\pt$ is the tank pressure, $\ploss$ is the pressure loss through the feed system, $\Wo$ is the oxidizer molar weight, and $\nul$ is the oxidizer liquid molar volume. As a result of assumption~\ref{item:raoult}, the liquid instantaneously evaporates such that it remains saturated, and $\pt = \psat(\Tt)$, where $\psat(\Tt)$ is the saturated vapor pressure of \nitrous{} evaluated at $\Tt$. As well, $\nul = \nulsat(\Tt)$, where $\nulsat(\Tt)$ is the saturated liquid molar volume of \nitrous{} at $\Tt$. Correlations for $\psat$ and $\nulsat$ as a function of $\Tt$ are provided in Appendix~\ref{a1:thermo}. Combining Eq.~(\ref{eq:cv1mass}) and (\ref{eq:ndotox}) yields:
\beq \label{eq:gov1}
-\Ci \Ni \Ai \sqrt{\f{2(\pt - \ploss - \pc)}{\Wo \nul}} = \f{\d \nv}{\dt} + \f{\d \nl}{\dt}.
\eeq

An additional equation for CV1 is obtained from the volume constraint of the tank,
\beq \label{eq:tankvolume}
\Vt = \nv \nuv  + \nl \nul
\eeq
where $\Vt$ is the internal volume of the tank, and $\nuv$ is the oxidizer vapor molar volume. Following assumption~\ref{item:raoult}, $\nuv = \nuvsat(\Tt)$, where $\nuvsat(\Tt)$ is the saturated \nitrous{} vapor molar volume evaluated at $\Tt$; it is computed with a linear interpolation scheme from a look-up table (Appendix~\ref{a1:thermo}). Differentiating Eq.~\ref{eq:tankvolume} with respect to time yields:
\beq \label{eq:diffvt}
0 = \nv \f{\d \nuv}{\d \Tt} \f{\d T}{\dt} + \nuv \f{\d \nv}{\dt} + \nl \f{\d \nul}{\d \Tt}\f{\d \Tt}{\dt} + \nul \f{\d \nl}{\dt}.
\eeq
The property $\f{\d \nul}{\d \Tt}$ is obtained by differentiating the correlation for $\nulsat$, while $\f{\d \nuv}{\d \Tt}$ is obtained from a look-up table (Appendix~\ref{a1:thermo}). Equation~\ref{eq:diffvt} can be re-arranged to yield:
\beq \label{eq:gov2}
0 = \nuv \f{\d \nv}{\dt} + \nul \f{\d \nl}{\dt} + \bigg(\nv \f{\d \nuv}{\d \Tt} + \nl \f{\d \nul}{\d \Tt} \bigg) \f{\d \Tt}{\dt}.
\eeq

The conservation of energy in CV1 yields,
\beq \label{eq:energy}
\ddt (\nv \uv + \nl \ul) = -\ndoto \ho 
\eeq
where $\uv$ is the molar internal energy of the vapor, $\ul$ is the molar internal energy of the liquid, and $\ho$ is the molar enthalpy of the liquid oxidizer exiting the tank. Following assumption~\ref{item:raoult}, the abovementioned thermophysical properties are equal to their saturated values, $\uv = \uvsat(\Tt)$, $\ul = \ulsat(\Tt)$, and $\ho = \hosat(\Tt)$; they are computed from a look-up table (Appendix~\ref{a1:thermo}) with a linear interpolation scheme. Expanding Eq.~(\ref{eq:energy}) yields,
\beq \label{eq:gov3}
- \ndoto \ho = \uv \f{\d \nv}{\dt} + \ul \f{\d \nl}{\dt} + \bigg( \nl \f{\d \ul}{\d \Tt} + \nv \f{\d \uv}{\d \Tt} \bigg) \f{\d \Tt}{\dt}
\eeq
where the derivatives $\f{\d \ul}{\d \Tt}$ and $\f{\d \uv}{\d \Tt}$ are interpolated from a look-up table (Appendix~\ref{a1:thermo}) with a linear interpolation scheme.

Equations~(\ref{eq:gov1}), (\ref{eq:gov2}), and (\ref{eq:gov3}) represent a linear system of three equations and three unknowns, where the derivatives of the state variables $\f{\d \nv}{\dt}$, $\f{d \nl}{\dt}$, and $\f{\d \Tt}{\dt}$ are the unknowns. Hence, the system can be written in a linear form $\mbf{A} \f{\d \mbf{\xt}}{\dt} = \mbf{b}$, and resolved with the MATLAB command $\f{\d \mbf{\xt}}{\dt} = \mbf{A} \backslash \mbf{b}$.

\subsubsection{Gaseous Nitrous Oxide Blowdown}

The system described by Eq.~(\ref{eq:gov1}), (\ref{eq:gov2}), and (\ref{eq:gov3}) is valid when CV1 contains a liquid-vapor mixture. When the liquid \nitrous{} is depleted, a different set of equations is considered. As the liquid \nitrous{} is depleted, $\nl = 0$, and:
\beq \label{eq:gov1vap}
\f{\d \nl}{\dt}  = 0.
\eeq
Equation~(\ref{eq:gov1}) becomes,
\beq \label{eq:gov2vap}
\f{\d \nv}{\dt} =  -\Ci \Ni \Ai \sqrt{\f{2(\pt - \ploss - \pc)}{\Wo \nuv}} 
\eeq
where the vapor molar volume is $\nuv = \Vt / \nv$. Additionally, the vapor is assumed to follow a polytropic expansion process, hence, 
\beq
C = \pt (\nuv)^m
\eeq
where $C$ is a constant, and $m$ is the polytropic exponent. As $\pt$, $\nuv$, and $m$ are strictly positive, logarithmic differentiation can be applied, resulting in,
\beq \label{eq:logdiff1}
0 = \f{1}{\pt} \f{\d \pt}{\dt} + \f{m}{\nuv} \f{\d \nuv}{\dt} \Rightarrow \f{1}{\pt} \f{\d \pt}{\dt} = - \f{m}{\nuv} \f{\d \nuv}{\dt}
\eeq
where $m$ is assumed to be a constant. Now using the definition $\nuv = \Vt/\nv$, and applying logarithmic differentiation:
\beq \label{eq:logdiff2}
\f{1}{\nuv} \f{\d \nuv}{\dt} = \f{-1}{\nv} \f{\d \nv}{\dt}.
\eeq
Additionally, the real gas law yields:
\beq
\pt = \f{Z \Ru \Tt}{\nuv}.
\eeq
where $Z$ is the vapor compressibility factor. It is assumed to remain approximately constant during the polytropic expansion process. Applying logarithmic differentiation one more time:
\beq \label{eq:logdiff3}
\f{1}{\pt} \f{\d \pt}{\dt} = \f{1}{\Tt} \f{\d \Tt}{\dt} - \f{1}{\nuv} \f{\d \nuv}{\dt}.
\eeq
Equations~(\ref{eq:logdiff1}), (\ref{eq:logdiff2}), and (\ref{eq:logdiff3}) can be combined to show:
\beq \label{eq:gov3vap}
\f{\d \Tt}{\dt} = \f{\Tt(m-1)}{\nv} \f{\d \nv}{\dt}.
\eeq

Equations~(\ref{eq:gov1vap}), (\ref{eq:gov2vap}), and (\ref{eq:gov3vap}) are an explicit system for $\f{\d \mbf{\xt}}{\dt}$. At the end of the liquid-vapor blowdown, the compressibility factor $Z$ is calculated with the real gas law, given the current values of $\pt$, $\Tt$, and $\nuv$. Additionally, the polytropic exponent is approximated as the heat capacity ratio, $m = \cpv / \cvv$, where $\cpv$ is the heat capacity at constant pressure of the vapor, and $\cvv$ is the heat capacity at constant volume of the vapor. The heat capacities are taken to be the saturated values and are obtained from a look-up table (Appendix~\ref{a1:thermo}).


\subsection{Combustion Chamber} \label{sec:chambermodel}

\begin{figure}[H]
    \centering
    \includegraphics[width=0.35\textwidth]{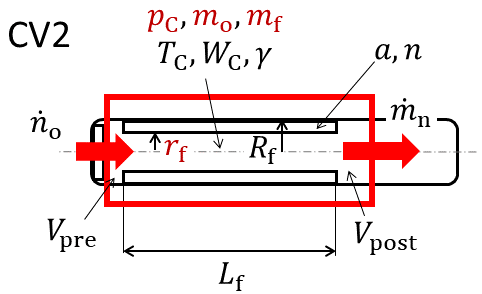}
    \caption{Control volume 2: unsteady combustion chamber.}
    \label{fig:cv2unsteady}
\end{figure}

The semi-empirical fuel grain regression rate theory \cite{genevieve2013} is applied to model the combustion process of the solid fuel grain in CV2, as depicted in Fig~\ref{fig:cv2unsteady}.  Assumptions used in the formulation include:
\begin{enumerate}
\item The fuel grain regression is axially uniform.
\item The combustion process is quasi-steady, hence the chamber mixture instantaneously attains chemical equilibrium at each time step.
\item The combustion gases obey the ideal gas law.
\item The combustion chamber walls are adiabatic.
\item The gaseous mixture is homogeneous across the chamber.
\item There is no pressure drop across the combustion chamber.
\end{enumerate}

Using the mass and energy conservation laws, a set of differential equations is derived for the chamber state variables $\xc = \bbm \rf & \mo & \mf & \pc \ebm^\mbf{T}.$

\subsubsection{Fuel Grain Regression}
The semi-empirical expression for the fuel grain regression rate \cite{genevieve2013} is,
\beq \label{eq:rdotf}
\f{\d \rf}{\dt} = a \Go^n \equiv a \bigg( \f{ \mdotoin}{\pi \rf^2} \bigg)^n
\eeq
where $\Go$ is the flux rate of oxidizer entering the combustion chamber, $\mdotoin = \Wo \ndoto$ is the mass flow rate of oxidizer entering the chamber, $a$ is the regression rate scaling constant, and $n$ is the regression rate exponent.

\subsubsection{Chamber Filling and Emptying}

The conservation of mass in CV2 yields,
\beqarr
\f{\d \mf}{\dt} &=& \mdotfin - \mdotfout \\
\f{\d \mo}{\dt} &=& \mdotoin - \mdotoout 
\eeqarr
where $\mdotfin$ is the mass flow rate of fuel entering the chamber, $\mdotfout$ is the mass flow rate of fuel exiting the chamber, and $\mdotoout$ is the mass flow rate of oxidizer exiting the chamber. The variable $\mdotfin$ can be related to the growth rate of the chamber volume:
\beq
\mdotfin = \rhof \f{\d \Vc}{\dt}.
\eeq
The chamber volume is $\Vc = \Vpre + \Vpost + \Vport$, where $\Vpre$ is the pre-chamber volume, $\Vpost$ is the post-chamber volume, and $\Vport = \pi \rf^2 \Lf$ is the fuel grain port volume. Hence:
\beq 
\mdotfin = \rhof \bigg(2 \pi \rf \Lf \f{\d \rf}{\dt}\bigg).
\eeq

Since the gaseous mixture in CV2 is assumed to be homogeneous, the OF ratio describes the storage of oxidizer and fuel in CV2, as well as the contribution of each propellant in the mass flow rate exiting the chamber:
\beq \label{eq:ofcv2}
\of = \f{\mo}{\mf} \equiv \f{\mdotoout}{\mdotfout}.
\eeq
Additionally, the total mass flow rate of gas exiting through the nozzle is,
\beq \label{eq:mdotncv2}
\mdotn = \mdotoout + \mdotfout
\eeq
where $\mdotn$ is computed with the nozzle gas dynamics (Section \ref{sec:nozzlemodel}). Equations~(\ref{eq:ofcv2}) and (\ref{eq:mdotncv2}) can be used to show:
\beqarr
\label{eq:dmfdt} \f{\d \mf}{\dt} &=& \mdotfin - \f{\mdotn}{1 + \of} \\ 
\label{eq:dmodt} \f{\d \mo}{\dt} &=& \mdotoin - \f{\mdotn}{1 + 1/\of} 
\eeqarr
In each equation, the first term represents the mass flow rate of propellant entering the chamber--from the fuel grain pyrolosis, or from the oxidizer tank--and the second term represents the contribution of each propellant in the mass flow rate exiting through the nozzle.

\subsubsection{Combustion Products and Chemical Equilibrium}
The ideal gas law applied to the combustion chamber gaseous products yields:
\beq
\pc = \f{\mc \Ru \Tc}{\Vc \Wc}
\eeq
where $\mc = \mo + \mf$ is the total mass of the gases stored in the combustion chamber, $\Tc$ is the chamber temperature, and $\Wc$ is the molar weight of the combustion gases. Applying logarithmic differentiation results in,
\beq \label{eq:cv2pc}
\f{1}{\pc} \f{\d \pc}{\dt} = \f{1}{\mc} \f{\d \mc}{\dt} + \f{1}{\Tc} \f{\d \Tc}{\dt} - \f{1}{\Vc} \f{\d \Vc}{\dt} - \f{1}{\Wc} \f{\d \Wc}{\dt}
\eeq
and the conservation of mass yields:
\beq \label{eq:cv2dmc}
\f{\d \mc}{\dt} = \mdotoin + \mdotfin - \mdotn.
\eeq
The PROPEP code \cite{brown1995} is used to compute $\Tc$ and $\Wc$. The inputs to the PROPEP code are $\pc$ and OF, hence, $\Tc = \Tc(\pc, \of)$, and $\Wc = \Wc(\pc, \of)$. The chain rule therefore yields:
\beqarr
\label{eq:cv2tc} \f{\d \Tc}{\dt} &=& \f{\del \Tc}{\del(\of)} \f{\d (\of)}{\dt} + \f{\del \Tc}{\del \pc} \f{\d \pc}{\dt} \\
\label{eq:cv2wc} \f{\d \Wc}{\dt} &=& \f{\del \Wc}{\del(\of)} \f{\d (\of)}{\dt} + \f{\del \Wc}{\del \pc} \f{\d \pc}{\dt}.
\eeqarr
Equations~(\ref{eq:cv2tc}) and (\ref{eq:cv2wc}) are a consequence of the conservation of energy and of the quasi-steady chemical equilibrium assumption. Using the definition $\of = \mo/\mf$ and logarithmic differentiation, the following can be shown:
\beq \label{eq:cv2dof}
\f{\d (\of)}{\dt} = \f{1}{\mf} \bigg(\f{\d \mo}{\dt} - \of \f{\d \mf}{\dt} \bigg).
\eeq
The derivatives $\f{\del \Tc}{\del (\of)}$, $\f{\del \Tc}{\del \pc}$, $\f{\del \Wc}{\del (\of)}$, and $\f{\del \Wc}{\del \pc}$ are approximated with a forward finite difference scheme with the PROPEP code. 

Substituting Eq.~(\ref{eq:cv2tc}) and (\ref{eq:cv2wc}) in Eq.~(\ref{eq:cv2pc}) and re-arranging yields,
\beq
\f{\d \pc}{\dt} = \f{\f{1}{\mc} \f{\d \mc}{\dt} - \f{1}{\Vc} \f{\d \Vc}{\dt} + \f{\d (\of)}{\dt} \Big( \f{1}{\Tc} \f{\del \Tc}{\del (\of)} - \f{1}{\Wc} \f{\del \Wc}{\del (\of)} \Big)}{\f{1}{\pc} - \f{1}{\Tc} \f{\del \Tc}{\del \pc} + \f{1}{\Wc} \f{\del \Wc}{\del \pc}}.
\eeq
where $\f{\d \mc}{\dt}$ and $\f{\d (\of)}{\dt}$ are computed from Eq.~(\ref{eq:cv2dmc}) and (\ref{eq:cv2dof}), respectively. The PROPEP code also outputs $\gamma$, the heat capacity ratio of the combustion gases, which is used as an input to the nozzle gas dynamics model in CV3.


\subsection{Nozzle Gas Dynamics} \label{sec:nozzlemodel}

\begin{figure}[H]
    \centering
    \includegraphics[width=0.4\textwidth]{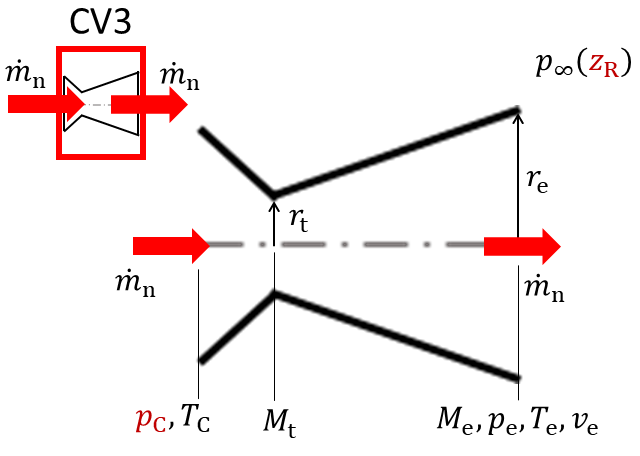}
    \caption{Control volume 3: unsteady nozzle.}
    \label{fig:cv3unsteady}
\end{figure}

In CV3, a simple converging-diverging nozzle is considered, as shown in Fig.~\ref{fig:cv3unsteady}. The nozzle gas dynamics are modeled using a one-dimensional flow model. Assumptions used in the formulation include:
\begin{enumerate}
\item The gaseous mixture is homogeneous.
\item The nozzle gas is in frozen equilibrium, hence its composition does not change across the nozzle axis.
\item The fluid obeys the ideal gas law and is calorically perfect, hence the heat capacity ratio is constant.
\item The nozzle walls are adiabatic.
\item The inlet gas velocity is negligible compared to the exit gas velocity.
\item The boundary layer flow effects are neglected.
\item Except if there is a shock wave in the nozzle diverging section, the nozzle flow is isentropic.
\item Shock waves in the nozzle diverging section are normal standing shock waves.
\end{enumerate}

The nozzle gas dynamics are coupled to the combustion chamber and the rocket ascent models. To compute the engine thrust $F$, the gas flow velocity and pressure at the nozzle exit plane, $\ve$ and $\pe$, must be resolved. As well, the nozzle mass flow rate, $\mdotn$, is constrained by the nozzle throat, hence the Mach number at the throat, $\Mt$, must be resolved.

\subsubsection{Flow Regimes}
The nozzle flow regime is determined by: the nozzle area expansion ratio, $\Ae/\At$, where $\Ae$ and $\At$ are the nozzle exit and throat areas, respectively; the inlet gas conditions, which are the chamber conditions (stagnation conditions); and the ambient pressure at the nozzle exit plane, $\pinf$, evaluated as a function of $\zr$ with the NASA atmospheric model \cite{hall2021} (Appendix~\ref{a1:thermo}). Two critical points delimiting the flow regimes are identified.

The first critical point is associated with the ambient pressure $p_1$ required such that the flow is choked at the nozzle throat, but is fully subsonic in the nozzle diverging section. The isentropic flow relations relate the gas Mach number at the exit plane associated to the first critical point, $M_1$, to the expansion area ratio \cite{higgins2021}:
\beq \label{eq:M1}
\f{\Ae}{\At} = \f{1}{M_1} \bigg[ \f{2}{\gamma+1} \bigg(1 + \f{\gamma-1}{2} M_1^2 \bigg) \bigg] ^ {\f{\gamma+1}{2(\gamma-1)}}.
\eeq
Equation~(\ref{eq:M1}) can be solved numerically for the subsonic solution $M_1<1$. The corresponding critical pressure is then related to the chamber stagnation pressure by \cite{higgins2021}:
\beq \label{eq:p1}
\f{\pc}{p_1} = \bigg[ 1 + \f{\gamma-1}{2} M_1^2 \bigg] ^ {\f{\gamma}{\gamma-1}}.
\eeq
If $\pinf \geq p_1$, the nozzle diverging section is fully subsonic and the gas exit pressure matches the ambient pressure, $\pe = \pinf$. In the case $\pinf = p_1$, the flow is choked at the throat, hence $\Mt = 1$, and $\Me = M_1$. In the case $\pinf > p_1$, the flow is subsonic throughout the nozzle, and the following relations can be solved numerically for $\{\Me,~\Mt\} < 1$ \cite{higgins2021}:
\beqarr
\f{\pc}{p_e} &=& \bigg[ 1 + \f{\gamma-1}{2} M_e^2 \bigg] ^ {\f{\gamma}{\gamma-1}} \\
\f{\At}{\Ae} &=& \f{\Me}{\Mt} \bigg[ \bigg( 1 + \f{\gamma-1}{2} \Mt^2 \bigg) \bigg/ \bigg( 1 + \f{\gamma-1}{2} \Me^2 \bigg) \bigg] ^ \f{\gamma+1}{2(\gamma-1)}.
\eeqarr

The second critical point is associated with a normal standing shock wave located exactly at the nozzle exit plane. The flow entering the shock wave is described by the subscript $2x$, while the flow exiting it is described by the subscript $2$. Since shock waves can only occur in supersonic flow, the nozzle throat is choked and the diverging section of the nozzle is fully supersonic. Equation~(\ref{eq:M1}) can be used by replacing $M_1$ with $M_{2x}$, and it can be solved numerically for $M_{2x} > 1$. Similarly, Eq.~(\ref{eq:p1}) can be used by replacing $p_1$ with $p_{2x}$ and $M_1$ with $M_{2x}$, and $p_{2x}$ can be solved. The pressure and Mach number after the shock wave are then described by the relations \cite{higgins2021}:
\beqarr
\f{p_2}{p_{2x}} &=& \f{2 \gamma M_{2x}^2 - (\gamma - 1)}{\gamma + 1} \\
M_2 &=& \sqrt{\f{2 + (\gamma - 1) M_{2x}^2}{2 \gamma M_{2x}^2 - (\gamma-1)}}.
\eeqarr
If $p_1 > \pinf \geq p_2$, the flow is choked at the throat ($\Mt = 1$), and there is a normal standing shock wave somewhere in the diverging nozzle section. Since after the shock wave the flow becomes subsonic, the gas exit pressure must match the ambient pressure, $\pe = \pinf$. In the case $\pinf = p_2$, the shock wave is located exactly at the outlet plane, and $\Me = M_2$. In the case $p_1 > \pinf > p_2$: let the subscript $x$ describe the flow entering the shock wave at some location along the nozzle diverging section, and the subscript $y$ describe the flow exiting it. As well, let $p_{0y}$ be the stagnation pressure of the flow exiting the shock wave. The following relations then apply \cite{higgins2021}:
\beqarr
\f{p_{0y}}{\pc} &=& \f{\At}{A_y^*} = \bigg[\f{2 \gamma M_x^2 - (\gamma - 1)}{\gamma + 1} \bigg]^\f{-1}{\gamma-1} \bigg[\f{(\gamma+1)M_x^2}{2 + (\gamma-1) M_x^2} \bigg]^\f{\gamma}{\gamma-1} \\
\f{\Ae}{A_y^*} &=& \f{1}{\Me} \bigg[\f{2}{\gamma+1} \bigg(1 + \f{\gamma-1}{2}\Me^2 \bigg) \bigg]^\f{\gamma+1}{2(\gamma-1)} \\
\f{p_{0y}}{\pe} &=& \bigg[1 + \f{\gamma-1}{2}\Me^2 \bigg]^\f{\gamma}{\gamma-1} \\
\pe &=& \pinf.
\eeqarr
An iterative procedure can be applied to find $M_x > 1$ which satisfies the system of equations. The associated flow exit Mach number $\Me < 1$ can then be resolved.

If the ambient pressure is below the second critical pressure, $\pinf < p_2$, the nozzle throat is choked and the nozzle diverging section is fully supersonic, with no shock wave. As the flow exiting the nozzle is supersonic, the pressure of the exiting gas flow does not need to mach the ambient pressure. The gas exit conditions are $\pe = p_{2x}$ and $\Me = M_{2x}$. The shockwave beyond the nozzle outlet is assumed to not impact engine performance.

\subsubsection{Engine Thrust}

The parameters $\ve$, $\mdotn$, and $F$ can be resolved with the following relations \cite{higgins2021}:
\beqarr
\f{\Tc}{\Te} &=& 1 + \f{\gamma-1}{2} \Me^2 \\
\ve &=& \Me \sqrt{\f{\gamma \Ru \Te }{\Wc}} \\
\label{eq:mdotn_unsteady} \mdotn &=& \At \pc \Mt \sqrt{\f{\gamma \Wc}{\Ru \Tc}} \bigg[1 + \f{\gamma-1}{2}\Mt^2 \bigg] ^ \f{-(\gamma+1)}{2(\gamma-1)} \\
\label{eq:thrustunsteady} F &=& \mdotn \ve + (\pe - \pinf) \Ae.
\eeqarr


\subsection{Rocket Flight Path} 

\subsubsection{Powered and Coasting Ascent} \label{sec:unsteady_ascent}

The rocket ascent is modeled with a one-dimensional model. A set of differential equations are formulated for the rocket state variables, $\xr = \bbm \zr & \vr \ebm^\mbf{T} $. The reader can refer to Section~\ref{sec:steady_ascent} for the formulation of the rocket ascent model; the only modification is that a non-constant thrust is considered, as computed from Eq.~(\ref{eq:thrustunsteady}). Equations~(\ref{eq:zrsteady}) and (\ref{eq:vrsteady}) provide the time rate change of the state variables, while the rocket total mass as a function of time is:
\beq
\mr(t) = \mdry + (\nv + \nl)\Wo + \pi \rhof \Lf (\Rf^2 - \rf^2) + \mo + \mf.
\eeq

\subsubsection{Parachute Descent}

In the unsteady formulation, the rocket flight is extended beyond apogee, and the descent of the rocket under parachutes is modeled. A dual-deployment scheme is considered. The drogue parachute is deployed at apogee, giving rise to a drag force $\Dchute$ in the upwards direction:
\beq \label{eq:chuteforce}
\Dchute = \f{1}{2}\Cddrogue \rhoa \vr^2 \Adrogue
\eeq
where $\Cddrogue$ is the drogue parachute drag coefficient, $\Adrogue$ is its surface area, and $\rhoa$ is the air density evaluated as a function of $\zr$ with the NASA atmospheric model \cite{hall2021} (Appendix~\ref{a1:thermo}). The other forces acting on the rocket are: the gravitational force $F_g$ in the downward direction, computed as a function of $\zr$ with Eq.~(\ref{eq:gravity}); and the rocket drag force $D$ in the upward direction, computed with Eq.~(\ref{eq:drag}). The time rate change of the state variables is then:
\beqarr
\label{eq:cv4unsteady1} \f{\d \zr}{\dt} &=& \vr \\
\label{eq:cv4unsteady2} \f{\d \vr}{\dt} &=& \f{1}{\mdry} \big( -F_g + D + \Dchute \big).
\eeqarr

Equations~(\ref{eq:cv4unsteady1}) and (\ref{eq:cv4unsteady2}) are solved numerically until the deployment altitude of the main parachute is reached, at which point $\Cddrogue$ and $\Adrogue$ are substituted by $\Cdmain$ and $\Amain$, the drag coefficient and surface area of the main parachute, in Eq.~(\ref{eq:chuteforce}). The equations are then solved numerically until the rocket hits the ground.


\subsection{Initialization of State Variables}

\subsubsection{Tank State Variables and Dip Tube Length} \label{sec:tank_sizing}

\begin{figure}[h]
    \centering
    \includegraphics[width=0.38\textwidth]{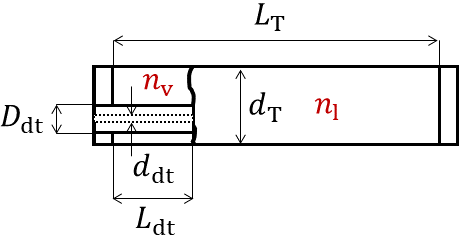}
    \caption{Control volume 1: oxidizer tank dimensions.}
    \label{fig:cv1diptube}
\end{figure}

A design initial tank temperature $\Tt$ and total oxidizer mass in CV1,
\beq \label{cv1dimeq1}
\motot = \Wo(\nv + \nl)
\eeq
are selected. The tank is assumed to be cylindrical and fitted with a dip tube, as illustrated in Fig.~\ref{fig:cv1diptube}. Given: the liquid molar volume $\nul$ at $\Tt$; the vapor molar volume $\nuv$ at $\Tt$; the tank internal diameter $\dTank$; the dip tube external diameter $\DDip$; and the dip tube internal diameter $\dDip$, the initial values of $\nl$ and $\nv$ can be computed, along with the dip tube length, $\LDip$, in two different ways.

The first method consists in defining the ullage factor of the tank $U $ and computing the required tank internal length $\LTank$. The definition of the ullage factor is:
\beq 
U = \f{\Vv}{\Vl}
\eeq
where $\Vv$ is the volume occupied by the vapor, and $\Vl$ is the volume occupied by the liquid. These parameters are related to the initial moles in the tank by:
\beqarr 
\label{cv1dimeq2} 0 &=& \nl \nul - \Vl \\
\label{cv1dimeq3} 0 &=& \nv \nuv - \Vv \equiv \nv \nuv - U \Vl.
\eeqarr
The volume occupied by the vapor can also be related to the dip tube length by:
\beq
\label{cv1dimeq4} U \Vl = \f{\pi}{4} \dTank^2 \LDip - \f{\pi}{4}(\DDip^2 - \dDip^2) \LDip.
\eeq
The volume occupied by the liquid is:
\beq
\label{cv1dimeq5} \Vl = \f{\pi \dTank^2}{4}(\LTank - \LDip).
\eeq
Equations~(\ref{cv1dimeq1}) and (\ref{cv1dimeq2}) to (\ref{cv1dimeq5}) can be re-arranged in a linear system of five equations, five unknowns,
\beq
\bbm \f{\motot}{\Wo} \\ 0 \\ 0 \\ 0 \\ 0 \ebm = \bbm 0 & 1 & 1 & 0 & 0 \\ -1 & \nul & 0 & 0 & 0 \\ -U & 0 & \nuv & 0 & 0 \\ \f{-4U}{\pi} & 0 & 0 & 0 & \dTank^2-\DDip^2+\dDip^2 \\ \f{-4}{\pi} & 0 & 0 & \dTank^2 & -\dTank^2 \ebm \bbm \Vl \\ \nl \\ \nv \\ \LTank \\ \LDip \ebm
\eeq
which can be solved analytically or numerically to determine the tank initial state variables and tank dimensions.

The second method consists in fixing the tank internal length $\LTank$ instead of the ullage factor. Equations~(\ref{cv1dimeq1}) and (\ref{cv1dimeq2}) to (\ref{cv1dimeq5}) still apply, but $\Vv$ is not substituted by $U \Vl$, and $\LTank$ is now treated as a known quantity. The system of equations becomes:
\beq
\bbm \f{\motot}{\Wo} \\ 0 \\ 0 \\ 0 \\ \dTank^2 \LTank \ebm = \bbm 0 & 1 & 1 & 0 & 0 \\ -1 & \nul & 0 & 0 & 0 \\ 0 & 0 & \nuv & -1 & 0 \\ 0 & 0 & 0 & \f{-4}{\pi} & \dTank^2-\DDip^2+\dDip^2 \\ \f{4}{\pi} & 0 & 0 & 0 & \dTank^2 \ebm \bbm \Vl \\ \nl \\ \nv \\ \Vv \\ \LDip \ebm.
\eeq

\subsubsection{Chamber State Variables}

A design initial total fuel mass $\mftot$ is selected, and the fuel length $\Lf$, external radius $\Rf$, and density $\rhof$ are known. The initial fuel port internal radius is then:
\beq
\rf = \sqrt{\Rf^2 - \f{\mftot}{\pi \rhof \Lf}}
\eeq
The initial storage of propellants in the chamber are $\mf = \mo = 0$. The initial chamber pressure is equal to the ambient air pressure evaluated as a function of $\zr$ with the NASA atmospheric model \cite{hall2021} (Appendix~\ref{a1:thermo}). 

\subsubsection{Rocket State Variables}

The state variable $\zr$ is the absolute altitude--elevation above sea level, not above ground level (AGL). The initial value of $\zr$ is determined from the launch site location. The initial velocity of the rocket is $\vr = 0$.


\section{Results: Performance Evaluation}


\subsection{Injector Model Parameters}

As can be seen from Eq.~(\ref{eq:ndotox}), the mass flow rate of \nitrous{} from the tank to the combustion chamber varies linearly with the injector discharge parameters $\Ci$, $\Ni$, and $\Ai$. The injector discharge coefficient $\Ci$ must in practice be determined experimentally. At the moment of writing the first iteration of the unsteady model, the MRT did not have an experimentally-characterized injector. As such, it was assumed that an injector which can provide the desired performance would be designed and manufactured at a future stage. In the current chapter, an ideal design allowing to attain desired performance is considered. In Chapter~\ref{chapter:hotfire}, an approach to determine $\Ci$ from hot fire testing data is outlined. 

A design tank temperature $\Ttzero = 25$~\super{$\circ$}C is selected, and the resulting oxidizer molar volume $\nul = 59.3$~L/kmol is obtained. The feed pressure loss is assumed, $\ploss = 0.345$~bar (5~psia). The target design chamber pressure is $\pc = 27.6$~bar (400~psia), and the target oxidizer flow rate is $\mdoto = 2.50$~kg/s, hence $\ndoto = \mdoto/\Wo = 0.0568$~kmol/s. Isolating for the product $\Ci \Ni \Ai$ in Eq.~(\ref{eq:ndotox}) results in $\Ci \Ni \Ai = 3.79~\times~10^{-5}$~m\super{2}. The parameters chosen are $\Ci = 1$, $\Ni = 10$, $\Ai = 3.79~\times~10^{-6}$~m\super{2}.

\subsection{Engine Transient Performance} \label{sec:unsteady_performance}

The fuel length, nozzle dimensions, and pad OF ratio determined in the steady-state model (Table~\ref{tab:steady_design}) are provided as inputs to the unsteady model. Due to the transient performance decrease of the engine, the $\motot$ and $\mftot$ determined in Table~\ref{tab:steady_design} are insufficient to bring \rocket{} to target apogee; they are used as a lower bound to determine propellant requirements. The unsteady model is resolved iteratively with different total propellant masses, while $\ofbar = 4.29$ is kept constant,  to determine propellant requirements to reach target apogee. At each iteration, the flight tank is sized with the ullage factor method. The model inputs and parameters are shown in Table~\ref{tab:unsteady_design}, and the resolved performance metrics are shown in Table~\ref{tab:unsteady_performance}.

\begin{table}
    \small
	\centering
	\caption{Unsteady model design point and parameters.}
	\label{tab:unsteady_design}
	\begin{tabular}{c|c|c|c}
	
	& \textbf{Description} & \textbf{Symbol} & \textbf{Value} \\ \hline

	\multirow{4}{*}{\textbf{Rocket}}
	& Dry mass & $\mdry$ & 52.5~kg \\
	& Drag coefficient & $C_d$ & 0.529 \\
	& Frontal area & $\Ar$ & 19,478~mm\super{2} (30.2~in\super{2}) \\
	& Launch angle & $\theta$ & 6\super{$\circ$} \\	
	\hline
	
	\multirow{2}{*}{\textbf{Drogue parachute}}
	& Drag coefficient & $\Cddrogue$ & 1.55 \\
	& Frontal area & $\Adrogue$ & 1.17~m\super{2} \\ \hline
	
	\multirow{3}{*}{\textbf{Main parachute}}
	& Deployment altitude & - & 457~m (1500~ft) \\
	& Drag coefficient & $\Cdmain$ & 2.20 \\
	& Frontal area & $\Amain$ & 10.5~m\super{2} \\ \hline
		
	\multirow{10}{*}{\textbf{Oxidizer tank}}
	& Oxidizer mass & $\motot$ & 7035~g \\
	& Initial temperature & $\Ttzero$ & 25~$^{\circ}$C \\
	& Initial pressure & $\ptzero$ & 57.3~bar (831~psia) \\
	& Internal diameter & $\dTank$ & 128~mm (5.047~in) \\
	& Internal shell length & $\LTank$ & 838~mm (33.009~in)\\
	& Internal volume & $\Vt$ & 10.8~L \\  
	& Ullage factor & $U$ & 20\% \\
	& Dip tube external diameter & $\DDip$ & 6.35~mm (0.25~in) \\
	& Dip tube internal diameter & $\dDip$ & 4.572~mm (0.18~in) \\
	& Dip tube length & $\LDip$ & 140~mm (5.507~in) \\ \hline
	
	\multirow{4}{*}{\textbf{Injector and feed}}
	& Discharge coefficient & $\Ci$ & 1 \\ 
	& Number of holes &$\Ni$ & 10 \\ 
	& Hole area & $\Ai$ & $3.79~\times~10^{-6}$~m\super{2}\\
	& Feed pressure loss & $\ploss$ & 0.345 bar (5~psia) \\ \hline
	
	\multirow{8}{*}{\textbf{Chamber}}
	& Fuel mass & $\mftot$ & 1639~g \\
	& Fuel length & $\Lf$ & 610~mm (24~in) \\
	& Fuel density & $\rhof$ & 900~kg/m\super{3} \\
	& Fuel external radius & $\Rf$ & 39.497~mm \\
	& Regression rate scaling constant & $a$ & 1.32 x 10\super{-4} \cite{genevieve2013} \\
	& Regression rate exponent & $n$ & 0.555 \cite{genevieve2013} \\
	& Pre-chamber volume & $\Vpre$ & 0.218~L \\ 
	& Post-chamber volume & $\Vpost$ & 0.885~L \\  \hline
	
	\multirow{2}{*}{\textbf{Nozzle}}
	& Throat radius & $\rt$ & 23.3~mm (0.916~in) \\
	& Exit radius & $\rexit$ & 38.1~mm (1.50~in) \\ 	
	\end{tabular}
\end{table}

\begin{table}
	\small
	\centering
	\caption{Unsteady model engine performance metrics.}
	\label{tab:unsteady_performance}
	\begin{tabular}{c|c|c}
	\textbf{Description} & \textbf{Symbol} & \textbf{Value} \\ \hline 
	Burntime & $\tburn$ & 3.34~s \\	
	Total impulse & $\itot$ & 16,845~Ns \\
	Specific impulse & $\isp$ & 198~s \\
	Peak chamber temperature & $\Tcmax$ & 2810~K \\
	Peak thrust & $\Fmax$ & 6.74~kN \\
	Average thrust & $\Favg$ & 5.05~kN \\
	Rocket wet mass & $\mrzero$ & 61.2~kg (135~lbf) \\
	Pad thrust-to-weight & - & 11.2 \\ 
		\end{tabular}
\end{table}

\begin{figure}[h]
	\centering 
	\begin{subfigure}{.49\textwidth}
	\centering 
	\includegraphics[width=\textwidth]{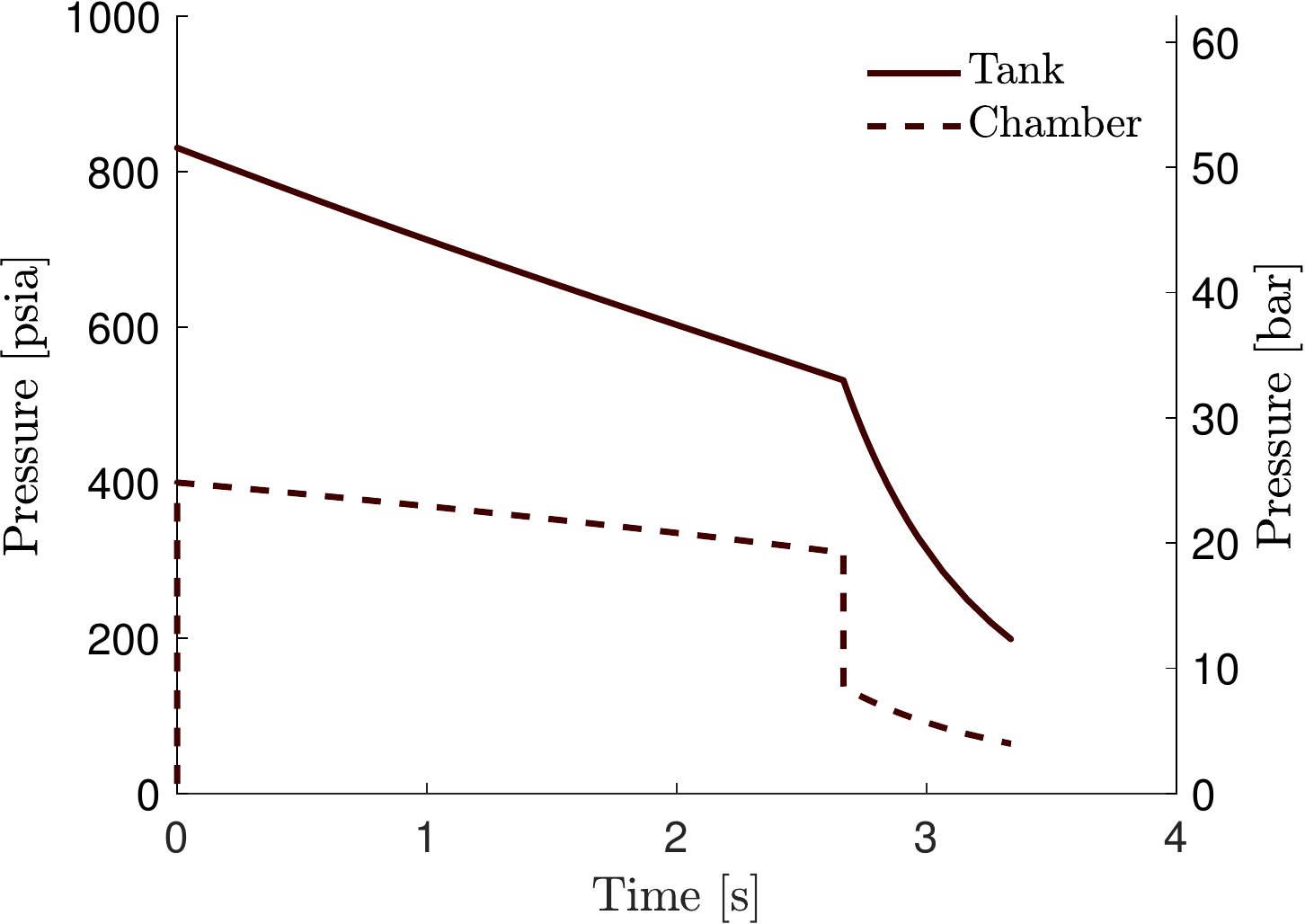}
	\caption{Tank and chamber pressures.}
	\label{fig:unsteady_pressures} 
	\end{subfigure} \hfill
	\begin{subfigure}{.49\textwidth}
	\centering 
	\includegraphics[width=\textwidth]{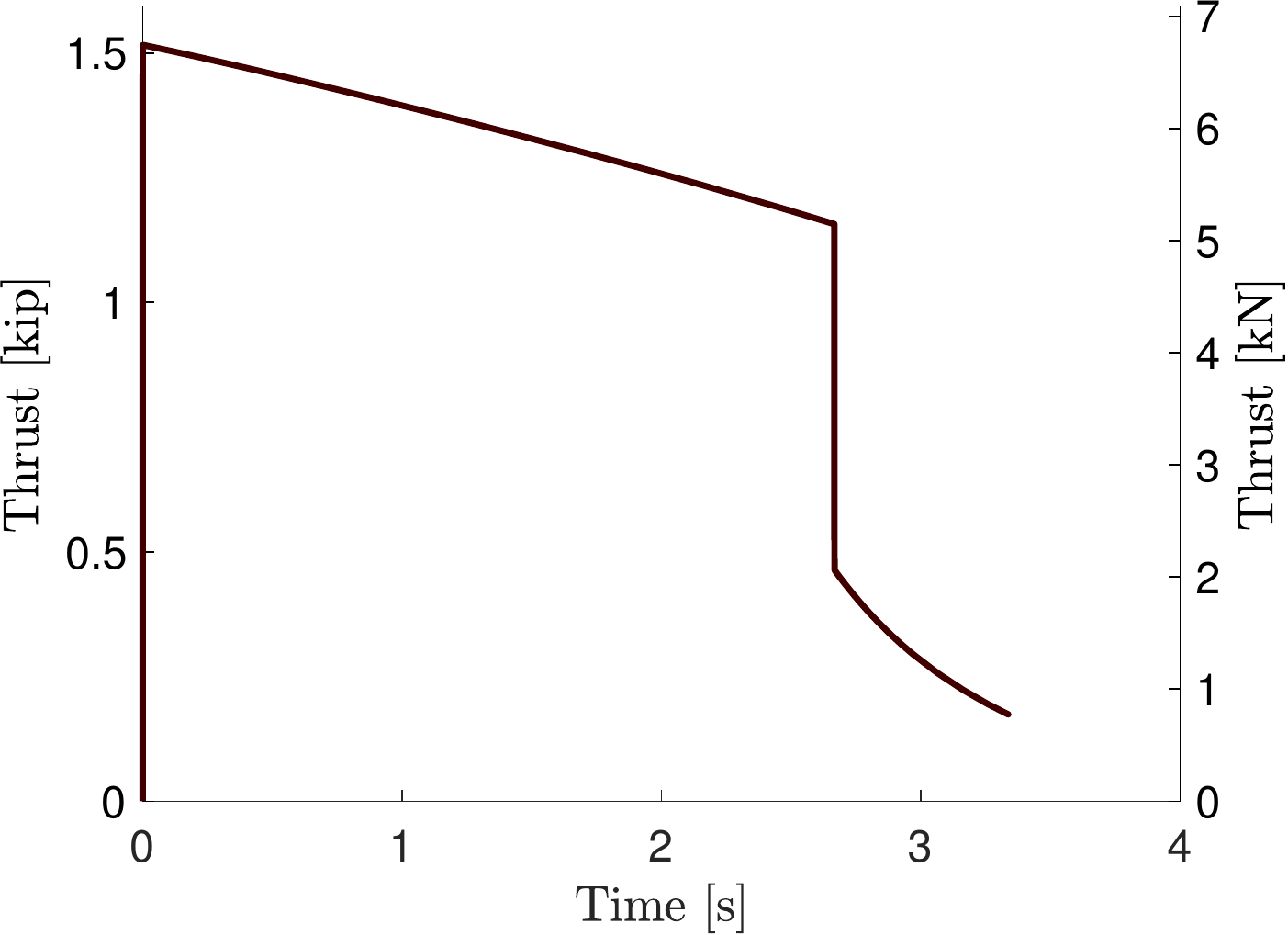}
	\caption{Thrust curve.}
	\label{fig:unsteady_thrust} 
	\end{subfigure}
	\caption{Time-resolved unsteady model engine transient performance.}
	\label{fig:unsteady_engine}
\end{figure}

The transient behavior of the engine is shown in Fig.~\ref{fig:unsteady_engine}. The hybrid propulsion system goes through different phases. Initially, the flight tank contains a liquid-vapor \nitrous{} mixture. As the dense liquid is delivered to the injector and it exits the tank, the volume available to the vapor increases, causing it to expand and lose pressure. Liquid \nitrous{} then evaporates at the liquid-vapor interface, in an attempt to maintain the saturation pressure of the propellant in the tank. Due to the endothermic nature of evaporation, the tank temperature decreases, causing  a decrease in the saturation vapor pressure, which leads to a transient decay of the tank pressure, as shown in Fig.~\ref{fig:unsteady_pressures}. As the \nitrous{} is initially delivered to the combustion chamber and the combustion gases accumulate due to the nozzle choking, the chamber pressure increases to eventually reach a peak of 27.6~bar (400~psia). Then, as the upstream tank pressure decreases in time, the chamber pressure as well decreases. Similarly, the nozzle thrust increases to a peak of 6.74~kN and eventually decreases as the chamber feed pressure gradually decreases, as shown in Fig.~\ref{fig:unsteady_thrust}. At $t \approx$~2.67~s, the liquid \nitrous{} is depleted from the flight tank, and the vapor occupies the entire tank volume. The ensuing process is the gaseous oxidizer blowdown, as the vapor expands polytropically in the tank while it is being delivered to the combustion chamber. Since the vapor has a much lower density than the liquid, the oxidizer mass flow rate through the injector significantly decreases. This results in an appreciable decrease of the chamber pressure and nozzle thrust. At $t \approx$~3.34~s, the fuel in the combustion chamber is depleted, and this characterizes engine burnout. There remains $\approx$~677~g of \nitrous{} vapor in the tank, which does not directly contribute to combustion. However, loading an excess of oxidizer is necessary to maintain tank pressure throughout the combustion.

\begin{figure}[h]
	\centering 
	\includegraphics[width=0.49\textwidth]{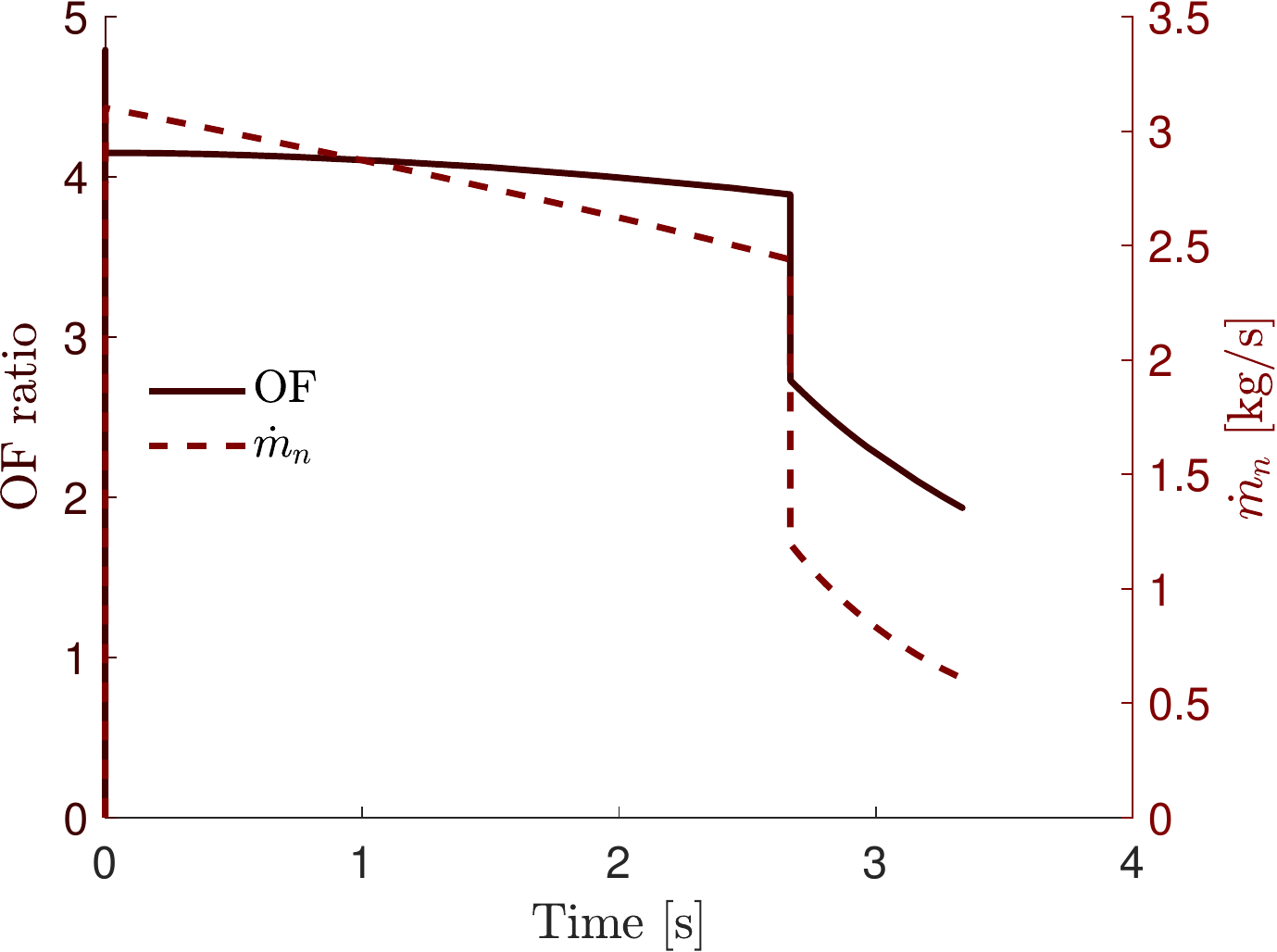}
	\caption{Shifts of the oxidizer-to-fuel mass ratio and nozzle mass flow rate resolved by the unsteady model.}
	\label{fig:unsteady_engineshift}
\end{figure}

Table~\ref{tab:unsteady_design} shows the propellant masses required to bring the vehicle to target apogee. The total propellant mass is 8673~g, which is 15.1\% superior to the 7535~g of propellant determined in Table~\ref{tab:steady_design} from the steady-state model. In fact, the specific impulse decreases from 223~s to 198~s. This appreciable performance decrease is the result of the transient performance decay of the hybrid engine, due to the shifting OF ratio during the combustion, and the decreasing chamber pressure causing a decrease in nozzle mass flow rate, as shown in Fig.~\ref{fig:unsteady_engineshift}. 

The shifting OF ratio is an inherent behavior of hybrid engines, due to the transient fuel grain regression behavior. The shift in OF ratio causes the combustion gases composition to change. Since the nozzle area ratios are optimized for only one specific gas composition, the performance of the nozzle decreases. The underlying physical process of the OF shift is described as follows: As the fuel grain port internal radius increases, the flux rate of oxidizer entering the chamber, $\Go = \mdoto / (\pi \rf^2)$, decreases. This results in a decrease of the erosion rate of the fuel grain, and therefore of the mass flow rate of fuel entering the chamber, which tends to increase the OF ratio. However, the decaying tank pressure results in a decrease of oxidizer mass flow rate entering the chamber, which tends to decrease the OF ratio. In the specific case of \engine{}, these competing effects lead to a negligible shift of the OF ratio during the majority of the burntime, as can be seen in Fig.~\ref{fig:unsteady_engineshift}. This is due to the relatively small dimensions of the fuel grain and the short burntime. 

Hence, a much larger portion of the performance penalty comes from the decreasing chamber pressure and resulting decreasing nozzle mass flow rate, as shown in Fig.~\ref{fig:unsteady_engineshift}. From Eq.~(\ref{eq:mdotn_unsteady}), the nozzle mass flow varies linearly with chamber pressure, and Eq.~(\ref{eq:thrustunsteady}) shows the engine thrust is directly proportional to $\mdotn$. Hence, the decreasing chamber pressure as seen in Fig.~\ref{fig:unsteady_pressures} results in an appreciable decrease in nozzle thrust. This performance penalty could partially be addressed by using an external inert pressurant in the oxidizer tank, as opposed to relying purely on the self-pressurizing capabilities of \nitrous{}. This would allow to better maintain the oxidizer mass flow rate to the chamber, hence to maintain chamber pressure and resulting engine thrust. Future engine designs of the MRT should consider this option for higher apogee targets.

\subsection{Flight Profile}

\begin{figure}[h]
	\centering 
	
	\begin{subfigure}{.49\textwidth}
	\centering 
	\includegraphics[width=\textwidth]{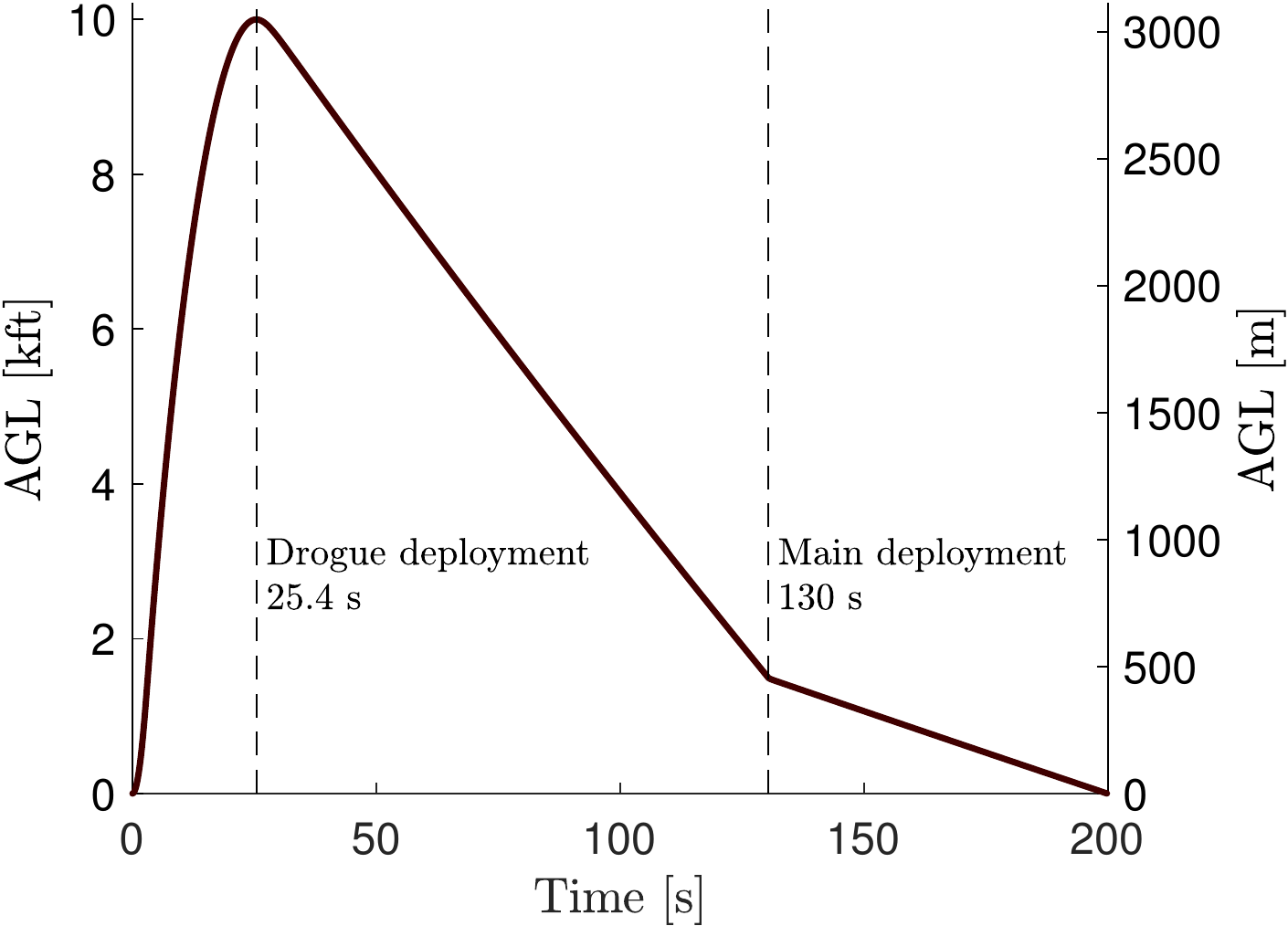}
	\caption{Elevation above ground level.}
	\label{fig:unsteady_agl} 
	\end{subfigure} \hfill
	\begin{subfigure}{.49\textwidth}
	\centering 
	\includegraphics[width=\textwidth]{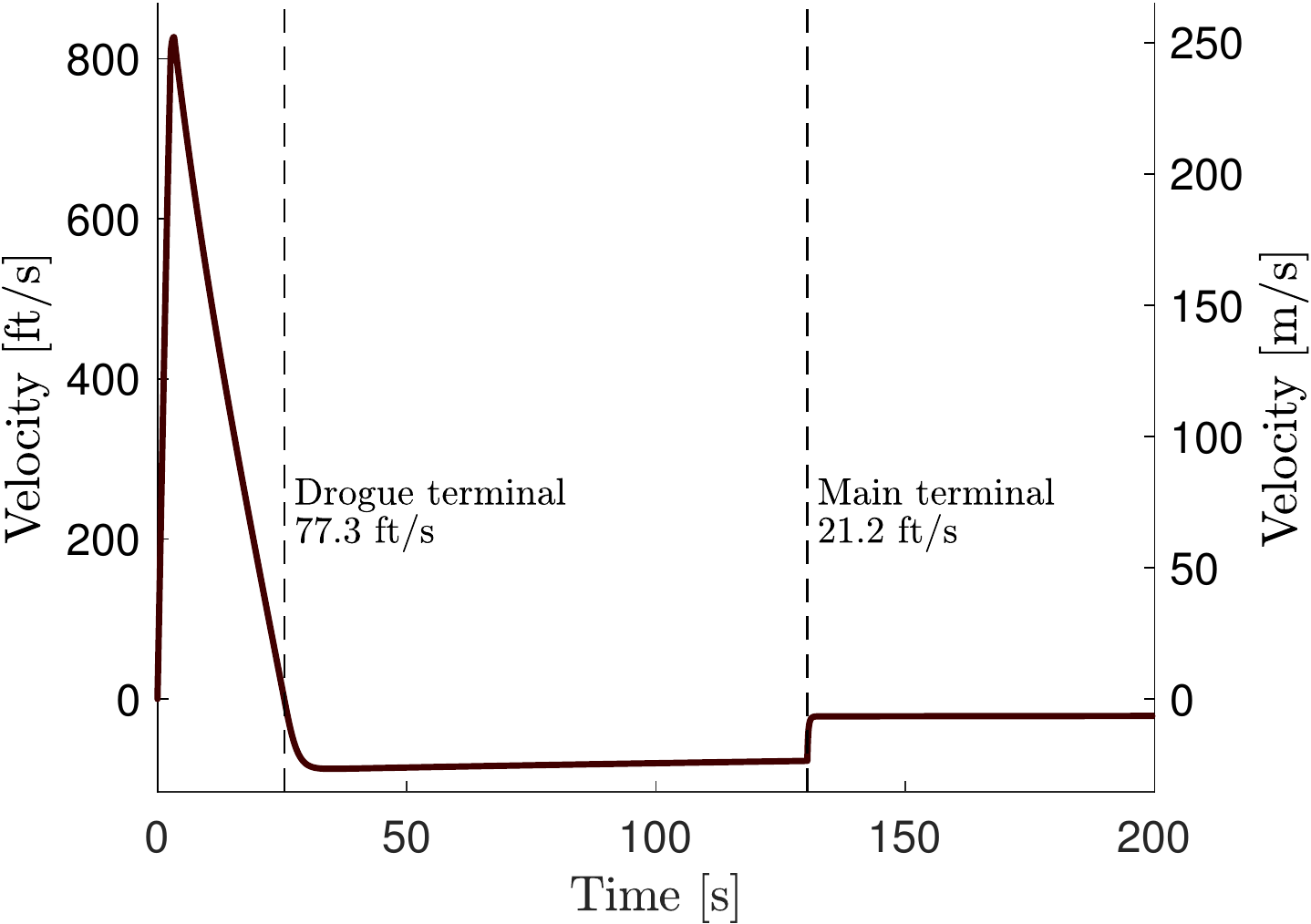}
	\caption{Total velocity.}
	\label{fig:unsteady_vtotal} 
	\end{subfigure}
	
	\begin{subfigure}{.49\textwidth}
	\centering 
	\includegraphics[width=\textwidth]{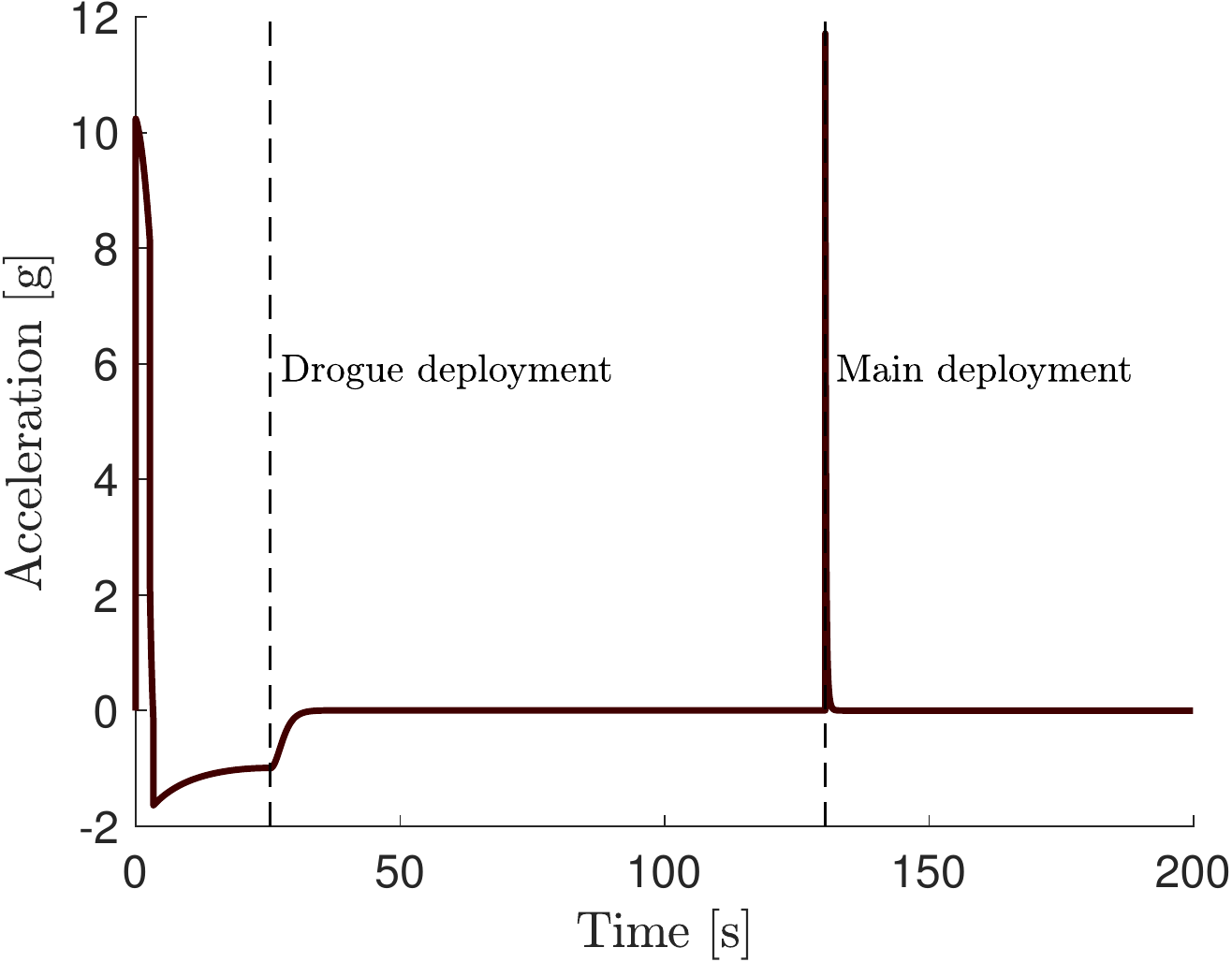}
	\caption{Total acceleration.}
	\label{fig:unsteady_gs} 
	\end{subfigure} \hfill
	\begin{subfigure}{.49\textwidth}
	\centering 
	\includegraphics[width=\textwidth]{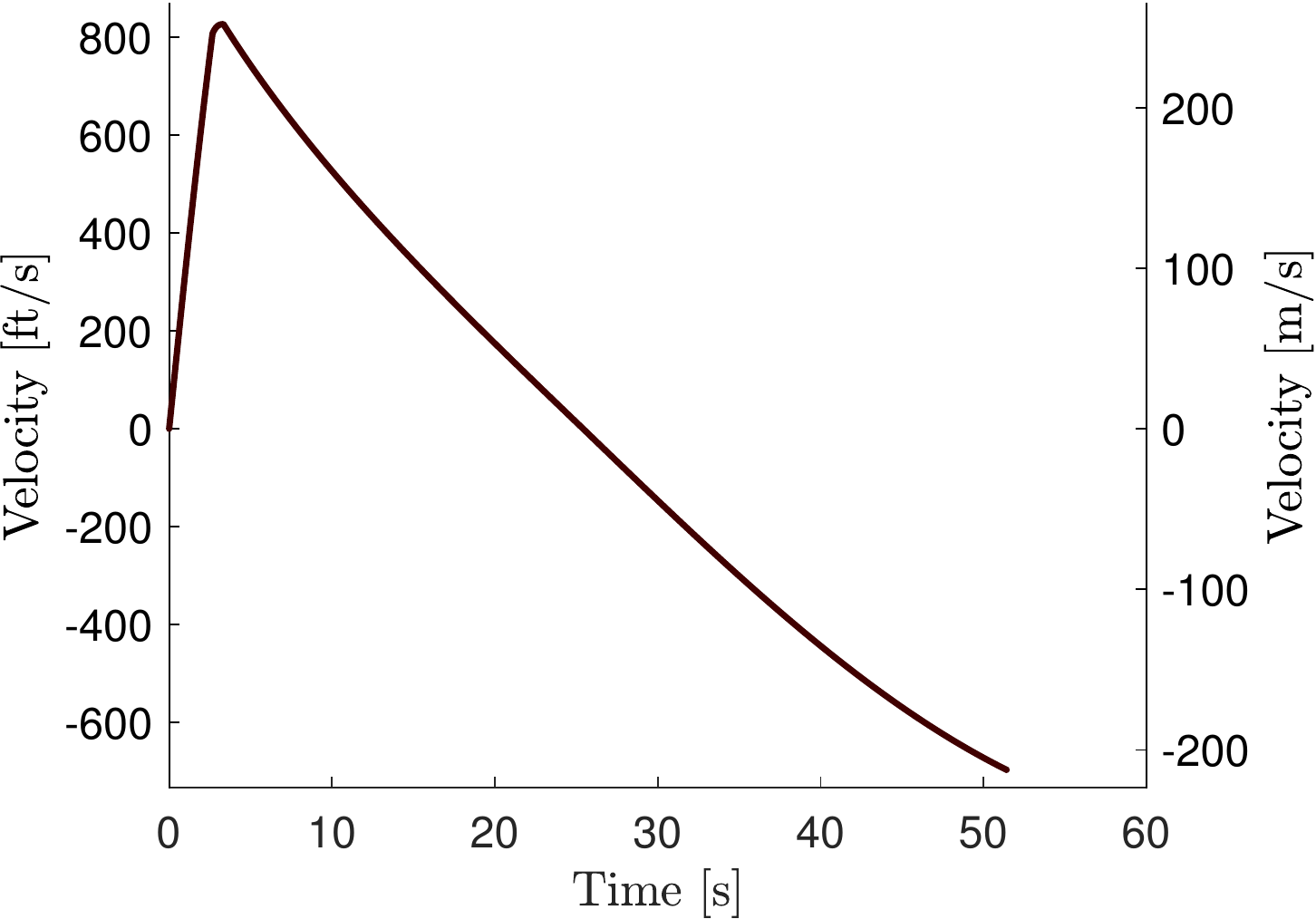}
	\caption{Total velocity, unsuccessful parachutes deployment.}
	\label{fig:unsteady_freefall} 
	\end{subfigure}	
	
	\caption{Time-resolved rocket flight profile.}
	\label{fig:unsteady_flight}
\end{figure}

\begin{table}
	\small
	\centering
	\caption{Unsteady model flight path parameters.}
	\label{tab:unsteady_flight_path}
	\begin{tabular}{c|c}
	\textbf{Description} & \textbf{Value} \\ \hline
	Effective rail length & 6.05~m (238~in) \\	
	Rail departure velocity & 34.0~m/s (111~ft/s) \\
	Predicted apogee &  3048~m (10,000~ft) \\
	Peak velocity & 252~m/s (827~ft/s) \\
	Peak Mach number & 0.757 \\
	Peak acceleration & 11.7$\gsl$ \\
	Terminal velocity under drogue & 23.6~m/s (77.3 ft/s) \\
	Terminal velocity under main & 6.45~m/s (21.2~ft/s) \\ 
	\end{tabular}
\end{table}

Figure~\ref{fig:unsteady_flight} shows the flight profile of the rocket, and Table~\ref{tab:unsteady_flight_path} compiles some flight path performance parameters. During the 3.34~s of engine burn, Fig.~\ref{fig:unsteady_vtotal} shows the rocket reaches a peak velocity of 252~m/s (827~ft/s), or Mach 0.757. \rocket{} therefore remains in the subsonic flight regime and does not encounter a shock wave in-flight, which is an important consideration for rocket structural design and fin design. The peak acceleration provided by \engine{} is shown in Fig.~\ref{fig:unsteady_gs} to be 10.2$\gsl$. After engine burnout, the rocket undergoes a coasting decelerating ascent until apogee, which is attained 25.4~s after engine ignition, as shown in Fig.~\ref{fig:unsteady_agl}. The predicted apogee is 3048~m (10,000~ft). The drogue parachute is then deployed, and ensures the rocket descent does not exceed a terminal velocity of 23.6~m/s (77.3~ft/s), as shown in Fig.~\ref{fig:unsteady_vtotal}. Then, 130~s after ignition and at 457~m AGL (1500~ft), the main parachute is deployed, further slowing down the rocket to a touchdown terminal velocity of 6.45~m/s (21.2~ft/s). Figure~\ref{fig:unsteady_gs} shows the acceleration provided by the main parachute deployment of 11.7$\gsl$ exceeds the peak acceleration provided by the engine; hence, this parameter should be considered for rocket structural design.

In the catastrophic case where both the drogue and the main parachute deployments fail, Fig.~\ref{fig:unsteady_freefall} shows the rocket hits the ground at a terminal velocity of 213~m/s (698~ft/s), ensuring \rocket{} would be shattered to pieces.

\subsection{Stability Analysis} \label{sec:unsteady_stability}

One model input shown in Table~\ref{tab:unsteady_flight_path} not previously mentioned is the effective rail length of 6.05~m (19.8~ft), with the associated output, the rocket rail departure velocity of 34.0~m/s (111~ft/s). The effective rail length corresponds to the distance traveled by the rocket on the launch rail at the moment it becomes free to move about the pitch, yaw, or roll axis. As stated in \cite{sac2022}, this typically occurs when the last rail guide forward of the center of gravity (CG) of the vehicle   separates from the launch rail. The effective rail length of 6.05~m (19.8~ft) corresponds to the design of the MRT. A rule of thumb is that a minimum rail departure velocity of 30.5~m/s (100~ft/s) should be attained to ensure a stable flight \cite{sac2022}; otherwise, detailed stability analysis should be conducted. The software \textit{OpenRocket} allows to conduct such detailed stability analysis. Although the rail departure velocity predicted in the current model exceeds 30.5~m/s (100~ft/s), such an analysis is conducted.

\subsubsection{Generation of Transient Engine File}

The transient engine test data is used to generate a .rse \textit{OpenRocket} engine file. Note this format differs from the .eng file generated in Section~\ref{sec:steady_orckt}: the .eng format is designed for solid rocket motors, and it assumes the motor CG is located at its middle point at all times. In contrast, the .rse format allows to specify the axial location of the engine CG as a function of time, CG\sub{total}$(t)$, which permits an accurate representation of hybrid engines. Since the stability analysis depends on the rocket CG location with respect to its  center of pressure (CP), the accurate representation of CG\sub{total}$(t)$ is critical.

\begin{figure}
    \centering
    \includegraphics[width=0.8\textwidth]{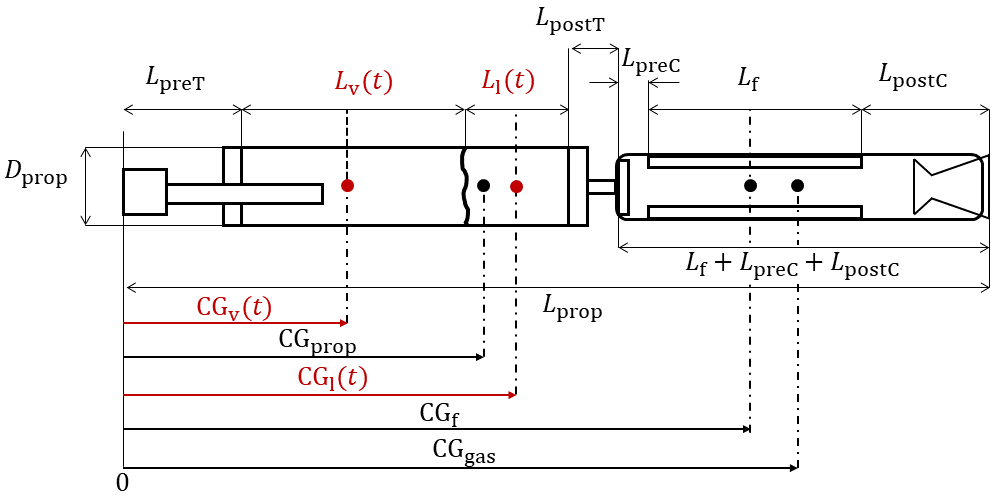}
    \caption{Geometric definition of the propulsion system center of gravity (CG). Dimensions and the location of $\cgprop$ are obtained from the CAD representation of \engine{}. The locations of $\cgv$, $\cgl$, $\cgf$, and $\cggas$ are calculated during engine modeling.}
    \label{fig:rse_engine}
\end{figure} 

The .rse engine file format is described in \cite{openrocket2021a}, and the parameters used to calculate CG\sub{total}$(t)$ are shown in Fig.~\ref{fig:rse_engine}. The mass of \nitrous{} vapor in the tank as a function of time is $\mv(t) = \Wo \nv(t)$, and the mass of \nitrous{} liquid is $\ml(t) = \Wo \nl(t)$. The lengths occupied by each phase in the tank are $\Lv(t)$ and $\Ll(t)$, and can be obtained from the molar volumes of each phase at $\Tt(t)$, and from the tank internal dimensions. The CGs of the vapor and liquid \nitrous{}, $\cgv$ and $\cgl$, are located at the middle point of the respective length occupied by the phases in the tank; they travel downwards during the combustion. The mass of the fuel grain over time is $\pi \rhof \Lf (\Rf^2 - [\rf(t)]^2)$, and its center of gravity $\cgf$ is located at the middle point of the fuel grain length. The mass of gases stored in the chamber is $\mc(t) = \mf(t) + \mo(t)$, and its center of gravity $\cggas$ is located in the middle of the engine casing. The locations of $\cgf$ and $\cggas$ do not change over time. The mass of the dry engine (no \nitrous{} and no fuel) is $\mprop$, located at $\cgprop$; these parameters are obtained from computer-aided design (CAD) software. Then,
\beq
\text{CG}_\text{total}(t) = \f{\sum_j m_j(t) [\text{CG}_j(t)]}{\sum_j m_j(t)}
\eeq 
where $j~\in~\{\text{v, l, f, gas, prop}\}$.

\subsubsection{Stability at Different Wind Speeds}

\begin{figure}[h]
	\centering 
	\begin{subfigure}{.49\textwidth}
	\centering 
	\includegraphics[width=\textwidth]{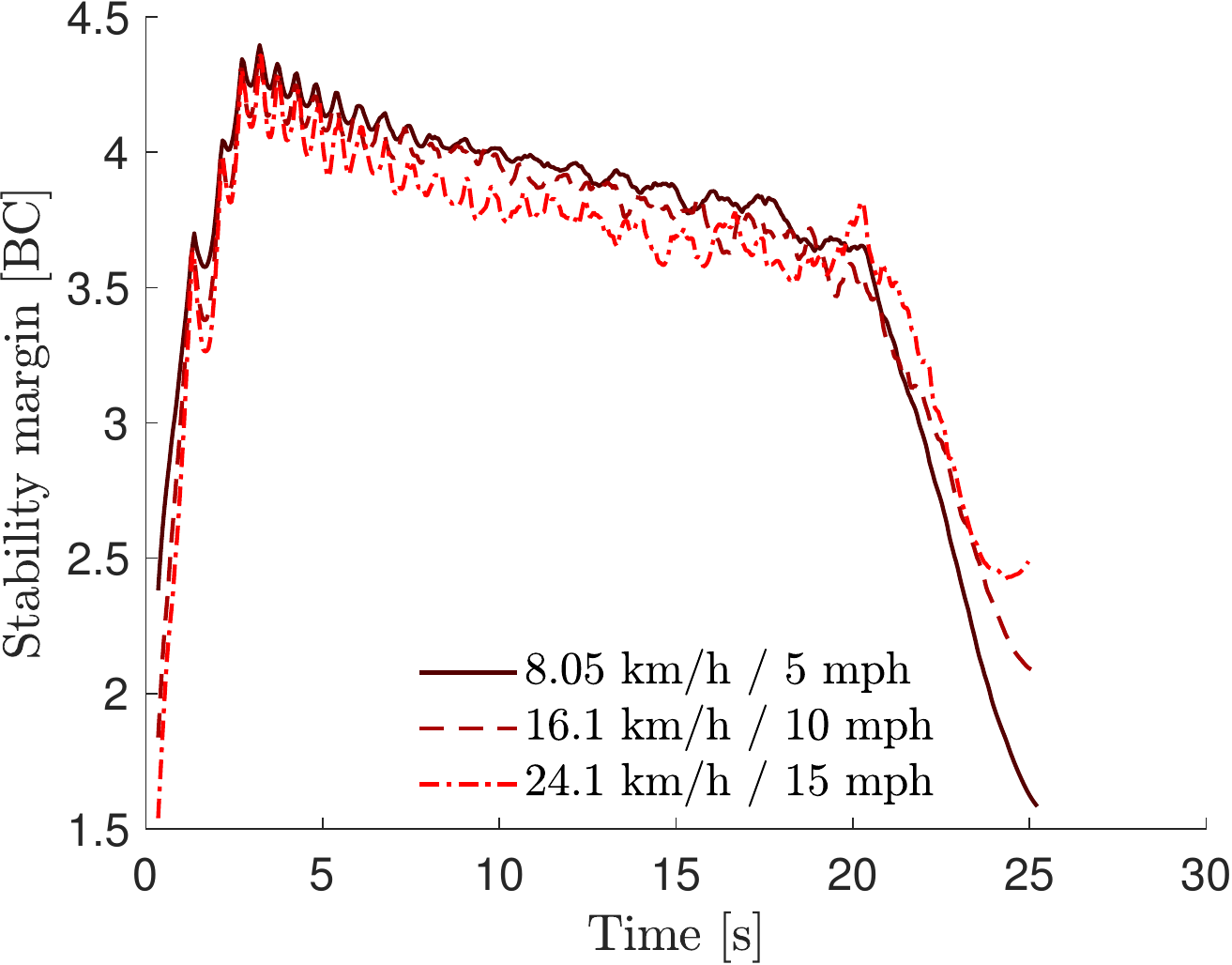}
	\caption{Ascent path.}
	\label{fig:stability} 
	\end{subfigure} \hfill
	\begin{subfigure}{.49\textwidth}
	\centering 
	\includegraphics[width=\textwidth]{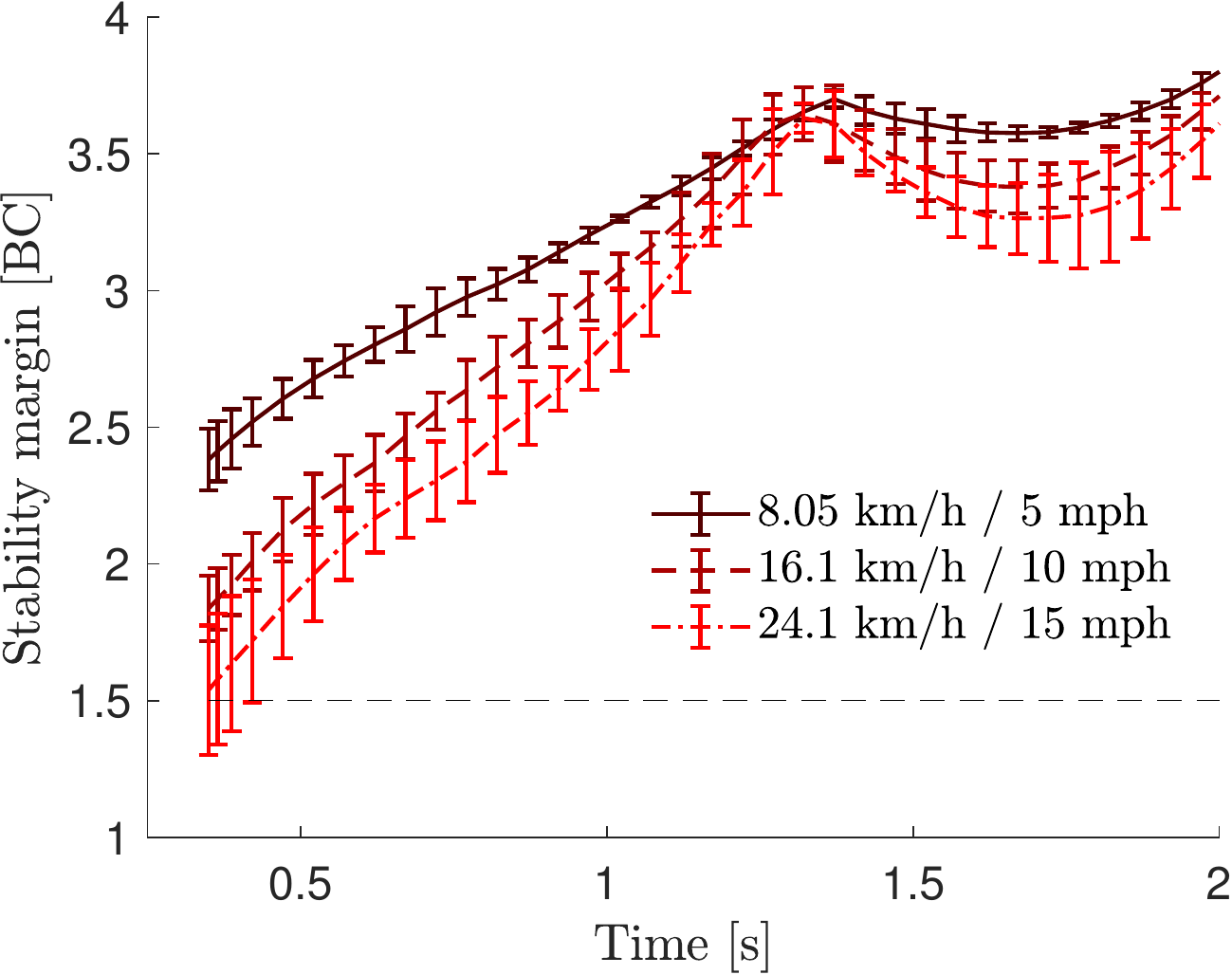}
	\caption{Zoomed-in near rail departure.}
	\label{fig:stabilityzoom} 
	\end{subfigure}
	\caption{Time-resolved stability margin body calipers of \rocket{}.}
	\label{fig:stabilityanalysis}
\end{figure}

The .rse engine file is input into \textit{OpenRocket} and the flight path of \rocket{} is modeled for three different wind speeds, varying between 8.05-24.1~km/h (5-15~mph), which is representative of wind speeds at the SAC launch site. Figure~\ref{fig:stabilityanalysis} shows the time-resolved stability margin body calipers (BC) of \rocket{}. To account for the stochastic behavior of wind turbulence models, the simulation is carried five times for each wind speed, and mean results are reported. Error bars shown in Fig.~\ref{fig:stabilityzoom} correspond to two standard deviations on each side of the mean, which statistically represents 95\% of the data.

To ensure stability, a minimum BC of 1.5 must be ensured at all times, and to avoid over-stability, a maximum margin of 6 is recommended \cite{sac2022}. Figure~\ref{fig:stability} shows the mean values remain between 1.5 and 4.5 for the entire ascent path for all wind speeds, which is satisfactory. Additionally, the most critical period is off-rail departure, which is shown with error bars in Fig.~\ref{fig:stabilityzoom}. The minimum BC of 1.5 is cleared for wind speeds between 8.05-16.1~km/h (5-10 mph), but the error bars lead to an under-stable departure for a wind speed of 24.1~km/h (15~mph). Hence, with the hereby-designed \engine{} engine, \rocket{} should be flown in wind speeds below 16.1~km/h (10~mph).


\chapter{Hot Fire Testing and Model Validation} \label{chapter:hotfire}

During the development campaign of \engine{}, a hybrid engine testing facility was designed and built by the MRT, and used to conduct live hot fire engine testing. In the current chapter, the test site plumbing installation is briefly described, and the unsteady model is calibrated against hot fire testing data.  


\section{Hybrid Engine Test Site}

\subsection{Plumbing Configuration}

\begin{figure}
    \centering
    \includegraphics[width=0.9\textwidth]{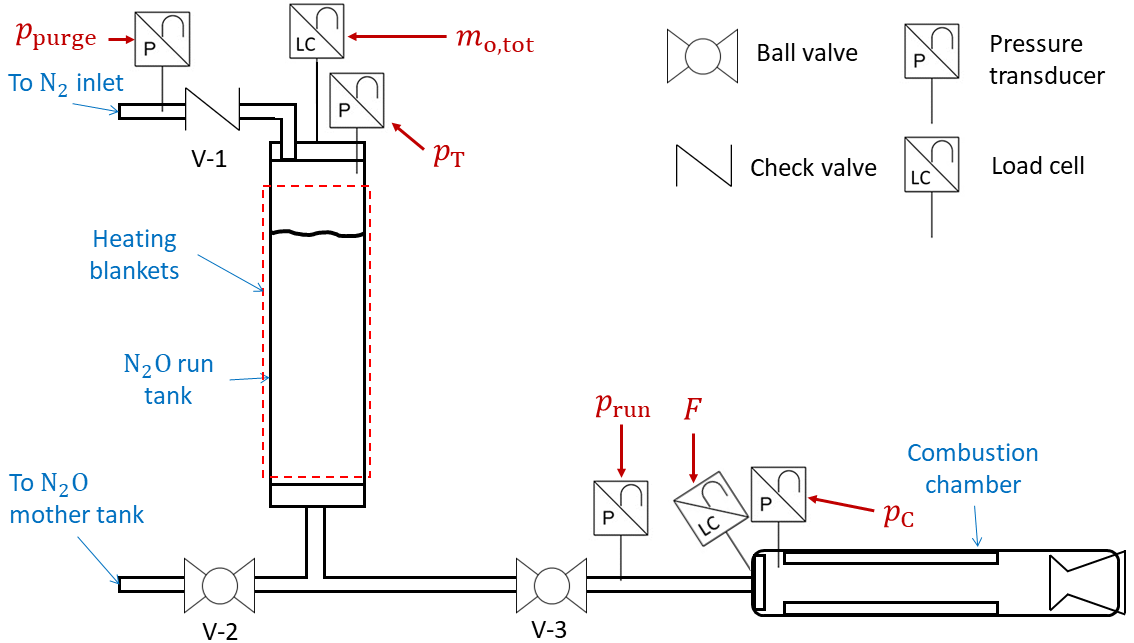}
    \caption{Simplified representation of test site layout and instrumentation.}
    \label{fig:testsite_schematic}
\end{figure}

A detailed description of the hybrid engine testing facility built by the MRT is beyond the scope of the current work. Instead, the plumbing system of the test site is briefly described, and some notable features which impact model validation are outlined. 

\begin{figure}

	\centering 
	\begin{subfigure}{\textwidth}
	\centering 
    \includegraphics[width=\textwidth]{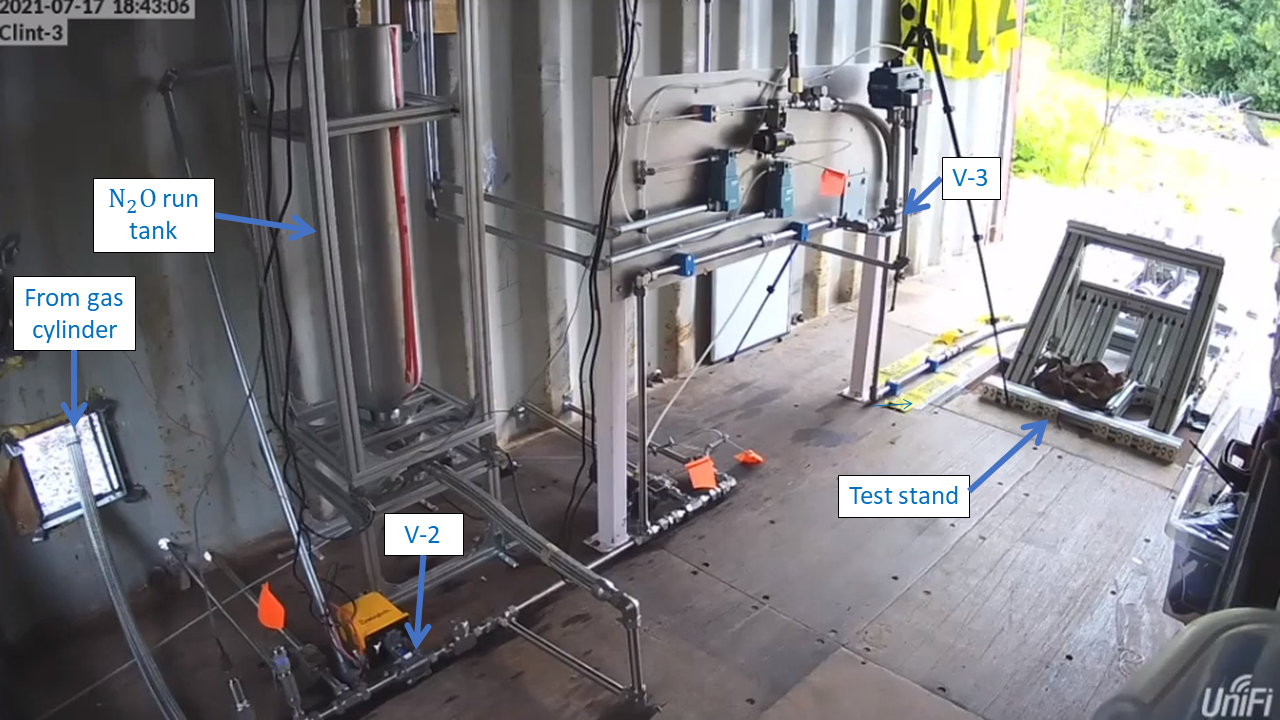}
    \caption{First iteration without insulation.}
    \label{fig:testsite_interior}
	\end{subfigure} 
	
	\begin{subfigure}{\textwidth}
	\centering 
    \includegraphics[width=\textwidth]{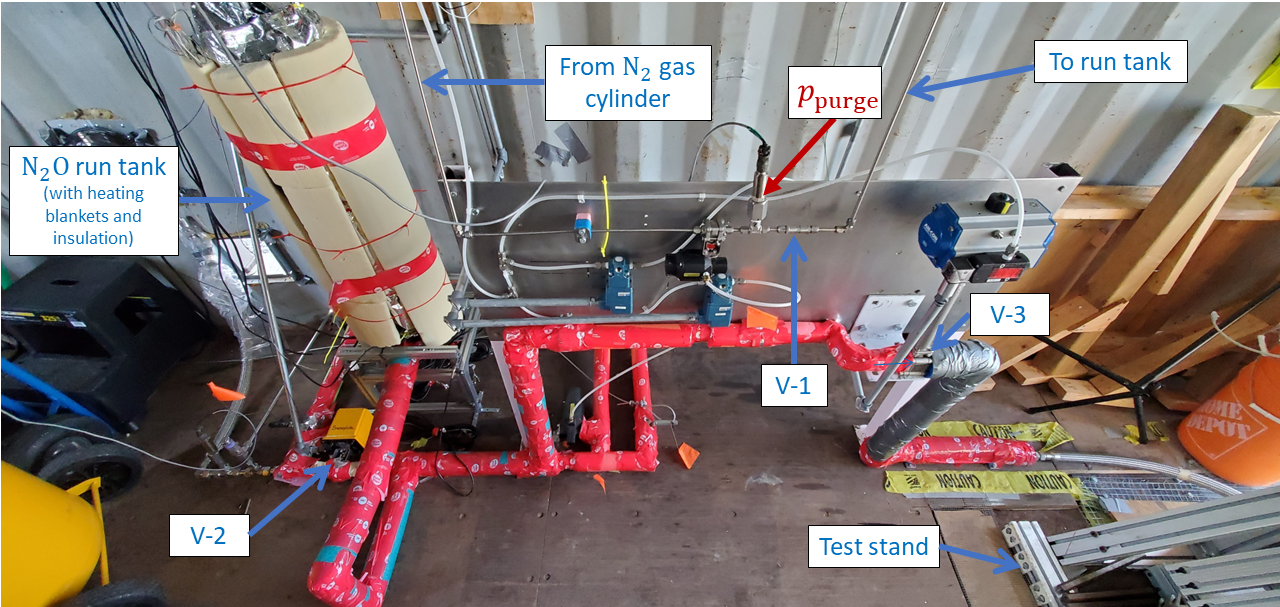}
    \caption{Insulated layout.}    \label{fig:testsite_insulated}
	\end{subfigure}
	\caption{Overview of container interior.}
	\label{fig:testsite_overview}
	
\end{figure}

The hybrid test site system is designed around a 6-meter (20-feet) outdoor shipping container. Figure~\ref{fig:testsite_schematic} shows a simplified schematic representing the container interior plumbing configuration, and Fig.~\ref{fig:testsite_overview} shows the installed infrastructure. A \nitrous{} K cylinder is located in a gas cylinder cage outside the container, and it is routed to the \nitrous{} run tank through valve V-2. The run tank is used during engine firing, and it is routed through valve V-3 to the combustion chamber, which is installed on the test stand. At the top of the run tank, an inlet of high-pressure nitrogen (\nitrogen{}) is installed and isolated from the run tank with check valve V-1. The inlet is connected to a \nitrogen{} K gas cylinder with a regulator.  The \nitrogen{} regulator set pressure is below the vapor pressure of \nitrous{} at ambient conditions, such that the pressurant is not used during the liquid-vapor blowdown process. However, \nitrogen{} is used during the gaseous blowdown to dilute the \nitrous{} vapor. This is a safety feature to avoid having a large amount of residual \nitrous{} vapor in the run tank, which is prone to decomposition. This safety feature impacts engine performance during the gaseous blowdown process.

\begin{figure}
    \centering
    \includegraphics[width=0.7\textwidth]{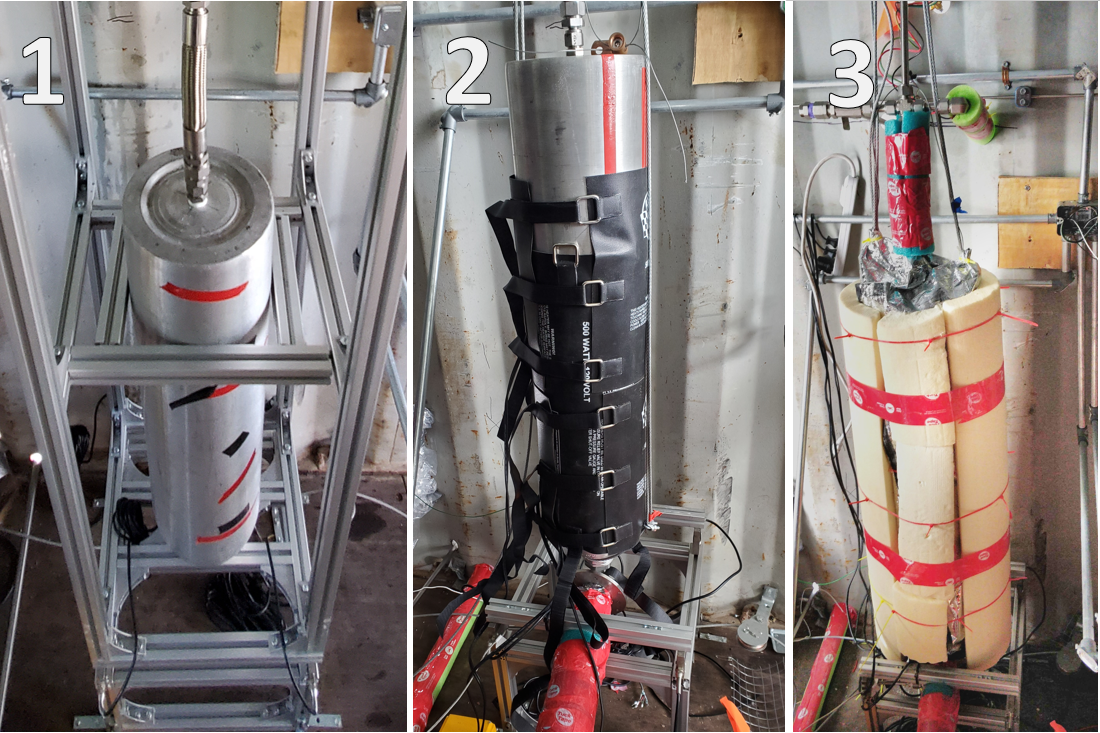}
    \caption{Nitrous run tank heating blankets and insulation.}
    \label{fig:tankheating}
\end{figure}

The target initial temperature of the \nitrous{} run tank is 20~\super{$\circ$}C, corresponding to a vapor pressure of 51.4~bar (745~psia). Achieving these conditions is a significant challenge during Quebec winter operation, where ambient temperatures routinely drop below -20~\super{$\circ$}C. To address this concern, the tank is fitted with two 500~W heating blankets on its exterior surface, as shown in Fig.~\ref{fig:tankheating}. The tank and plumbing lines are then insulated, as shown in Fig.~\ref{fig:testsite_insulated}. After filling procedures of the run tank and before engine firing, the heating blankets are activated to heat the \nitrous{} to the desired operating temperature through the tank walls. The blankets are then turned off just prior to engine firing. However, they reach a surface temperature above 120~\super{$\circ$}C during the heating. Consequently, the residual temperature gradient between the \nitrous{} and blankets results in transient heat addition to the \nitrous{} which is not captured by the unsteady model. The flight engine does not feature heating blankets, as it is designed for desert operation at the SAC.

\begin{figure}
	\centering 
	\begin{subfigure}{0.7\textwidth}
	\centering 
    \includegraphics[width=\textwidth]{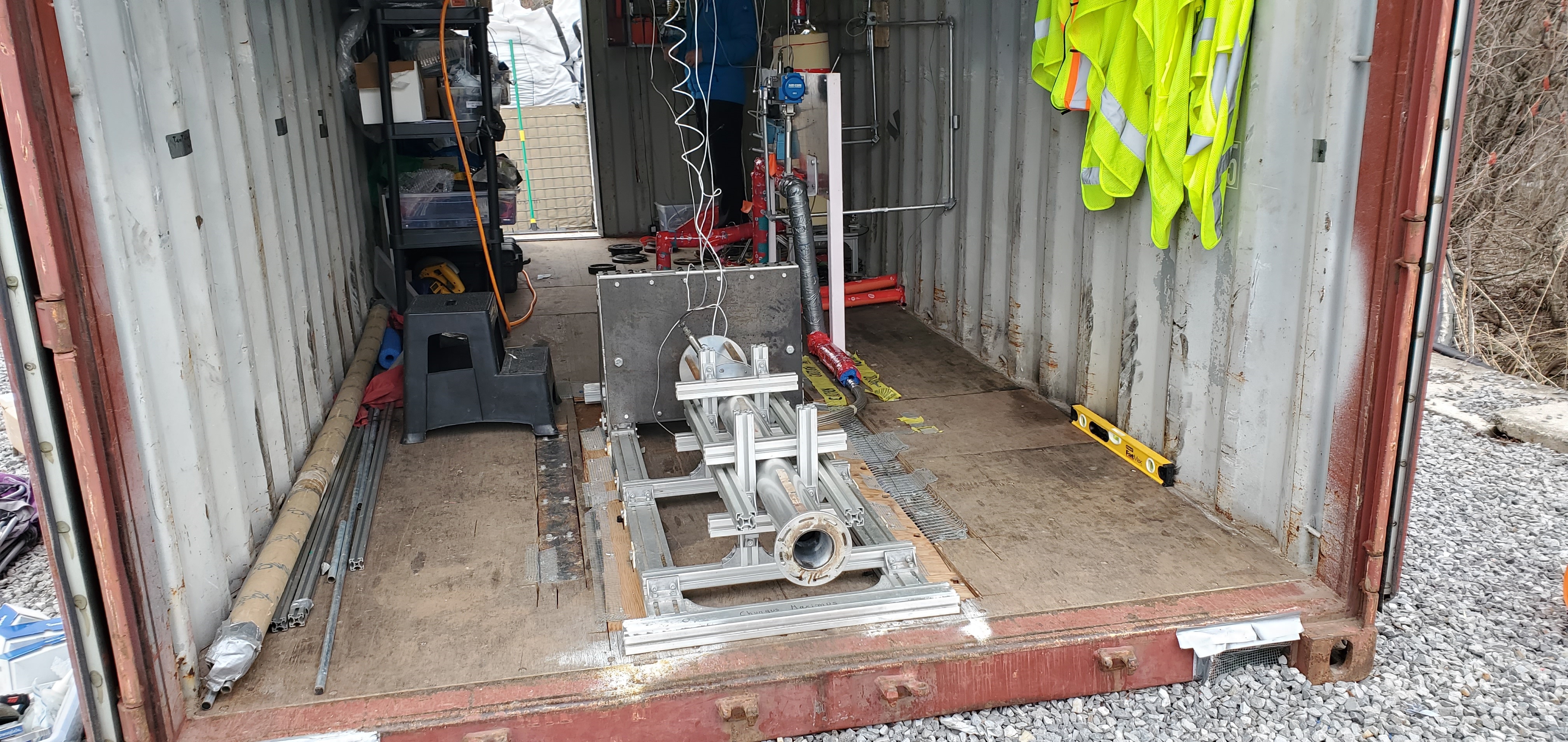}
    \caption{Test stand.}
	\end{subfigure} 
	\begin{subfigure}{0.7\textwidth}
	\centering 
    \includegraphics[width=\textwidth]{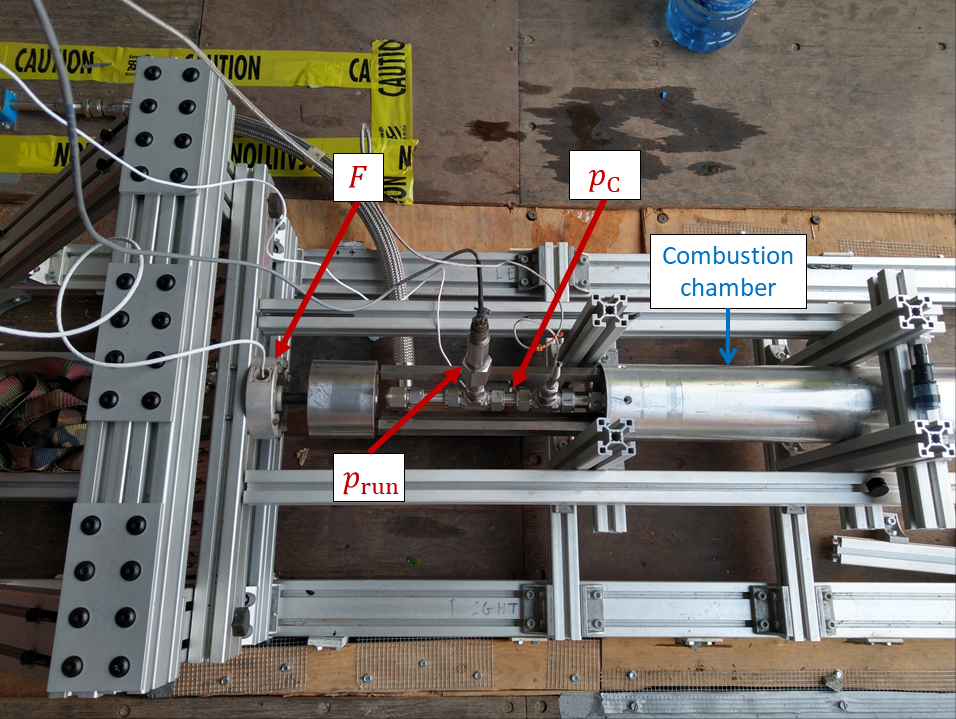}
    \caption{Line with pressure transducer and load cell.}    \label{fig:testsite_engineline}
	\end{subfigure}
	\caption{Test stand and engine overview.}
\end{figure}

Regarding instrumentation, the run tank does not have a dip tube, and the initial mass of \nitrous{} in the tank before firing, $\motot$, is measured through a load cell from which the tank is hung. The run tank is also fitted with a pressure transducer allowing to monitor its pressure, $\pt$, over time. As shown in Fig.~\ref{fig:testsite_overview}, there is a relatively long plumbing line between the run tank and the combustion chamber, with several elbows and bends, which create appreciable pressure losses. Hence, a pressure transducer is installed right before the injector inlet, as shown in Fig.~\ref{fig:testsite_engineline}, which allows to measure the pre-injector pressure, or run line pressure, $\prun$. An additional pressure transducer allowing to measure $\pc$ is installed on the combustion chamber, along with a load cell to measure the thrust $F$. Finally, a pressure transducer is installed before valve V-1 to monitor the \nitrogen{} purge line pressure.

\subsection{Engine Test Data}

\begin{figure}

	\centering 
	\begin{subfigure}{0.49\textwidth}
	\centering 
    \includegraphics[width=\textwidth]{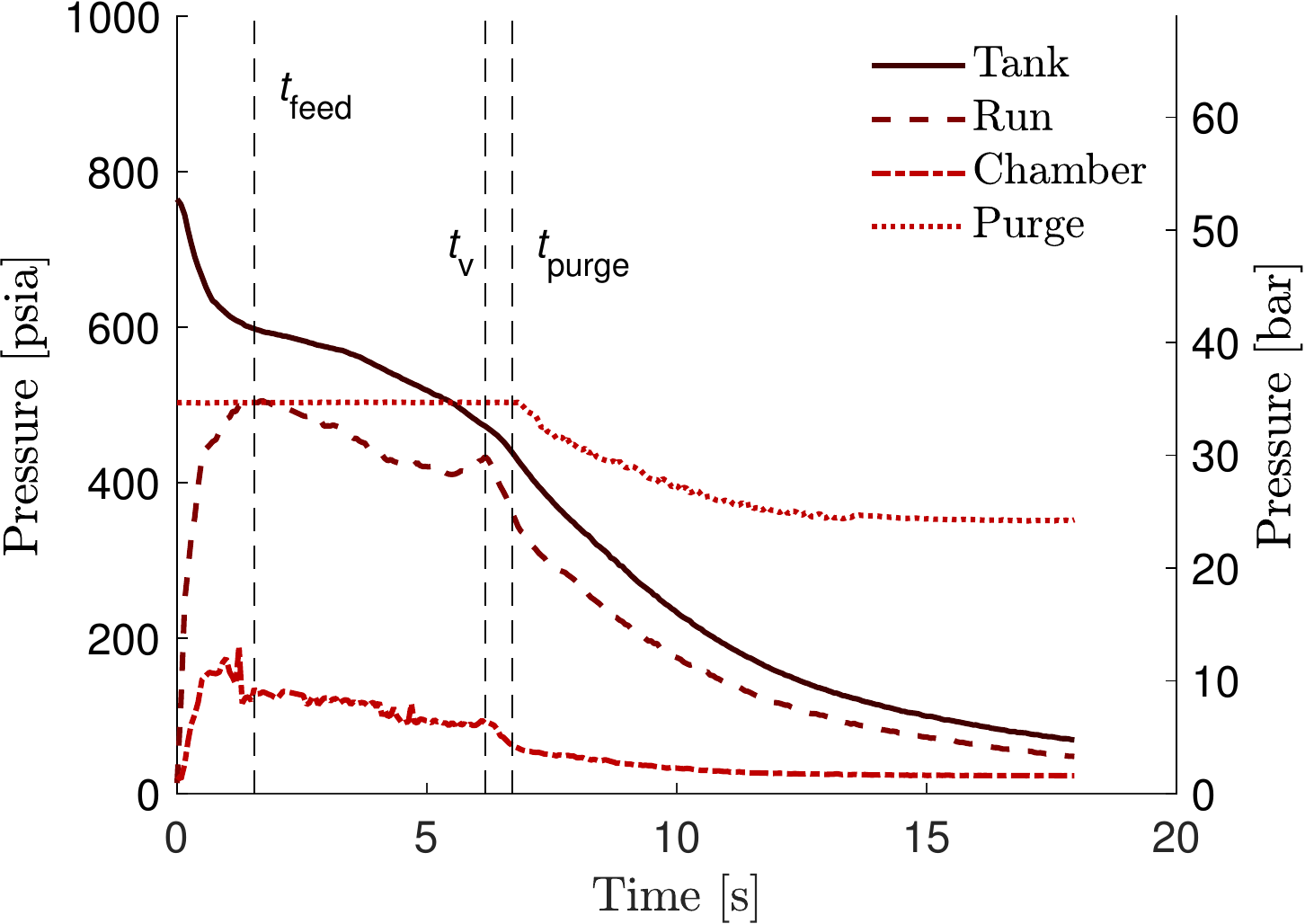}
    \caption{Hot Fire 4 pressures.}    
    \label{fig:hf4_pressure}
	\end{subfigure} 
	\begin{subfigure}{0.49\textwidth}
	\centering 
    \includegraphics[width=\textwidth]{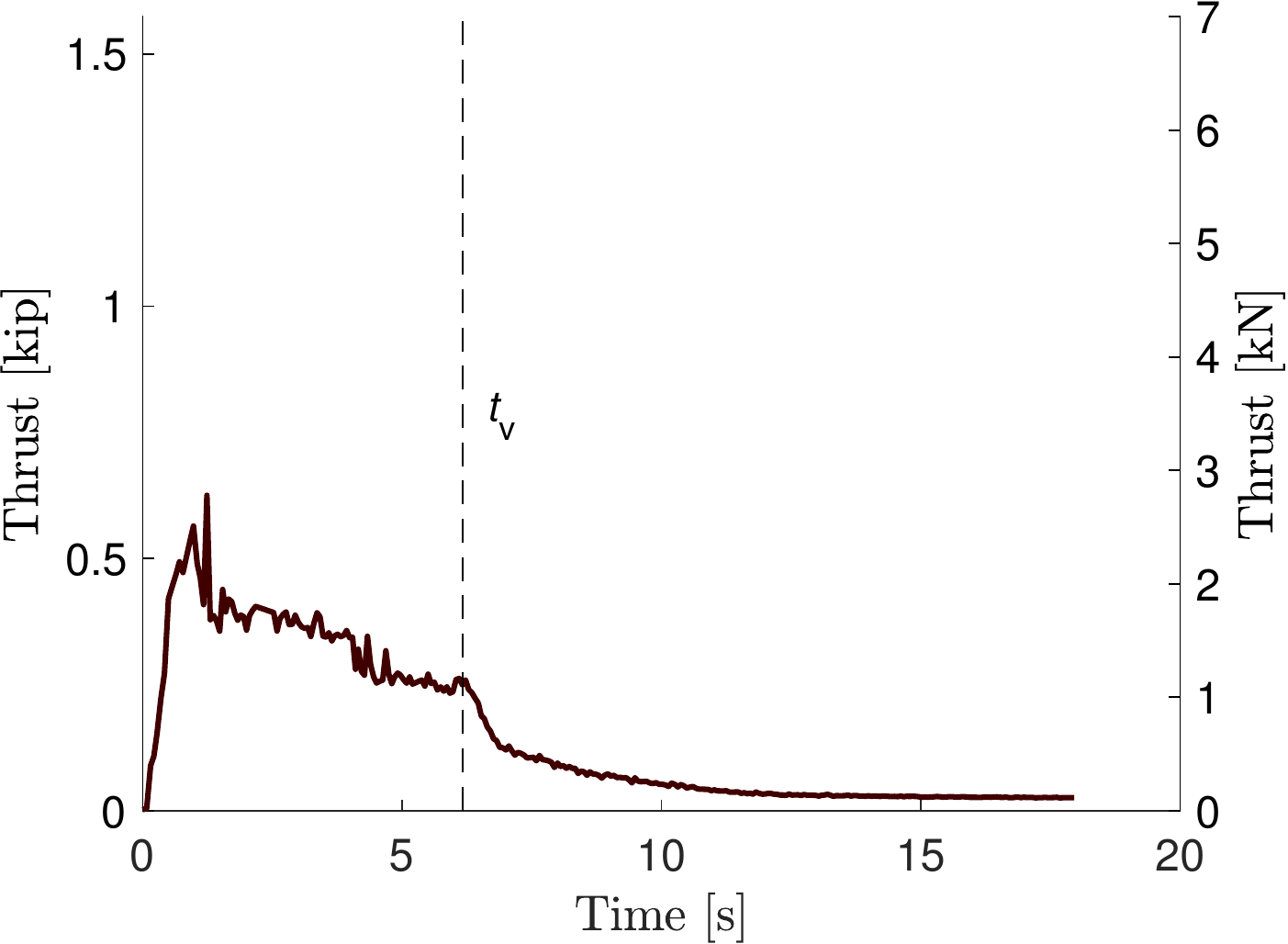}
    \caption{Hot Fire 4 thrust.}    
	\label{fig:hf4_thrust}
	\end{subfigure}
	
	\centering 
	\begin{subfigure}{0.49\textwidth}
	\centering 
    \includegraphics[width=\textwidth]{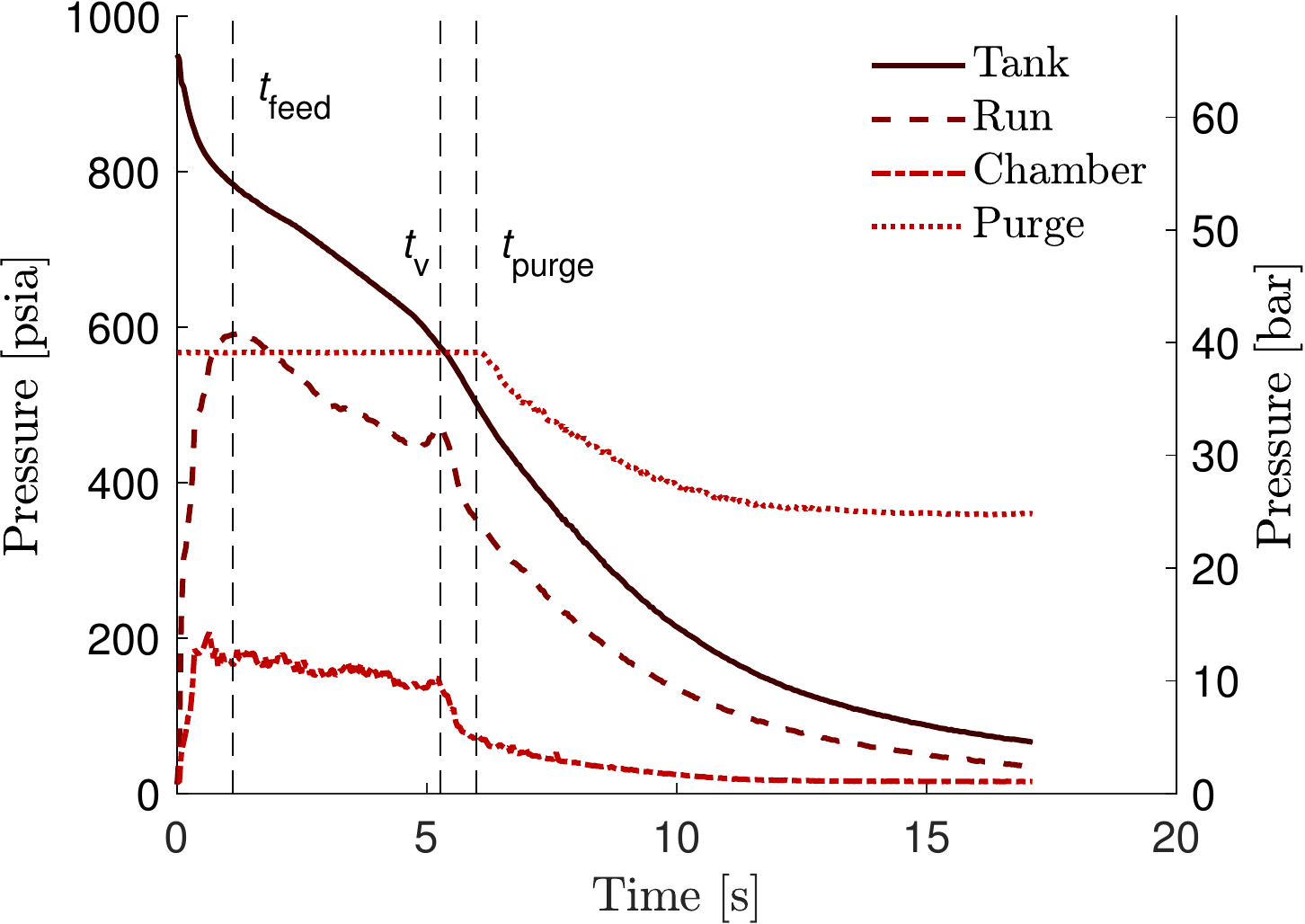}
    \caption{Hot Fire 5 pressures.}    
    \label{fig:hf5_pressure}
	\end{subfigure} 
	\begin{subfigure}{0.49\textwidth}
	\centering 
    \includegraphics[width=\textwidth]{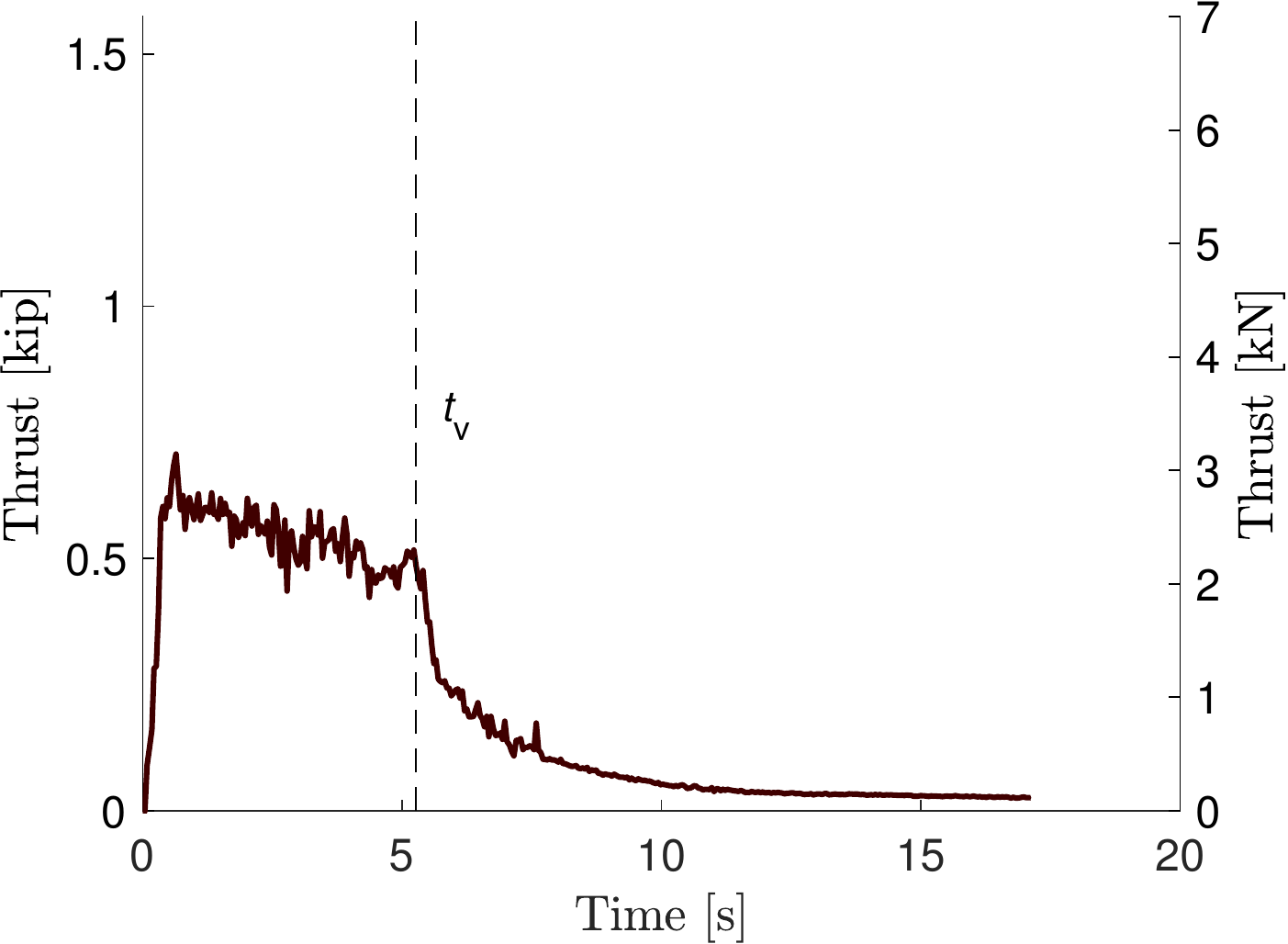}
    \caption{Hot Fire 5 thrust.}    
	\label{fig:hf5_thrust}
	\end{subfigure}

	\caption{Test data for Hot Fire 4, 2022-03-05, and Hot Fire 5, 2022-04-03.}
	\label{fig:hf45_data}
\end{figure}

Several hot fire tests were conducted during the 2021-2022 testing campaign of \engine{}; data for the two tests that produced useful thrust curves, Hot Fire 4 (HF4) and Hot Fire 5 (HF5), is shown in Fig.~\ref{fig:hf45_data}. The test site tank blowdown process can be separated in four phases. At $t = 0$, valve V-3 is opened, and the system enters phase~1. The \nitrous{} is discharged from the tank with a high flow rate to fill the volume between valve V-3 and the injector manifold, and starts entering the combustion chamber through the injector. During this period, the tank pressure slope is initially remarkably steep, as can be seen in Fig.~\ref{fig:hf4_pressure} and Fig.~\ref{fig:hf5_pressure} for $0 \leq t < \tfeed$, and it gradually decreases. The run line and chamber pressure slopes follow a similar trend. At $t = \tfeed$, the tank has lost $\approx 11$~bar (160~psia) of pressure from its starting value, because of the appreciable volume and large pressure drop between the tank and the injector manifold. In flight configuration, where the intermediate volume and pressure drop are negligible, it is expected that the initial pressure drop of the oxidizer tank will be significantly smaller. To account for this test site limitation realized from HF4 test data, the team over-pressurized the run tank for HF5 to 65.6~bar (951~psia), which is 14.2~bar (206~psia) above the target operating point.

As pressure builds up in the run line and the combustion chamber, the discharge rate from the tank to the run line decreases, and the tank-emptying eventually becomes limited by the discharge rate through the injector. The tank-emptying process enters phase~2, the liquid-vapor blowdown, for $\tfeed \leq t < \tv$. The flow in the run line is established, and the pressure drop $\ploss = \pt - \prun$ remains approximately constant. The tank, run line, and chamber pressures gradually decrease over time. During phases~1 and 2, Fig.~\ref{fig:hf4_thrust} and \ref{fig:hf5_thrust} show the engine thrust follows exactly the chamber pressure oscillations, with a high sensitivity. The peak chamber pressure attained is below 13.8~bar (200~psia) for both HF4 and HF5, with a resulting peak thrust around 3~kN. This indicates the injector provides an insufficient mass flow rate of \nitrous{} to the combustion chamber, compared to the target operating point.

At $t = \tv$, the liquid is depleted, and the system enters phase~3, the \nitrous{} vapor blowdown process, for $\tv \leq t < \tpurge$. The chamber pressure and engine thrust exhibit an appreciable decrease, as the fluid entering the combustion chamber has a much lower density. At $t = \tpurge$, the system enters phase~4, the \nitrous{}--\nitrogen{} gaseous blowdown process. At this moment, the purge line pressure $\ppurge$ starts decreasing, indicating \nitrogen{} is being fed to the run tank. Valve V-1 has a cracking pressure of $\approx$~4.14~bar (60~psia), hence it opens when the tank pressure has dropped below the purge line pressure by that amount. 

The unsteady model describes phase~2 and 3, and the engine behavior during these phases~qualitatively agree with the unsteady model predictions. In following sections, parameters used in the unsteady model are calibrated on the test data.


\section{Tank-Emptying Model Calibration}

\subsection{Estimation of Injector Discharge Coefficient}

\begin{figure}
	\centering 
	\begin{subfigure}{0.49\textwidth}
	\centering 
    \includegraphics[width=\textwidth]{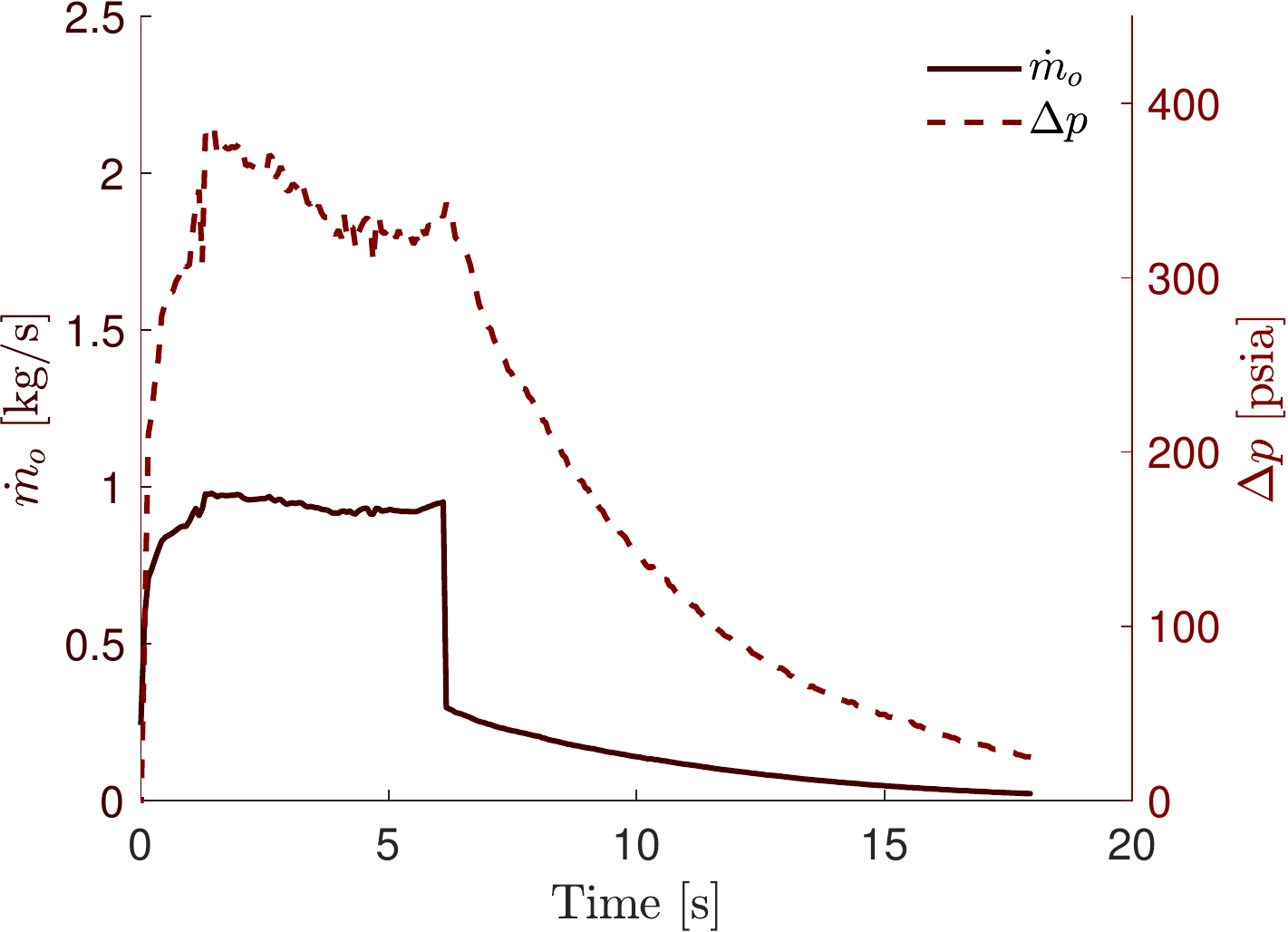}
    \caption{Hot Fire 4.}    
    \label{fig:hf4_mdot}
	\end{subfigure} 
	\begin{subfigure}{0.49\textwidth}
	\centering 
    \includegraphics[width=\textwidth]{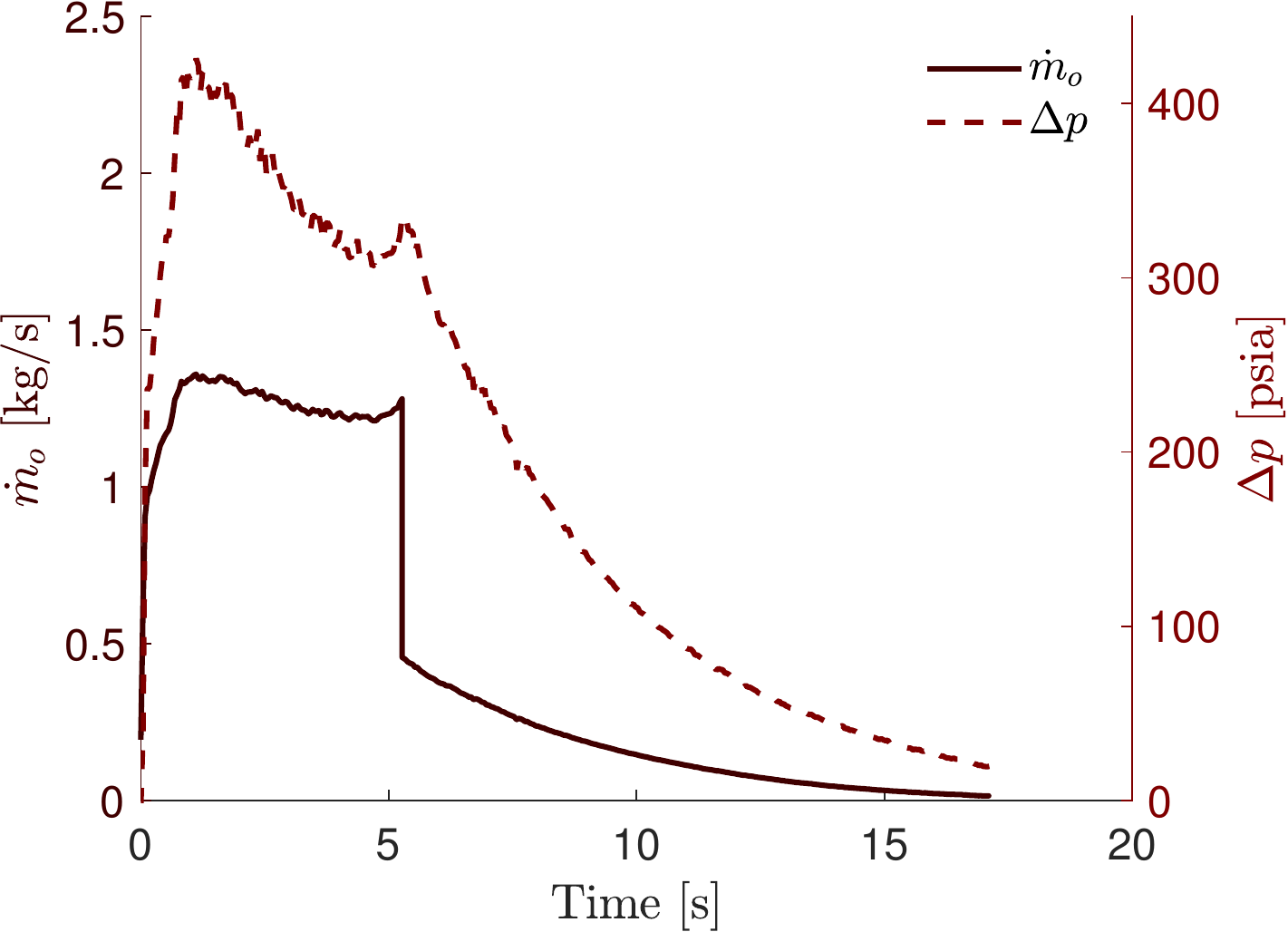}
    \caption{Hot Fire 5.}    
	\label{fig:hf5_mdoto}
	\end{subfigure}
	
	\caption{Pressure differential across the injector and resulting calculated oxidizer mass flow rate.}
	\label{fig:hf45_mdoto}
\end{figure}

The data measured during phases~1 and 2 of the tank blowdown process ($0 \leq t < \tv$) is used to calibrate the injector discharge coefficient $\Ci$, through Eq.~(\ref{eq:ndotox}). For each HF, the injector number of holes $\Ni$ and their area $\Ai$ are known. The tank pressure $\pt$, the feed line pressure loss $\ploss = \pt - \prun$, and the chamber pressure $\pc$ are measured over time. Assuming the liquid-vapor mixture remains saturated during phases~1 and 2, $\pt = \psat(\Tt)$, the tank temperature $\Tt$ is resolved numerically over time, and the corresponding liquid molar volume $\nul = \nulsat(\Tt)$ is computed (Appendix~\ref{a1:thermo}). Given an estimate of $\Ci$, the molar flow rate of \nitrous{} through the injector is calculated through Eq.~\ref{eq:ndotox}, and converted to a mass flow rate: $\mdoto(t) = \Wo \ndoto(t)$. The initial mass of \nitrous{} in the tank is measured through the tank load cell, and given the calculated values of $\mdoto(t)$, the total mass of \nitrous{} remaining in the tank and feed line over time, $\motot(t)$, is resolved through numerical integration with \textit{ode45}. Note the transient tank load cell measurements cannot be used to estimate $\motot(t)$ or $\mdoto(t)$, because engine firing causes extreme oscillations of the load cell measurements. Knowing the internal volume of the run tank and feed line, $\Vt$, and using the system of equations,
\beqarr
\bbm \f{\motot(t)}{\Wo} \\ \Vt \ebm = \bbm 1 & 1 \\ \nuv(t) & \nul(t) \ebm \bbm \nv(t) \\ \nl(t) \ebm 
\eeqarr
the moles of each phase~remaining in tank and feed system over time, $\nl(t)$ and $\nv(t)$, are calculated. The molar volumes are taken to be the saturated values (Appendix~\ref{a1:thermo}). Given a value for $\Ci$, the time taken such that $\nl = 0$ is calculated, and it is compared to the measured value of $\tv$. The parameter $\Ci$ is iterated until the $\tv$ predicted by the model equals the measured value. Parameters used in the analysis and the resulting $\Ci$ are shown in Table~\ref{tab:ci_model}, while the pressure differential across the injector and resulting predicted $\mdoto(t)$ are plotted in Fig.~\ref{fig:hf45_mdoto}. 

\begin{table}
	\small
	\centering
	\caption{Parameters used for the calibration of injector discharge coefficient, $\Ci$. Resulting values of $\Ci$.}
	\label{tab:ci_model}
	\begin{tabular}{c|c|c|c|c}
	\textbf{Description} & \textbf{Symbol} & \textbf{HF4} & \textbf{HF5} & \textbf{Units} \\ \hline
	Tank and feed internal volume & $\Vt$ & 16.5 & 16.5 & L \\
	Initial oxidizer mass & $\motot $ & 7.04 & 8.35 & kg \\
Vapor blowdown timestamp & $\tv$ & 6.164 & 5.272 & s \\ 
	Injection area per hole & $\Ai$ & 3.142 & 3.142 & mm\super{2} \\ 
Number of injection holes & $\Ni$ & 18 & 22  & - \\ \hline
Injector discharge coefficient & $\Ci$ & 0.259 & 0.294 & -
		\end{tabular}
\end{table}

Table~\ref{tab:ci_model} shows from HF4 to HF5, the number of injection holes was increased, resulting in an  increase in $\Ci$. However, calculated values of $\Ci$ remain below 0.30, while typical reported values for high-performance injectors range from 0.50 to 0.79. As such, it can be concluded that the injector plate design exhibits low performance. As a consequence, the peak oxidizer mass flow rate attained for HF4 is around 1~kg/s, and for HF5, around 1.3~kg/s, well below the target 2.50~kg/s. This explains why the chamber pressure peaks around 13.8~bar (200~psia) for both hot fires, as the nozzle was designed for a higher mass flow rate than what is provided by the injector. Future injector plate designs should aim to increase its performance.

\subsection{Estimation of Polytropic Exponent}

The data measured during phase~3 of the tank blowdown process ($\tv \leq t < \tpurge$) is used to calibrate the polytropic expansion exponent of the \nitrous{}, $m$. Through the methodology described in the previous section, $\Ci$ and $\nv$ at $t = \tv$ are calculated. The real gas law is used to calculate the compressibility factor $Z$ at $t = \tv$,
\beq
Z = \f{\pt \Vt}{\nv \Ru \Tt}
\eeq
where $\pt$ is the measured value at $\tv$, and $\Tt$ is the corresponding temperature assuming the vapor is saturated. The parameter $Z$ is then assumed to remain constant, and Eq.~(\ref{eq:gov2vap}) and (\ref{eq:gov3vap}) are numerically integrated with \textit{ode45}. At each time step, $\nuv = \Vt/\nv$, $\pt$ is resolved with the real gas law, and $\ploss$ and $\pc$ are resolved from measured data. The numerical integration is terminated when the calculated tank pressure equals the tank pressure measured at $t = \tpurge$ for each HF, which corresponds to the time valve V-1 opens and \nitrogen{} starts entering the system. The polytropic exponent is iterated until the modeled time equals the measured time. Parameters used in the analysis and resolved values of $m$ are shown in Table~\ref{tab:m_model}.

\begin{table}
	\small
	\centering
	\caption{Parameters used for the calibration of the polytropic exponent, $m$. Resulting values of $m$.}
	\label{tab:m_model}
	\begin{tabular}{c|c|c|c|c}
	\textbf{Description} & \textbf{Symbol} & \textbf{HF4} & \textbf{HF5} & \textbf{Units} \\ \hline
	Purge line pressure & $\ppurge $ & 34.7 (503) & 39.1 (567) & bar (psia) \\ 
	Check valve cracking pressure & - & 4.27 (62) & 4.27 (62) & bar (psia) \\ 
	Purge timestamp & $\tpurge$ & 6.704 & 5.993 & s \\ \hline
	Polytropic exponent & $m$ & 0.746 & 0.749 & - 
		\end{tabular}
\end{table}

The predicted values of $m$ are below 1, which means the tank temperature increases during the polytropic expansion process, as per Eq.~(\ref{eq:gov3vap}). This is a result of the hot heating blankets, which add residual heat to the \nitrous{} vapor during the expansion. Hence, the polytropic exponents calculated from HF test data cannot be extrapolated to flight tank configuration, due to intrinsic differences in the system.

\subsection{Tank Model Validation}

\begin{figure}
	\centering 
	\begin{subfigure}{0.49\textwidth}
	\centering 
    \includegraphics[width=\textwidth]{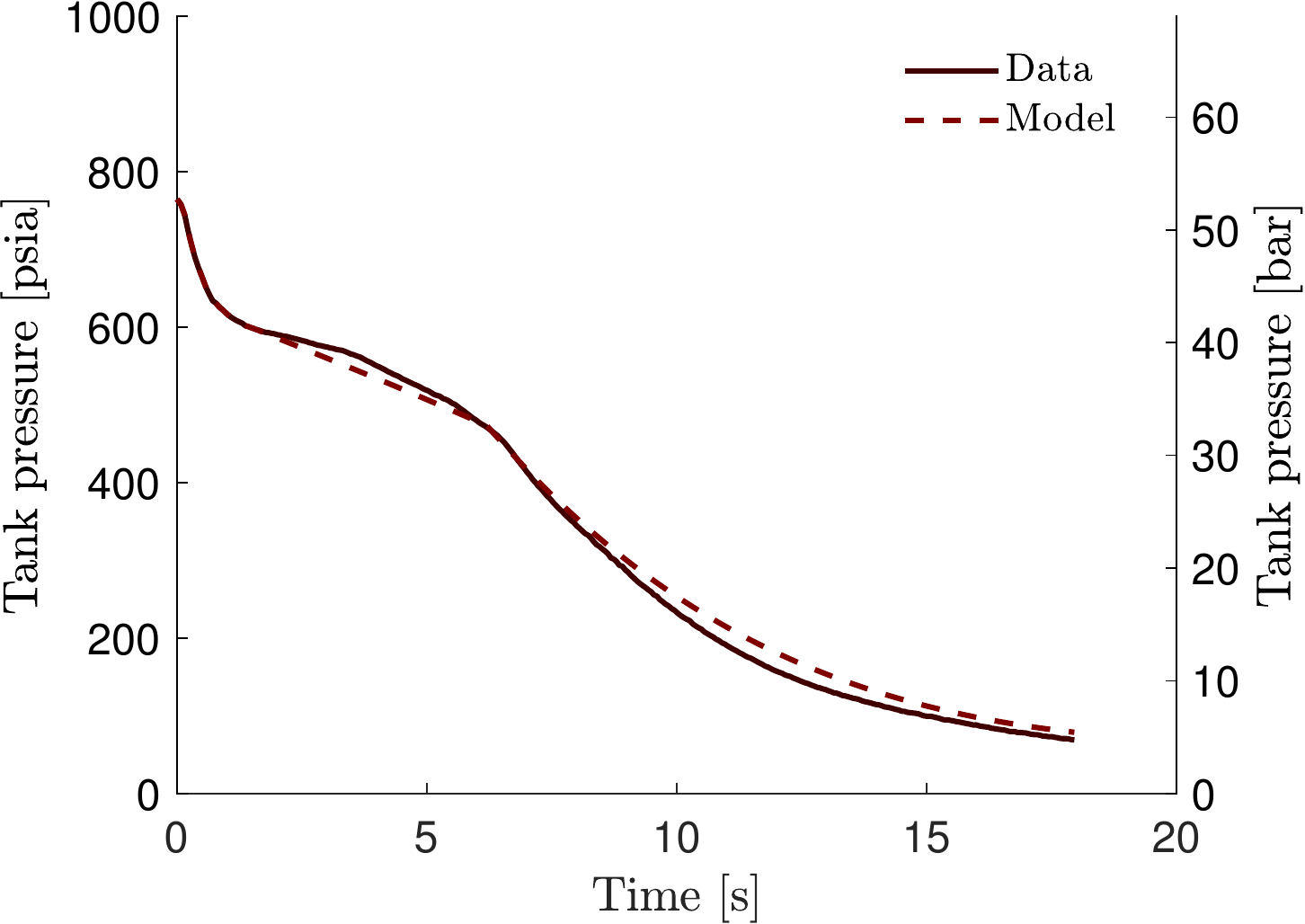}
    \caption{Hot Fire 4.}    
    \label{fig:hf4_freetank}
	\end{subfigure} 
	\begin{subfigure}{0.49\textwidth}
	\centering 
    \includegraphics[width=\textwidth]{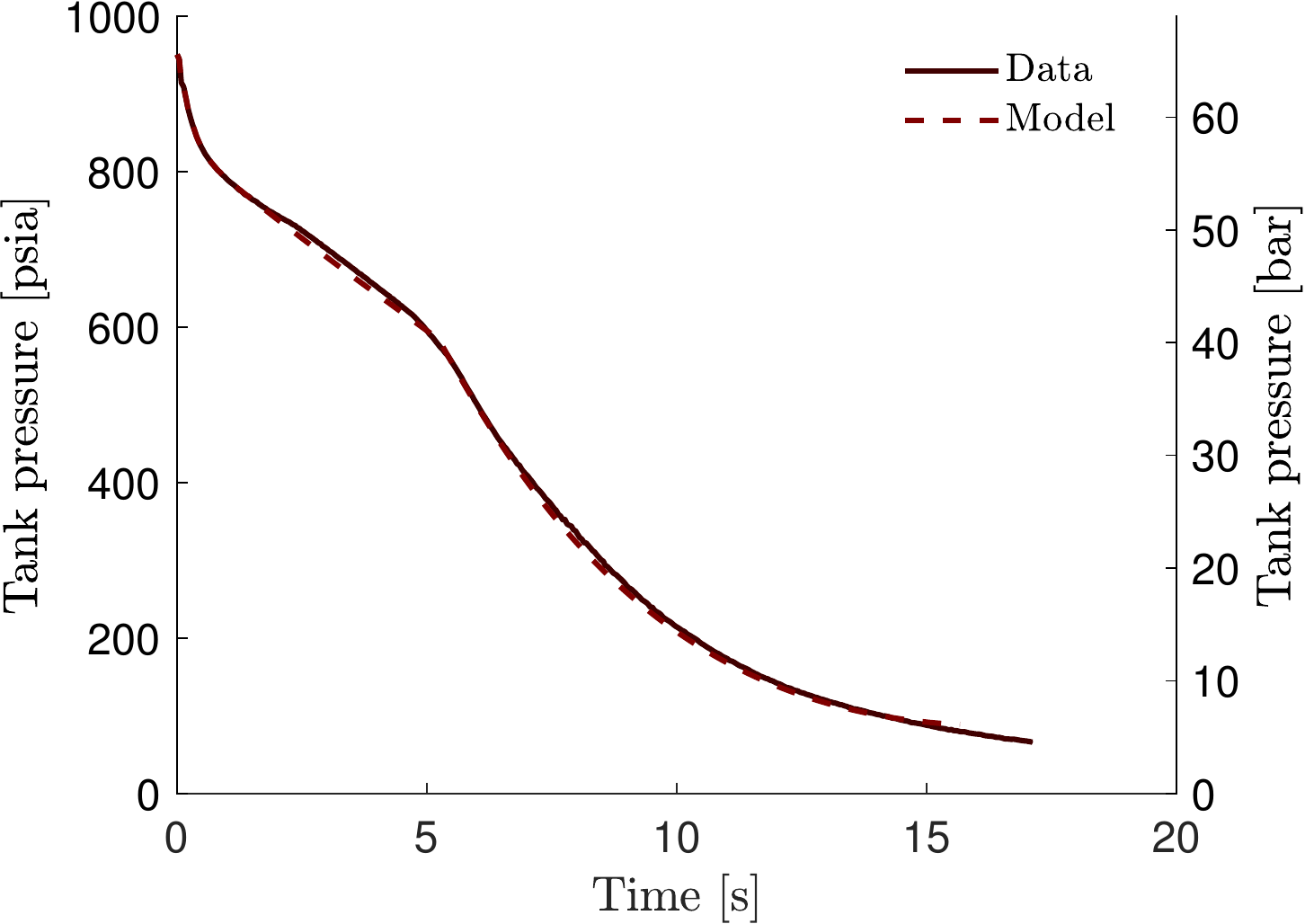}
    \caption{Hot Fire 5.}    
	\label{fig:hf5_freetank}
	\end{subfigure}
	
	\caption{Comparison of experimental and modeled transient tank pressure.}
	\label{fig:hf45_freetank}
\end{figure}

The unsteady tank-emptying model is validated without transient input of tank and run line pressure measurements. The $\Ci$ and $m$ shown in Tables~\ref{tab:ci_model} and \ref{tab:m_model} are input to the model. The pressure loss data is numerically averaged for each of phases~2 and 3 and input to the model, as shown in Table~\ref{tab:ploss}. The chamber measured transient pressure data is also input to the model. The model is then resolved for $t > \tfeed$; results are shown in Fig.~\ref{fig:hf45_freetank} and compared to experimental data. 

\begin{table}
	\small
	\centering
	\caption{Free tank-emptying model average pressure losses.}
	\label{tab:ploss}
	\begin{tabular}{c|c|c|c|c}
	\textbf{Description} & \textbf{Symbol} & \textbf{HF4} & \textbf{HF5} & \textbf{Units} \\ \hline
	Average feed pressure loss, liquid phase & $\ploss$ & 6.34 (92.0) & 11.9 (172) & bar (psia) \\ 
	Average feed pressure loss, vapor phase & $\ploss$ &  3.12 (45.3) & 4.98 (72.2)  & bar (psia) 
		\end{tabular}
\end{table}

Table~\ref{tab:ploss} shows the average pressure loss is higher for HF5 when compared to HF4. In fact, increasing the tank initial pressure also causes an increase of the \nitrous{} flow rate through the system, and therefore an increase in $\ploss$, limiting the gain in $\prun$. Hence, a different test site configuration which significantly reduces the intermediate length and pressure losses between the run tank and combustion chamber should be considered. 

Figure~\ref{fig:hf45_freetank} shows the measured and predicted results of transient tank pressure are in overall excellent agreement, which validates the unsteady tank-emptying model. Small discrepancies between model and experimental data can be attributed to components of the test site configuration which are not modeled: In the liquid-vapor blowdown phase, the tank heating helps maintain tank pressure, which leads to a slight under-prediction of tank pressure by the model. In the gaseous blowdown phase, the injection of \nitrogen{} after $\tpurge$ causes a change in the polytropic expansion curve of the system.


\section{Chamber Model Calibration}

\subsection{Estimation of Regression Rate Ballistic Coefficients} \label{sec:hf_fuelregression}

The non-classical equation for the fuel grain regression rate--Eq.~(\ref{eq:rdotf})--is semi-empirical, hence it does not describe the fundamental physics of the process with high accuracy \cite{genevieve2013}. Instead, it can be used to approximate the regression rate for specific chamber dimensions, propellants, and thermodynamic conditions. With time-resolved data of the fuel grain port size, an accurate determination of the constants $a$ and $n$ would be possible, for example through a nonlinear least squares fitting procedure. However, HF4 and HF5 did not feature such measurements, hence a different approach is used: Through an analysis of Eq.~(\ref{eq:rdotf}), it can be inferred that the scaling constant $a$ has a large impact on the chamber pressure, as it controls the growth rate of the chamber volume while combustion gases are produced. In addition, the exponent $n$ controls the fuel grain burntime, as it scales the impact of oxidizer mass flux on fuel grain consumption rate. These inferences are confirmed by numerical simulations and are used in conjunction with experimental data to estimate the ballistic coefficients $a$ and $n$.

\begin{figure}
	\centering 
	\begin{subfigure}{0.49\textwidth}
	\centering 
    \includegraphics[width=\textwidth]{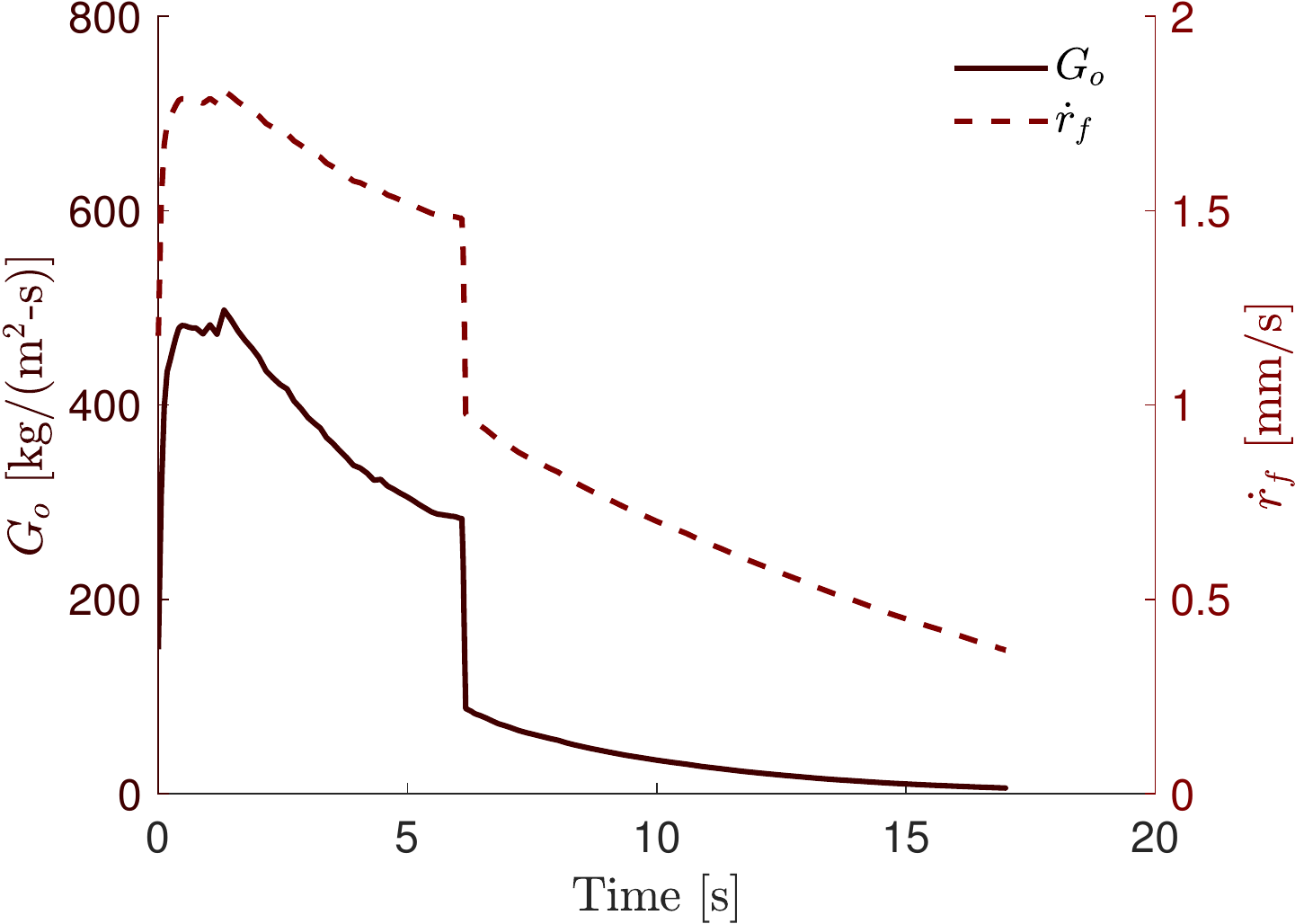}
    \caption{Hot Fire 4.}    
	\end{subfigure} 
	\begin{subfigure}{0.49\textwidth}
	\centering 
    \includegraphics[width=\textwidth]{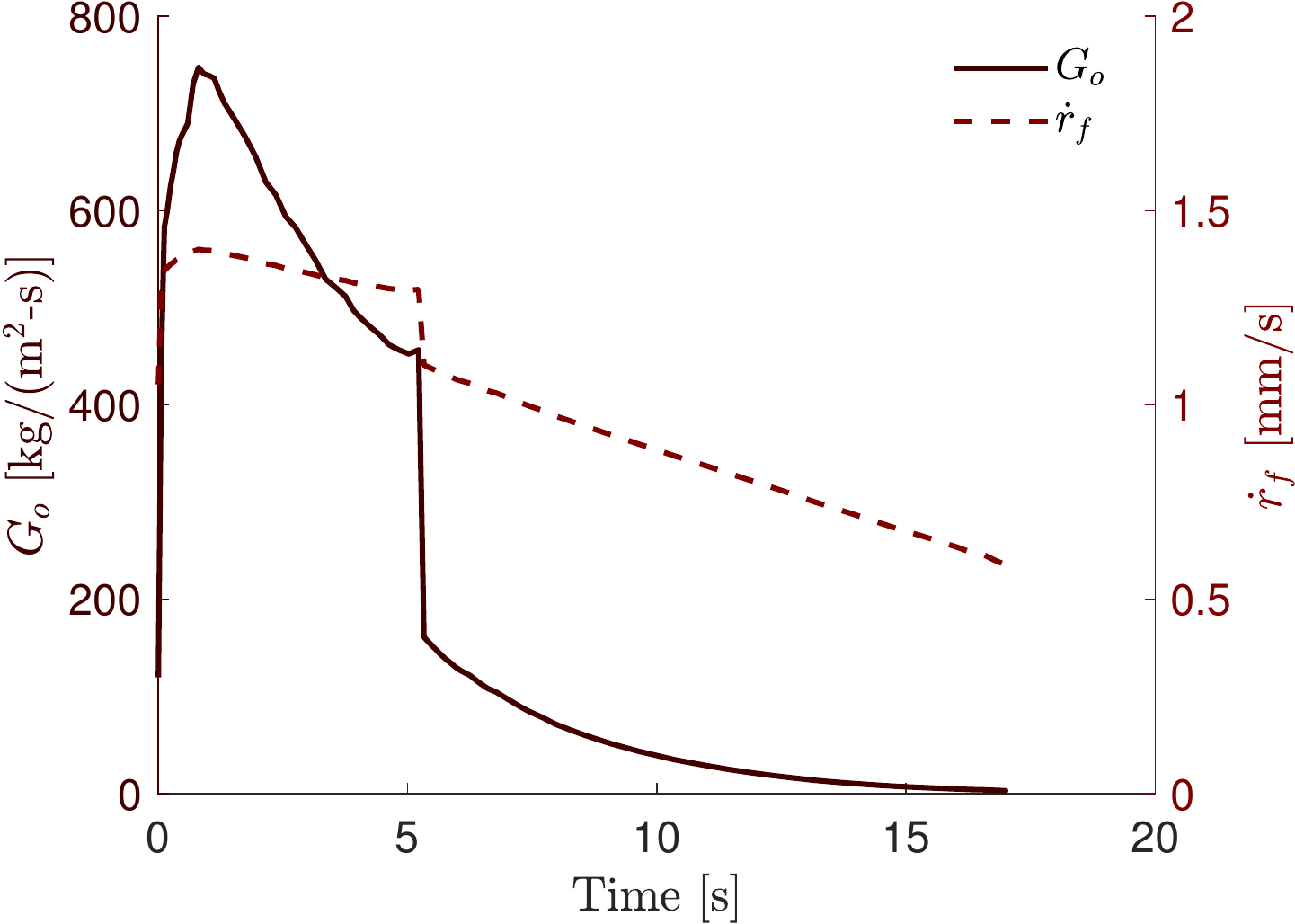}
    \caption{Hot Fire 5.}    
	\end{subfigure}
	
	\caption{Fuel grain regression rate.}
	\label{fig:hf45_rdotf}
\end{figure}

For each HF, the initial fuel grain dimensions and its mass are measured, and an estimate of the constant $a$ is provided. The oxidizer mass flow rate as a function of time, $\mdoto$, is calculated from the tank-emptying model (shown in Fig.~\ref{fig:hf45_mdoto}); it is input to Eq.~(\ref{eq:rdotf}), which is integrated numerically with \textit{ode45} until the fuel grain is depleted. For a fixed value of $a$, the parameter $n$ is iterated until the modeled and measured burntime from HF test videos are in agreement. This procedure decouples the fuel grain regression from the remainder of the unsteady chamber model, which is enabled by the independence of Eq.~(\ref{eq:rdotf}) on $\pc$. Subsequently, the unsteady model is resolved independently from hot fire test data (see Section~\ref{sec:hybrid_validation}). The resulting transient chamber pressure profile is compared to experimental data, to determine the accuracy of the parameter $a$. The value of $a$ is iterated, and the procedure is repeated until the model and the experiment are in reasonable agreement. Parameters used in the analysis and the resulting ballistic coefficients are shown in Table~\ref{tab:regression_model}. Figure~\ref{fig:hf45_rdotf} shows the regression rate data, and Fig.~\ref{fig:hf5_validation_thrust} shows the transient chamber pressure profile; further discussion of Fig.~\ref{fig:hf5_validation_thrust} is provided in Section~\ref{sec:hybrid_validation}.

\begin{table}
	\small
	\centering
	\caption{Fuel grain parameters.}
	\label{tab:regression_model}
	\begin{tabular}{c|c|c|c|c}
	\textbf{Description} & \textbf{Symbol} & \textbf{HF4} & \textbf{HF5} & \textbf{Units} \\ \hline
	Mass & $\mftot$ & 1594 & 1594 & g \\ 
	Length & $\Lf$ & 610 (24) & 610 (24) & mm (in) \\ 
	Density & $\rhof$ & 788.9 & 788.9 & kg/m\super{3} \\ 
	Burntime & $\tburn$ & 17 & 17 & s \\ \hline
	Scaling constant & $a$ & 2 x 10\super{-4} & 5 x 10\super{-4} & - \\ 
	Exponent & $n$ & 0.355 & 0.156  & - 
		\end{tabular}
\end{table}

\subsection{Quasi-Steady Chemical Equilibrium and Nozzle Gas Dynamics} \label{sec:nozzle_validation}

To validate the quasi-steady chemical equilibrium assumption and the nozzle gas dynamics model, the fuel grain regression is solved as described in the previous section. Subsequently, the chamber filling and emptying model is resolved through numerical integration of Eq.~(\ref{eq:dmfdt}) and (\ref{eq:dmodt}) with \textit{ode15s}. At each time step, the chamber pressure measurements are provided as inputs to the system of equations to calculate the thermodynamic properties of the combustion gases with the PROPEP code, while the OF ratio is calculated from the current values of $\mo$ and $\mf$ which are being integrated. The nozzle mass flow rate, $\mdotn$, is determined, along with the transient thrust profile.  

\begin{figure}
	\centering 
	\begin{subfigure}{0.49\textwidth}
	\centering 
    \includegraphics[width=\textwidth]{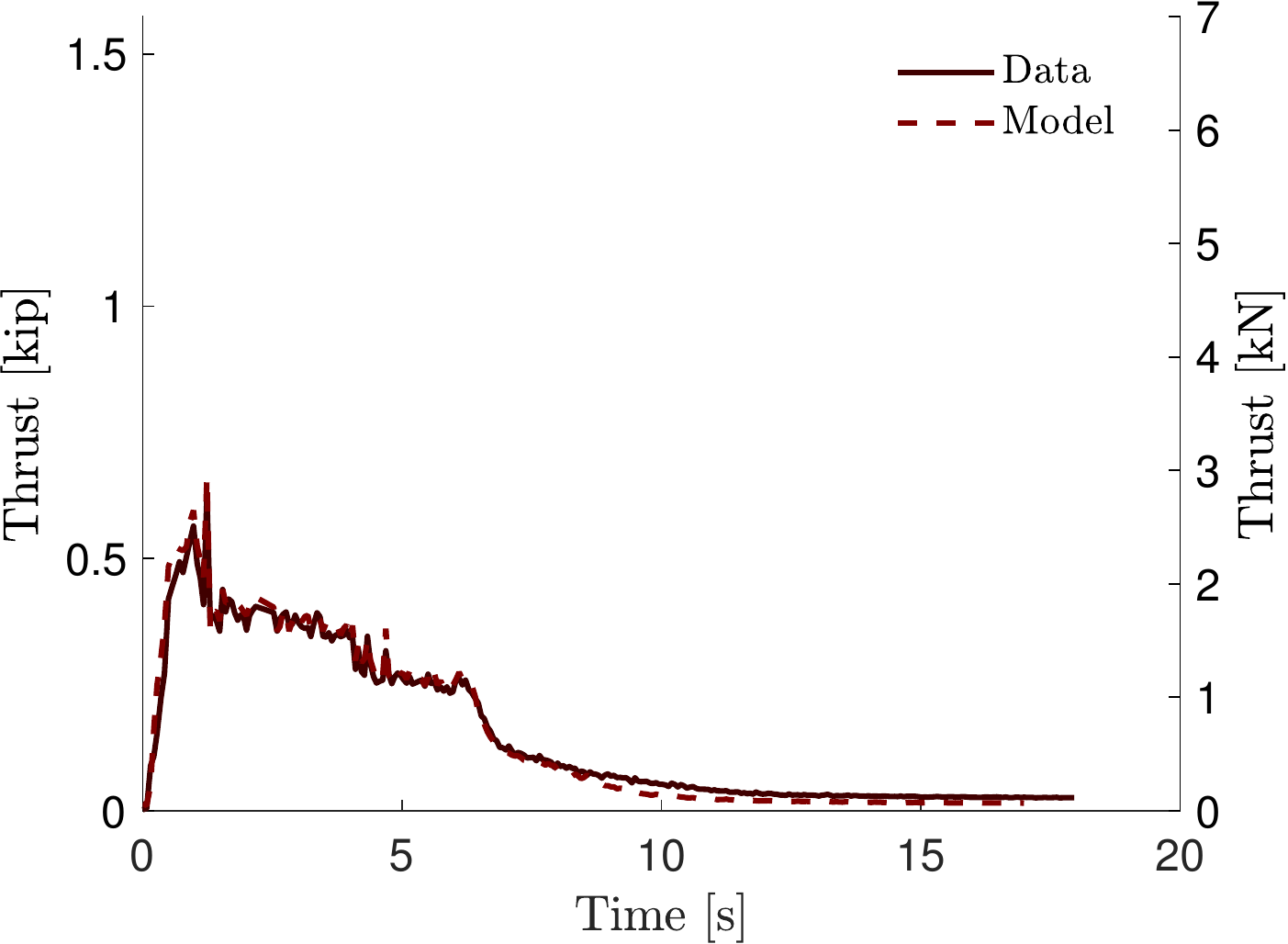}
    \caption{Hot Fire 4.}    
	\end{subfigure} 
	\begin{subfigure}{0.49\textwidth}
	\centering 
    \includegraphics[width=\textwidth]{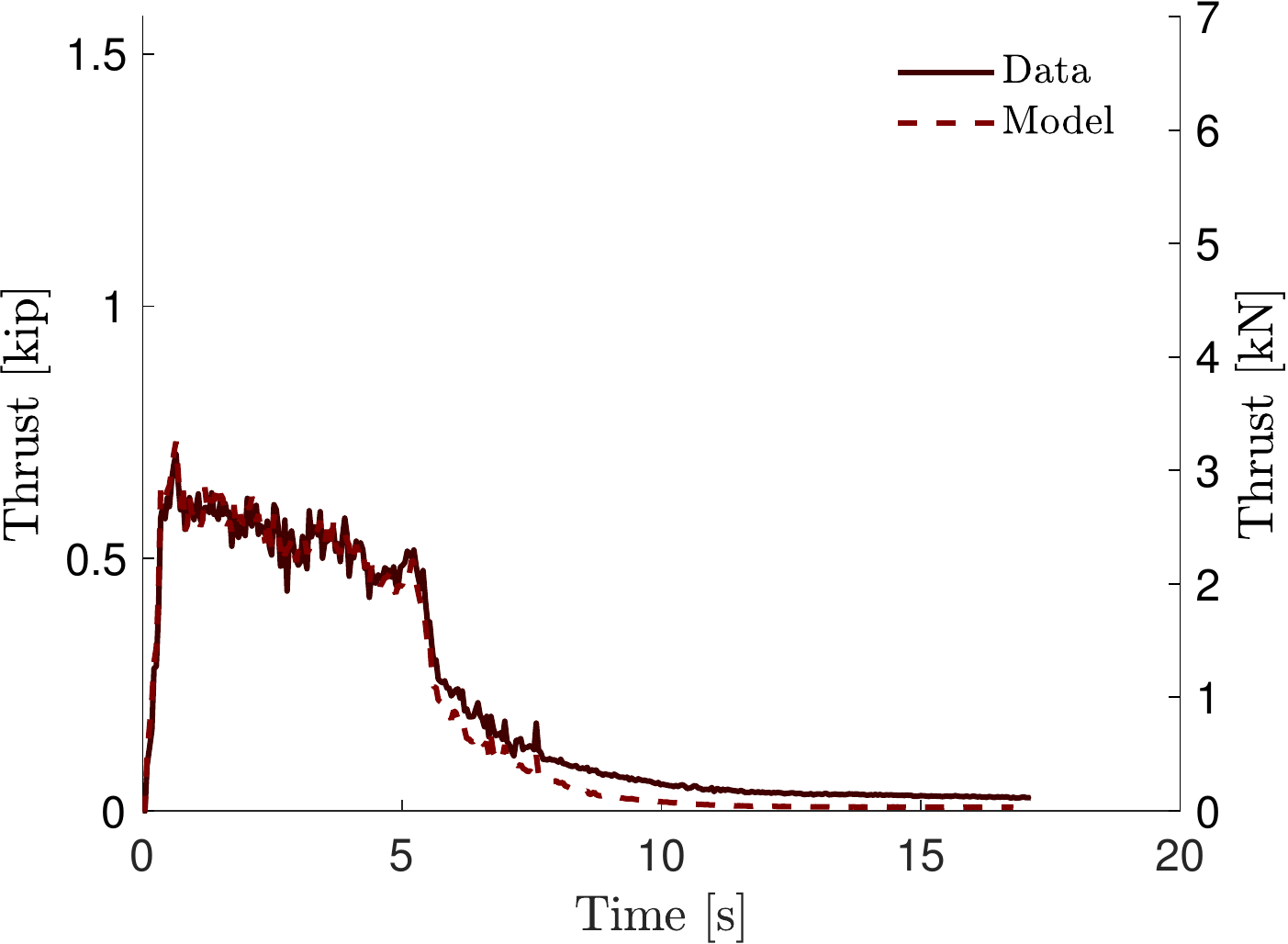}
    \caption{Hot Fire 5.}    
	\end{subfigure}
	
	\caption{Comparison of experimental and modeled thrust curve, where the model uses transient measurements of chamber pressure as inputs.}
	\label{fig:hf45_forcedthrust}
\end{figure}

The nozzle thrust predicted by the model is compared to experimental data in Fig.~\ref{fig:hf45_forcedthrust}, where the model and experimental data are found to be in excellent agreement. This validates the chemical equilibrium assumption and the one-dimensional unsteady nozzle model. Figure~\ref{fig:hf45_forcedthrust} shows an increasing discrepancy between model and experimental data is observed for $t > \tv$. This is because of the introduction of \nitrogen{} in the chamber through the purge system, which changes the composition of the combustion products and impacts the nozzle thrust. Since the \nitrogen{} is not modeled, this effect is not captured. 


\section{Hybrid Engine Model Validation} \label{sec:hybrid_validation}

With parameters determined from the calibration of the tank and chamber models, a full-scale validation of the hybrid engine model is considered, with no transient inputs of HF test data. In flight configuration, it is hypothesized that the initial pressure drop of the tank measured in phase~1 of HF test data would be negligible, hence the unsteady model does not capture this initial pressure drop. Given the oxidizer mass and tank volume for each HF, the initial pressure of the tank for the unsteady model is determined such that the $\pt$ measured at $t = \tfeed$ and $\pt$ predicted by the model after $\tfeed$ are in agreement. This is effectively an extrapolation of measured hot fire data during phase~2 to $t < \tfeed$.

\begin{figure}

	\centering 
	\begin{subfigure}{0.49\textwidth}
	\centering 
    \includegraphics[width=\textwidth]{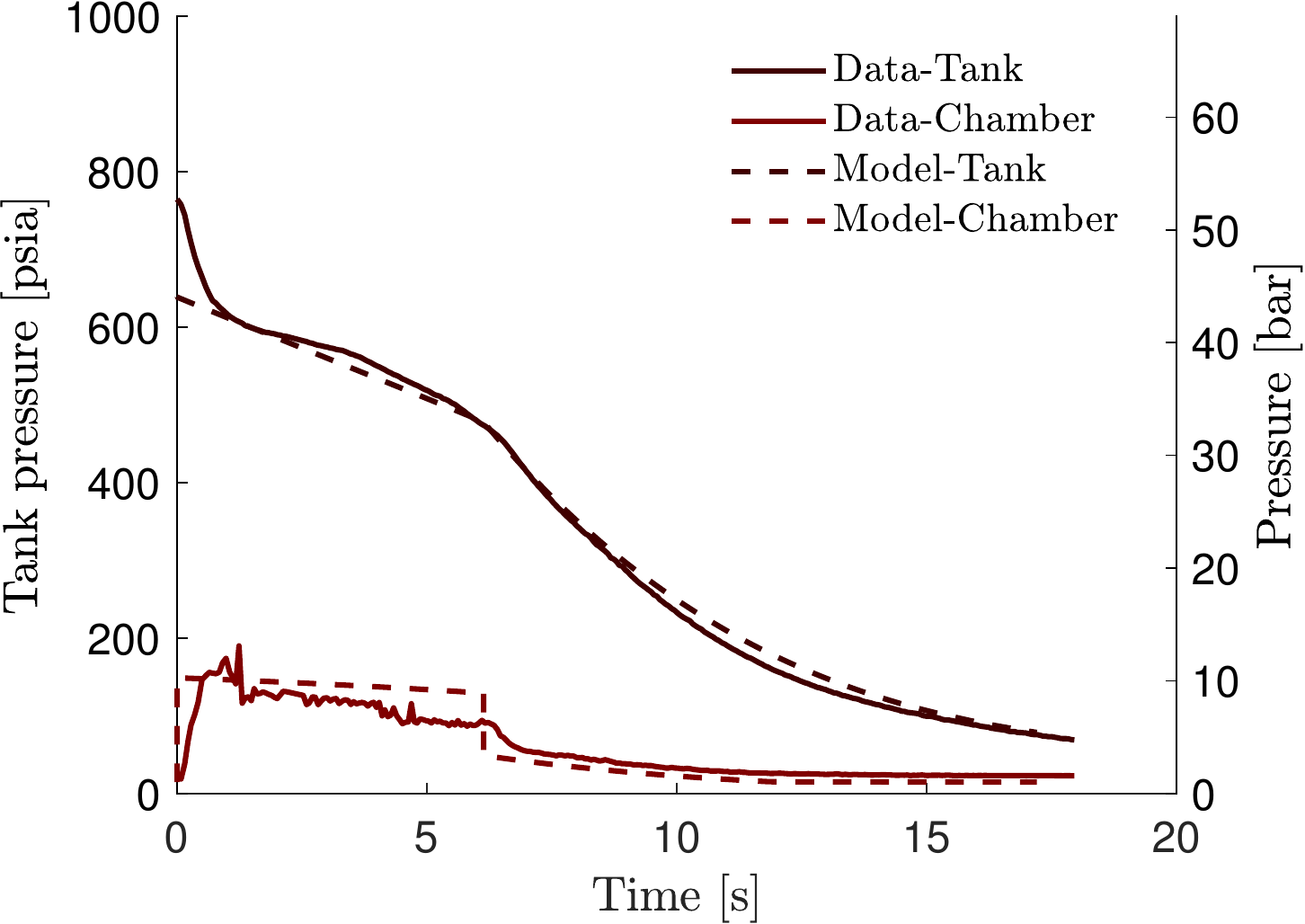}
    \caption{Hot Fire 4 pressures.}    
    \label{fig:hf4_validation_pressure}
	\end{subfigure} 
	\begin{subfigure}{0.49\textwidth}
	\centering 
    \includegraphics[width=\textwidth]{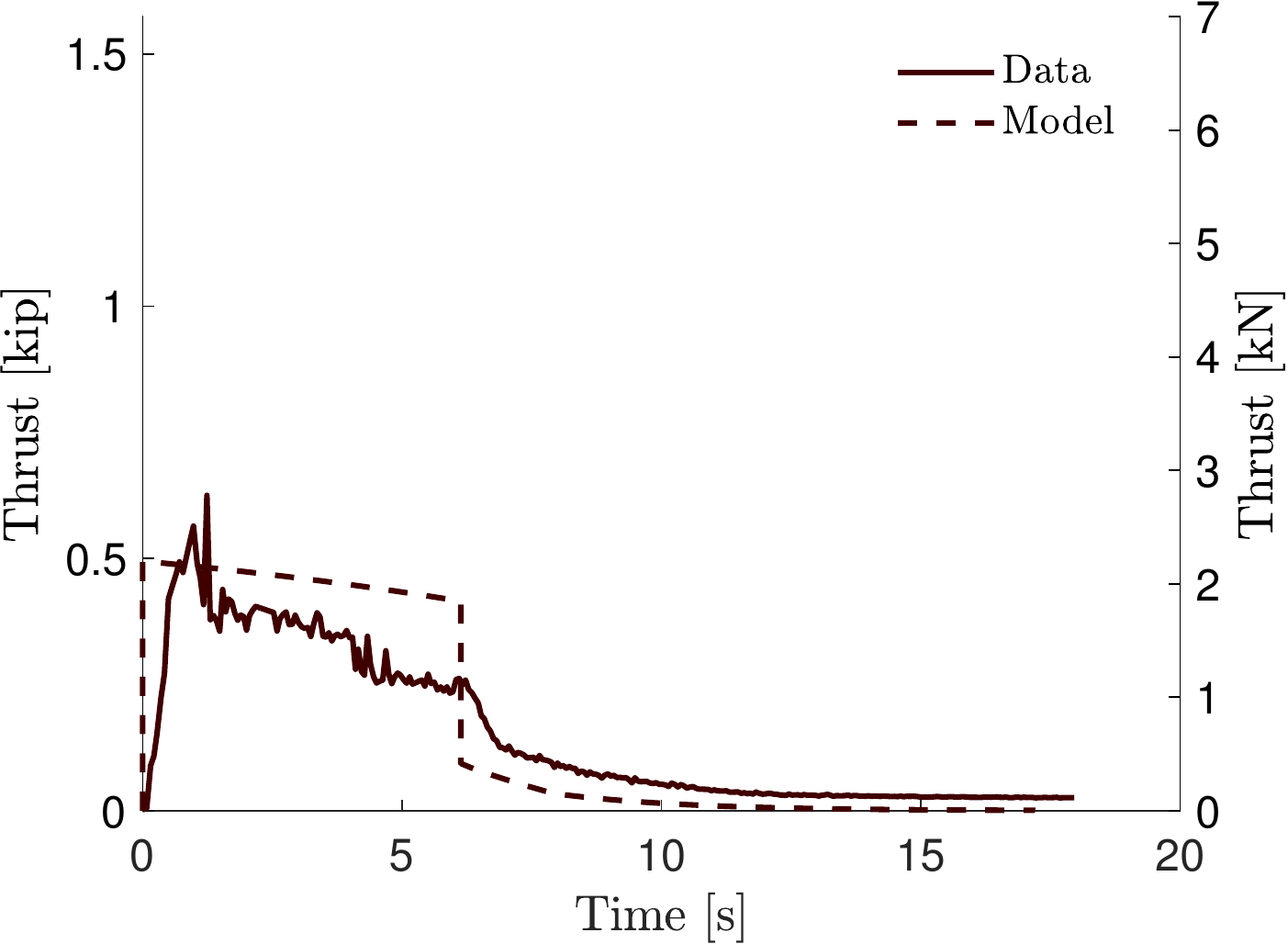}
    \caption{Hot Fire 4 thrust.}    
	\label{fig:hf4_validation_thrust}
	\end{subfigure}
	
	\centering 
	\begin{subfigure}{0.49\textwidth}
	\centering 
    \includegraphics[width=\textwidth]{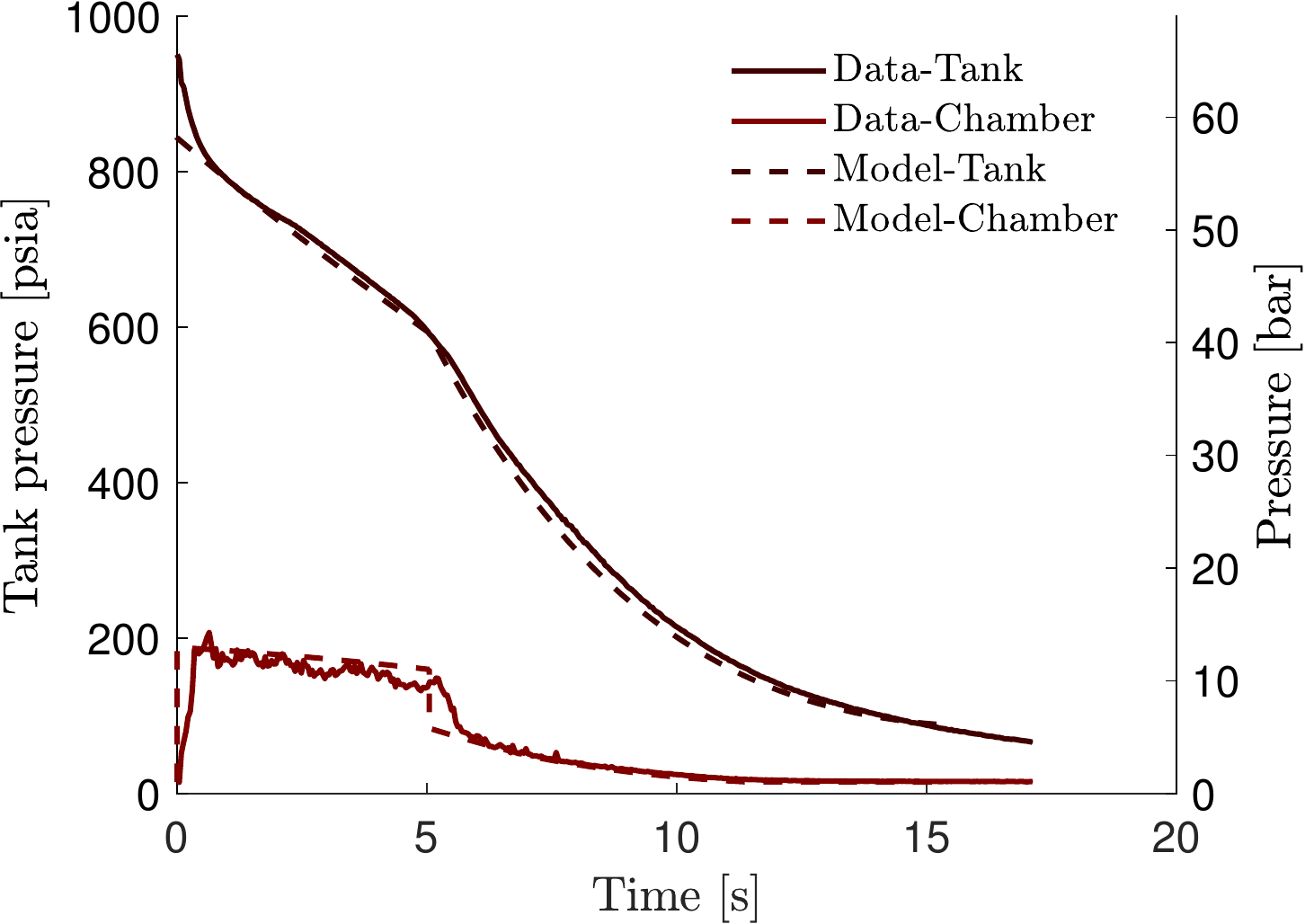}
    \caption{Hot Fire 5 pressures.}    
    \label{fig:hf5_validation_pressure}
	\end{subfigure} 
	\begin{subfigure}{0.49\textwidth}
	\centering 
    \includegraphics[width=\textwidth]{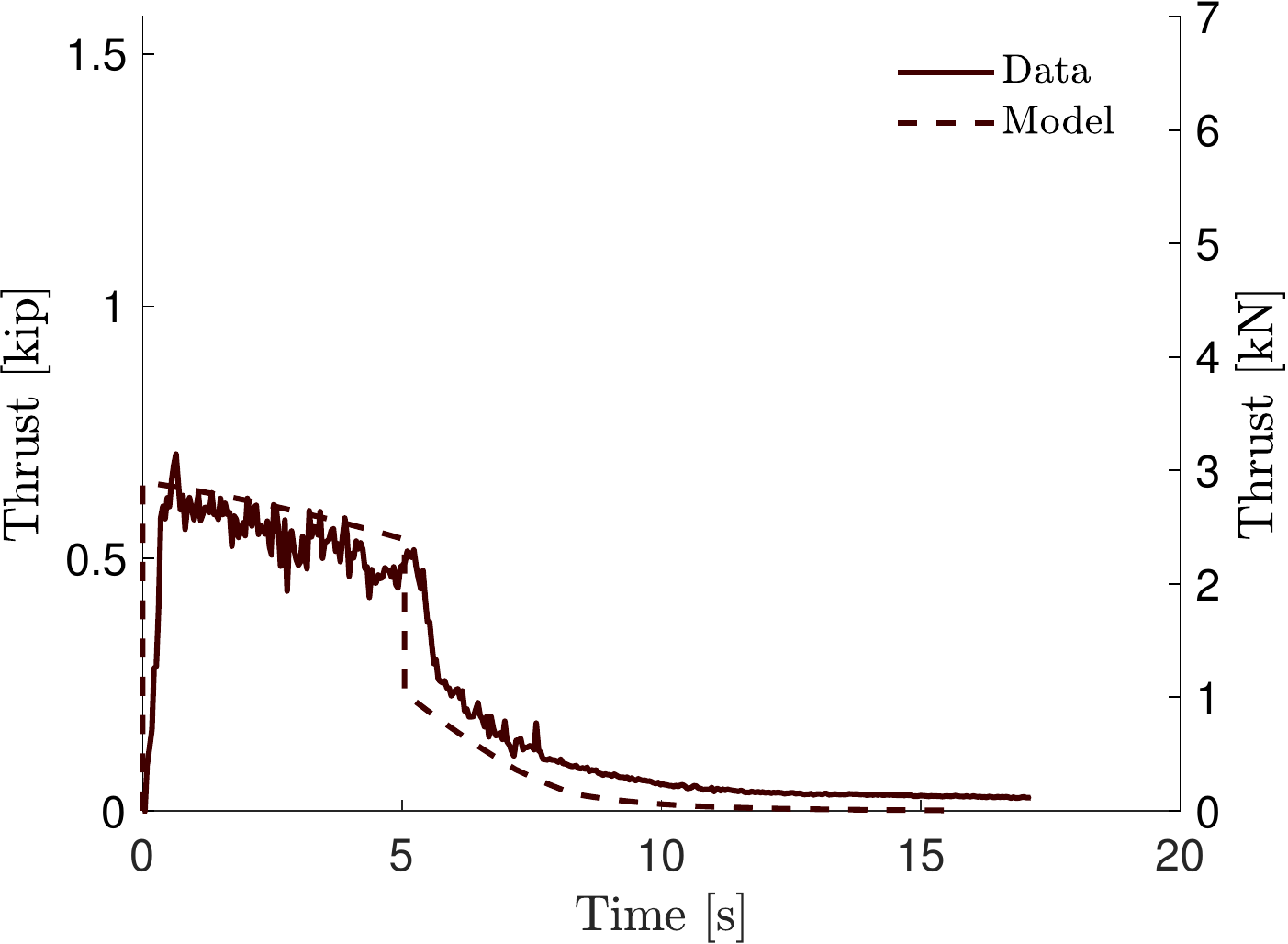}
    \caption{Hot Fire 5 thrust.}    
	\label{fig:hf5_validation_thrust}
	\end{subfigure}

	\caption{Unsteady model comparison to experimental data.}
	\label{fig:hf45_validation}
\end{figure}

Figure~\ref{fig:hf45_validation} shows the results of the calibrated unsteady model, and compares them to hot fire testing data. The tank pressure curves are found to be in excellent agreement for both HFs, and the chamber pressure and thrust curve are found to be in good agreement for HF5. For HF4, the initial pressure peak in the chamber followed by an abrupt decrease lead to an inability to capture the pressure curve accurately by the model. The ballistic coefficients for HF5 can therefore be trusted with higher accuracy than those for HF4. Nevertheless, an accurate determination of $a$ and $n$ requires transient measurements of regression rate data. 


\section{Flight Performance}

The unsteady ascent model is resolved (Section~\ref{sec:unsteady_ascent}), while using directly the thrust curve measured in each HF, to predict the flight performance of \rocket{} with the two \engine{} iterations. The quasi-steady chamber validation described in Section~\ref{sec:nozzle_validation} allows to determine the mass flow rate through the nozzle as a function of time, therefore allowing to estimate the mass of the vehicle at every point of its flight. 

\begin{figure}
	\centering 
    \includegraphics[width=0.49\textwidth]{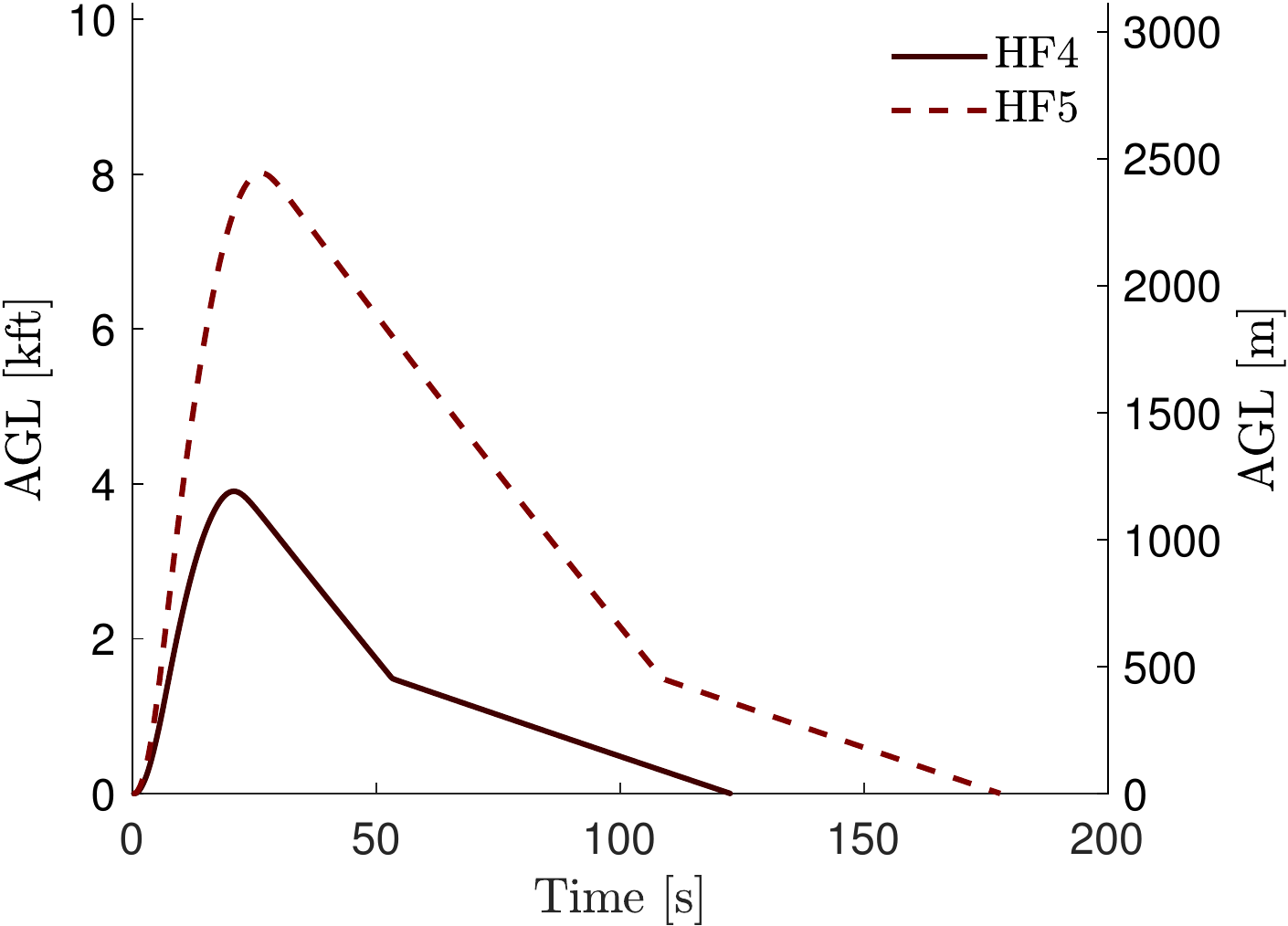}
	\caption{Flight profile with hot fire thrust curves.}
	\label{fig:hf45_flight}
\end{figure}

\begin{table}
	\small
	\centering
	\caption{Unsteady model engine performance metrics.}
	\label{tab:hf_performance}
	\begin{tabular}{c|c|c|c|c|c}
	\textbf{Description} & \textbf{Symbol} & \textbf{Target} & \textbf{HF4} & \textbf{HF5} & \textbf{Units} \\ \hline 
	Burntime & $\tburn$ & 3.34 & 17 & 17 & s \\	
	Oxidizer mass & $\motot$ & 7.04 & 7.04 & 8.35 & kg \\
	Fuel mass & $\mftot$ & 1639 & 1594 & 1594 & g \\
	Total impulse & $\itot$ & 16,845 & 12,068 & 16,423 & Ns \\
	Specific impulse & $\isp$ & 198 & 142 & 168 & s \\
	Peak chamber pressure & $\pc$ & 27.6 (400) & 13.1 (190) & 14.3 (207) & bar (psi) \\
	Peak thrust & $\Fmax$ & 6.74 & 2.78 &  3.14 & kN \\
	Average thrust & $\Favg$ & 5.05 & 0.672 & 0.959 & kN \\
	Rocket wet mass & $\mrzero$ & 61.2 & 61.2 & 62.5 & kg \\
	Pad thrust-to-weight & - & 11.2 & 2.47 & 3.31 & - \\
	Rail departure velocity & - & 34.0 (111) &  18.1 (59.3) & 20.5 (67.4) & \\
	Predicted apogee & - & 3048 (10,000) & 1190 (3904) & 2444 (8017) & m (ft) \\
	Peak velocity & - & 252 (827) & 99.4 (326) & 166 (544) & m (ft) \\
	Peak Mach number & - & 0.757 & 0.298 & 0.499 & - \\
	Engine acceleration & - & 11.7 & 3.13 & 4.11 & $\gsl$ \\ 
	\end{tabular}
\end{table}

Table~\ref{tab:hf_performance} compares the target performance metrics of the unsteady model determined in Chapter~
\ref{chapter:unsteady}, to the performance achieved during HF4 and HF5. \engine{}-HF4 only provided 12,068~Ns of impulse, 28.3~\% below the target value, leading to an apogee of only 1190~m (3904~ft). This low total impulse is a result of the inability to maintain chamber pressure during HF4. For HF5, a larger quantity of propellants was loaded in the system, and the chamber pressure was maintained for a higher duration. In fact, the total impulse provided by \engine{}-HF5 is comparable to the target value, falling only 2.51~\% short. However, $\approx$~1.3 kg more of propellants is required to achieve this, and the engine specific impulse is 15.2~\% below the target value. Hence, the overall vehicle weight during the flight is higher, leading to a lower apogee. Despite, an apogee of 2444~m (8017~ft) is attained, which is respectable. Due to time constraints before competition, \engine{}-HF5 was selected as the flight engine for \rocket{} at the SAC~2022. 

\begin{figure}
	\centering 
	\begin{subfigure}{.49\textwidth}
	\centering 
	\includegraphics[width=\textwidth]{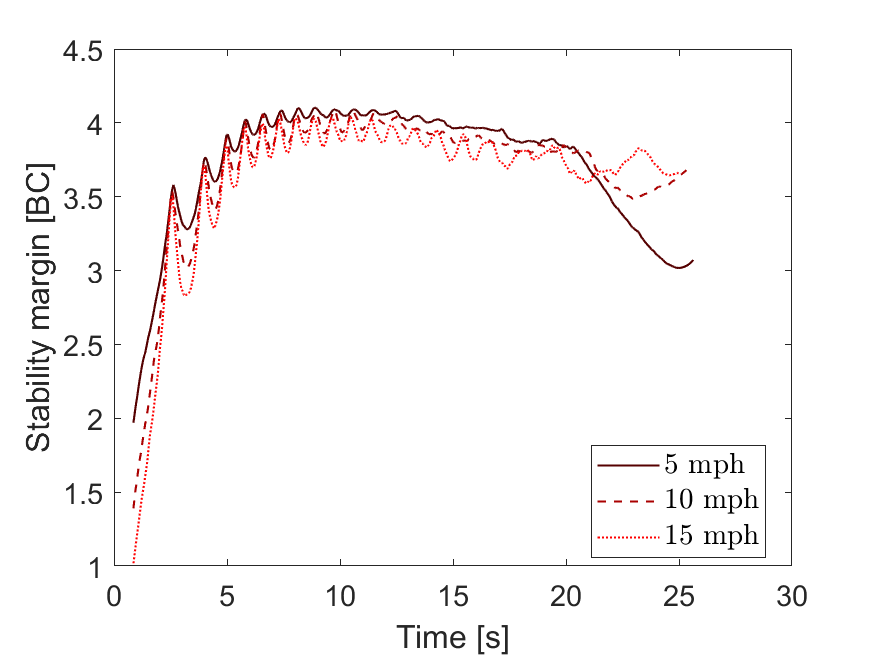}
	\caption{Ascent path.}
	\end{subfigure} \hfill
	\begin{subfigure}{.49\textwidth}
	\centering 
	\includegraphics[width=\textwidth]{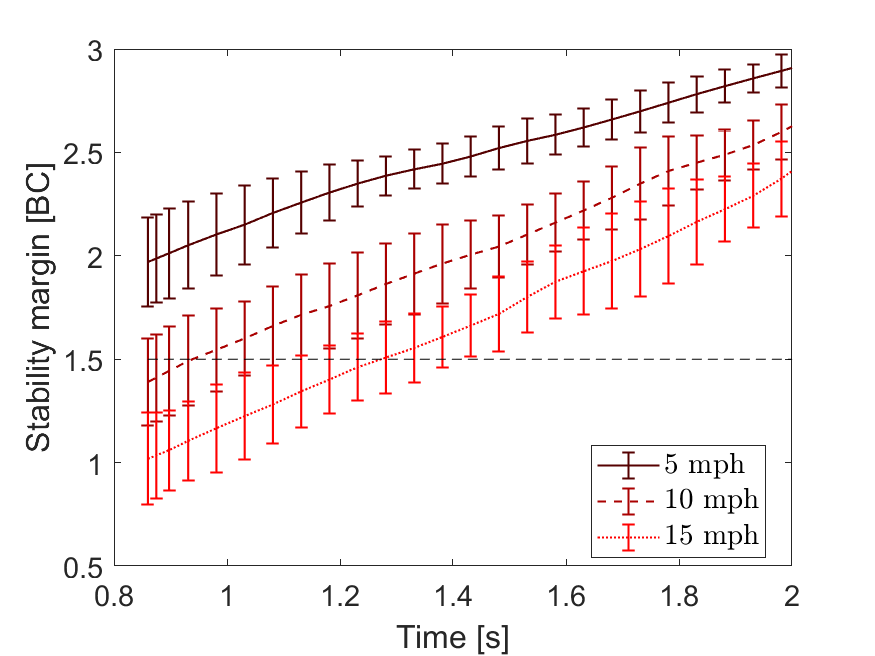}
	\caption{Zoomed-in near rail departure.}
	\end{subfigure}
	\caption{Time-resolved stability margin body calipers of \rocket{} with HF5 thrust curve.}
	\label{fig:hf5_stability}
\end{figure}

Table~\ref{tab:hf_performance} shows the burntime of \engine{}-HF5 is 17~s, while the target is 3.34~s. Although the total impulse provided is comparable, this is achieved over a much longer burn duration, leading to a low peak thrust of 3.14~kN, about half the desired value. The pad thrust-to-weight ratio decreases from a desired 11.2 to 3.31, and the rail departure velocity is 20.5~m/s (67.4~ft/s), not clearing the 30.5~m/s (100~ft/s) recommended minimum. To ensure off-rail stability, a stability analysis as described in Section~\ref{sec:unsteady_stability} is conducted with \engine{}-HF5; results are shown in Fig.~\ref{fig:hf5_stability}. For a wind speed below 8.05~km/h (5~mph), the stability remains between 2 and 4 BC, hence the low pad thrust-to-weight still leads to a stable flight. As achieving high chamber pressure and high peak thrust on a small-scale hybrid vehicle has been difficult, future engine designs could aim to lower these targets and optimize the nozzle for lower thrust values, while still clearing the required off-rail stability.


\chapter{Discussions and Conclusions}

\section{Sources of Error}

\begin{itemize}

\item \textbf{Adiabatic combustion chamber}: The hybrid engine model assumes an adiabatic combustion chamber, since the engine is insulated with a carbon phenolic liner. The adiabatic assumption is valid for sufficiently short burntimes, in which case the chamber heat flux has not significantly penetrated through the liner wall. However, with a burntime of 17~s as realized for HF4 and HF5, this assumption is not necessarily valid and can introduce errors in the model. A future hybrid engine model could implement heat transfer from the combustion chamber to the surroundings.

\item \textbf{Adiabatic oxidizer tank}: Similarly, the unsteady model assumes adiabatic oxidizer tank walls, and no heat transfer between the tank walls and the oxidizer, due to the low thermal conductivity of the propellant. This is a possible source of error for a long engine burntime, and future hybrid iterations of the model could implement heat transfer between the tank, oxidizer, and surroundings.

\item \textbf{Chemical equilibrium}: The PROPEP code allows to input the chamber pressure and the propellant composition, and the combustion products are equilibrated at constant enthalpy and pressure (HP equilibrium). However, the code does not permit to input the enthalpy of the propellants entering the combustion chamber: a set of standard conditions is pre-determined for the list of propellants supported by the software. PROPEP assumes \nitrous{} is provided in the ideal gas state at 25~\super{$\circ$}C, with a corresponding molar enthalpy of 82.1~MJ/kmol. The actual molar enthalpy of \nitrous{} in the saturated vapor state at 298.15~K is 16.6~MJ/kmol, while that of the saturated liquid is 10.1~MJ/kmol. Hence, PROPEP over-estimates the molar enthalpy of the propellant by nearly an order of magnitude, introducing sources of error in the model. Additionally, during the transient decay of the tank temperature, the molar enthalpy of the \nitrous{} saturated liquid entering the chamber further decreases. 

An alternative equilibrium code which allows to directly control propellant molar enthalpy is the NASA Chemical Equilibrium with Applications (CEA) code. Figure~\ref{fig:cea_vs_propep} compares the specific impulse predicted by NASA-CEA, to that predicted by PROPEP. It can be seen that the NASA-CEA code predicts lower specific impulses, due to the lower energy content of the propellants. Furthermore, NASA-CEA predicts the optimum OF ratio is $\approx$~8.5, while PROPEP predicts it is $\approx$~6.5. Hence, there is a difference in the thermodynamic assumptions of the two software. The NASA-CEA code is more widely used and could be a more accurate solver for the calculation of the chemical equilibrium of the mixture; future iterations of the hybrid engine model could implement NASA-CEA. 

\begin{figure}[h]
    \centering
    \includegraphics[width=0.49\textwidth]{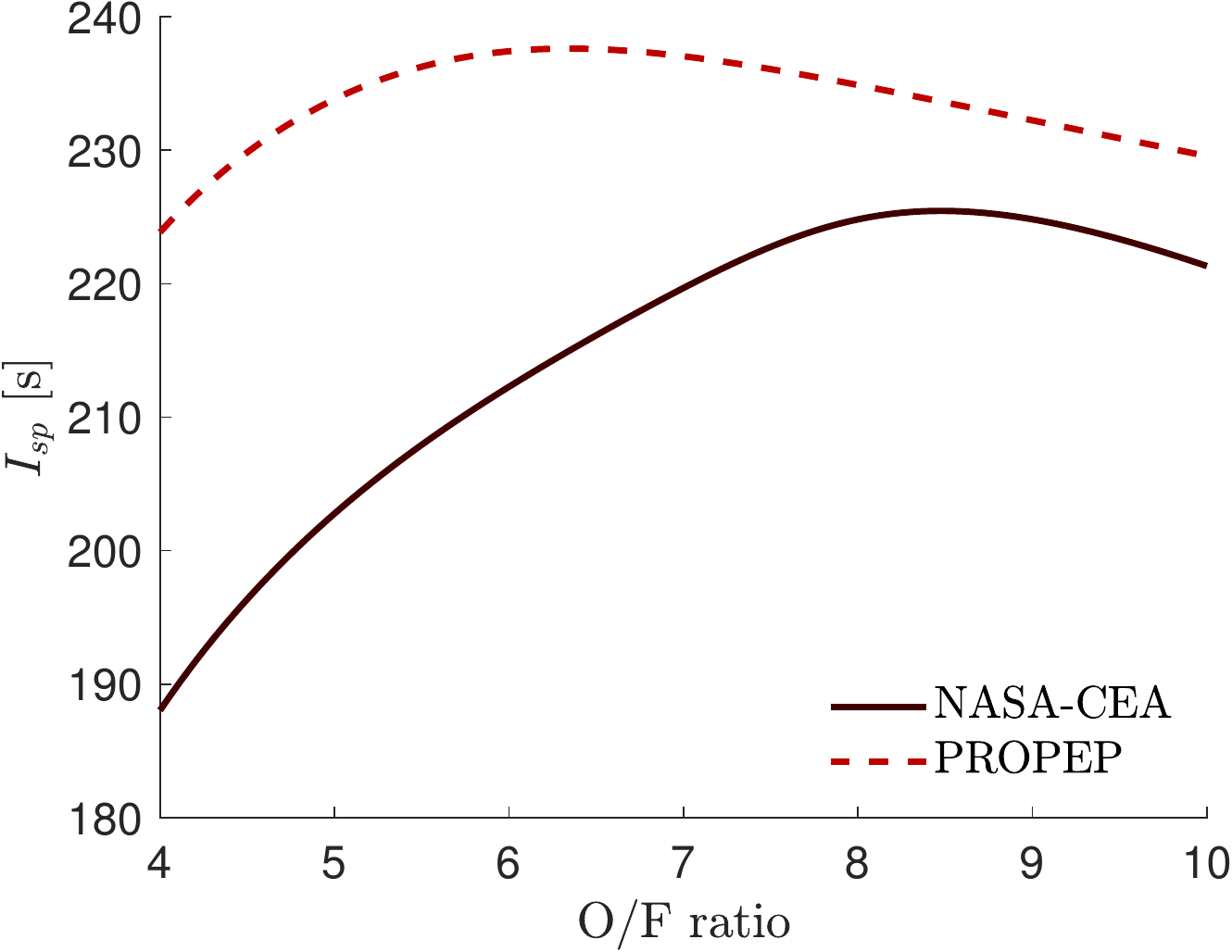}
    \caption{Comparison of the prediction of specific impulse by NASA-CEA and PROPEP for a \nitrous{}-paraffin wax mixture. Chamber pressure: 27.6~bar (400~psia). Nozzle gas exit pressure: 0.846~atm (12.4~psia). For NASA-CEA code: molar enthalpy of \nitrous{}: 9.44~MJ/kmol--saturated liquid at 293.15~K; molar enthalpy of paraffin wax: $h = h_\text{f,0} + c_p (T - T_\text{ref})= 557$~MJ/kmol, where: $h_\text{f,0} = -554$~MJ/kmol is the enthalpy of formation of the species at the reference temperature, $T_\text{ref} = 298.15$~K; $c_p = 602.5$~kJ/kmol is the heat capacity at constant pressure; And $T = 293.15$~K is the propellant inlet temperature.}  
    \label{fig:cea_vs_propep}
\end{figure} 

\item \textbf{Regression rate ballistic coefficients}: The regression rate ballistic coefficients were estimated through the unsteady model based on hot fire testing data, given the engine burntime and pressure profile. The ballistic coefficients found in HF4 do not agree with those found in HF5. The uncertainty in the ballistic coefficients is a source of error, which could be addressed by implementing transient measurements of the fuel grain dimensions during hot fires, therefore allowing to use a least squares fitting procedures to determine $a$ and $n$ in a repeatable and reliable manner. Alternatively, several more engine tests should be conducted to obtain repeatability in the ballistic coefficients.

\end{itemize}

\section{Further Model Improvements}

\begin{itemize}

\item \textbf{Drag coefficient and stability analysis}: The current work couples the hybrid engine model to the OpenRocket flight simulation software, to determine average rocket drag coefficient, and to conduct stability analysis. The model could be rendered more efficient and accurate by implementing a mathematical formulation to estimate the drag coefficient as a function of rocket geometry and wind turbulence models \cite{niskanen2009}. Additionally, the stability analysis could be conducted within the model, by implementing the Barrowman equations as a function of fin geometry. Future iterations of the model could therefore render it completely independent on external input.

\end{itemize}

\section{Closing Remarks}

In the current work, a computational model was developed to assist the design and performance analysis of the \engine{} engine, the first hybrid rocket engine of the MRT. The propulsion system used nitrous oxide as an oxidizer, taking advantage of its self-pressurizing capabilities at ambient conditions; and paraffin wax as a fuel. 

In the first stage of the modeling campaign, a steady-state model was developed, neglecting transient performance decrease effects of the hybrid engine. The propulsion system was designed to bring the launch vehicle \rocket{} with a dry mass of 52.5~kg to a target apogee of 3048~m (10,000~ft). The design chamber pressure selected was 27.6~bar (400~psia), with a theoretical maximum specific impulse of 238~s at an OF ratio of 6.34. The oxidizer mass flow rate, 2.50~kg/s, and fuel length, 610~mm (24.0~in), were selected to maximize the specific impulse of the system within physical and structural system constraints. The resulting average OF ratio calculated was 4.29, leading to a nozzle ideal throat radius of 23.3~mm (0.916~in). The nozzle (non-ideal) outlet radius was selected to be 38.1~mm (1.50~in), yielding a specific impulse of 223~s, while avoiding the additional design complexity of a nozzle insert. The total propellant mass determined was 7535~g, yielding a total impulse of 16,500~Ns at a peak thrust of 6.75~kN.

In the second stage of the modeling campaign, an unsteady model implementing transient performance decrease of the hybrid engine was developed. It was determined that to bring the vehicle to target apogee, the total propellant mass required was 8674~g, an increase of 15.1~\% when compared to the prediction of the steady-state model. The total impulse required increased to 16,845~Ns, while the specific impulse decreased to 198~s. The decreasing chamber pressure during the engine burn was determined to yield a significant performance penalty on the hybrid engine.

In the third stage of the modeling campaign, the unsteady model was validated against hot fire test data. The injector discharge coefficient, polytropic exponent of the tank, and regression rate ballistic coefficients were estimated. The calibrated model was found to provide very good agreement with experimental data, which validates the unsteady model. The \engine{}-HF5 engine achieved a total impulse of 16,423~Ns at a peak thrust of 3.14~kN, while requiring a total propellant mass of 9944~g at a specific impulse of 168~s, and leading to an apogee of 2444~m (8017~ft) for \rocket{}. The inability to attain target performance was attributed to the inability to achieve target chamber pressure: the peak pressure achieved was 14.3~bar (207~psia). This indicated the injector design required further optimization, and the target chamber pressure could be decreased to facilitate engine design. Additionally, the regression rate ballistic coefficients were found to require transient regression rate measurements, or several more engine tests to be validated.

The model developed in the current work provides a simple, yet sufficiently accurate formulation which allows to model amateur hybrid engines. The model can be extended to include a supercharging pressurant such as nitrogen, or to different propellant combinations. Hence, it can be used as a foundation for future hybrid engine designs of the MRT. In the current work, the development of the steady, unsteady, and hot fire models was conducted sequentially, and empirical parameters--such as injector discharge coefficient, fuel density, and regression rate ballistic coefficients--were adjusted along the way to match experimental data. To further optimize future hybrid engine designs, the steady, unsteady, and hot fire models could be used iteratively, with an aim to maximize engine performance while the rocket and engine parameters are defined throughout the design cycle.



\bibliographystyle{pci}
\bibliography{mrt_refs}


\appendix

\chapter{Thermophysical Properties} \label{a1:thermo}

\section{Nitrous Oxide Saturated Properties}

The saturated properties provided in the current section are valid for the temperature range $T = 182.33-309.52$~K. Correlations and look-up tables are re-transcribed from Ref.~\cite{green2008}. To lighten the notation, the subscript "sat" is dropped.

\subsection{Correlated Properties}

The saturated pressure of nitrous oxide is,
\beqarrnn
p(T) &=& \exp \bigg[ c_1 + \f{c_2}{T} + c_3 \ln T + c_4 T^{c_5} \bigg] \\
\Rightarrow \f{\d p(T)}{\d T} &=& p(T) \bigg( \f{-c_2}{T^2} + \f{c_3}{T} + c_4 c_5 T^{c_5-1} \bigg) 
\eeqarrnn
where $T$ is in Kelvin and $p$ is in Pa, and $c_1 = 96.512$, $c_2 = -4045$, $c_3 = -12.277$, $c_4 = 2.886 \times 10^{-5}$, $c_5 = 2$.

The saturated liquid molar volume of nitrous oxide is,
\beqarrnn
\nul(T) &=& \f{c_2^{1 + {\big(1-\f{T}{c_3}\big)}^{c_4}  } }{c_1} \\
\Rightarrow \f{\d \nul(T)}{\d T} &=& \f{c_4}{c_3} (\ln c_2) \nul(T) {\bigg(1 - \f{T}{c_3} \bigg)}^{c_4-1}
\eeqarrnn
where $T$ is in Kelvin and $\nul$ is in m\super{3}/kmol, and $c_1 = 2.781$, $c_2 = 0.27244$, $c_3 = 309.57$, $c_4 = 0.2882$.

\subsection{Look-Up Tables}

Table~\ref{tab:lookup1} compiles the saturated: vapor molar volume, $\nuv$; liquid internal energy, $\ul$; liquid enthalpy, $\hl$; vapor internal energy, $\uv$; vapor enthalpy, $\hv$; vapor heat capacity at constant volume, $\cvv$; and vapor heat capacity at constant pressure, $\cpv$. Table~\ref{tab:lookup2} compiles the derivatives calculated with finite difference schemes from data in Table~\ref{tab:lookup1}. The first two rows of Table~\ref{tab:lookup2} are computed with a forward finite difference, the last two rows, with a backward finite difference, and the intermediate rows, with a central finite difference. (Pay close attention to units reported in the tables.)

\begin{table}[H]
	\small
	\centering
	\caption{Saturated properties of nitrous oxide.}
	\label{tab:lookup1}
	\begin{tabular}{cccccccc}
	$T$ & $\nuv$ & $\ul$ & $\hl$  &  $\uv$ & $\hv$ & $\cvv$ & $\cpv$  \\ \hline
	K & m\super{3}/kmol & kJ/mol & kJ/mol & kJ/mol & kJ/mol & kJ/(mol-K) & kJ/(mol-K) \\ \hline
	182.33 & 16.823 & -0.18142 & -0.17830 & 14.931 & 16.411 & 0.025545 & 0.034993 \\ 
        185 & 14.503  & 0.020258  &0.023954& 14.984 &16.481& 0.025844 & 0.035439 \\ 
        190    &11.089   &0.39840   &0.40340  &15.082 &16.610 &0.026427 &0.036334 \\ 
        195    &8.6098   &0.77745   &0.78408  &15.176 &16.733 &0.027038 &0.037309 \\ 
        200    &6.7780   &1.1577    &1.1664   &15.268 &16.852 &0.027674 &0.038365 \\ 
        205    &5.4030   &1.5396    &1.5508   &15.356 &16.965 &0.028331 &0.039506 \\ 
        210    &4.3558   &1.9234    &1.9377   &15.440 &17.071 &0.029006 &0.040736 \\ 
        215    &3.5474   &2.3094    &2.3274   &15.521 &17.170 &0.029694 &0.042062 \\ 
        220    &2.9156   &2.6982    &2.7206   &15.596 &17.262 &0.030395 &0.043494 \\ 
        225    &2.4162   &3.0899    &3.1177   &15.666 &17.346 &0.031105 &0.045048 \\ 
        230    &2.0171   &3.4852    &3.5192   &15.730 &17.421 &0.031824 &0.046741 \\ 
        235    &1.6951   &3.8844    &3.9257   &15.789 &17.487 &0.032552 &0.048600 \\ 
        240    &1.4329   &4.2881    &4.3379   &15.840 &17.541 &0.033289 &0.050660 \\ 
        245    &1.2174   &4.6969    &4.7566   &15.883 &17.584 &0.034037 &0.052967 \\ 
        250    &1.0388   &5.1115    &5.1826   &15.917 &17.614 &0.043798 &0.055585 \\ 
        255    &0.88979  &5.5326    &5.6170   &15.941 &17.630 &0.035577 &0.058600 \\ 
        260    &0.76438  &5.9614    &6.0609   &15.953 &17.629 &0.036377 &0.062136 \\ 
        265    &0.65808  &6.3990    &6.5161   &15.952 &17.609 &0.037205 &0.066370 \\ 
        270    &0.56730  &6.8471    &6.9843   &15.935 &17.567 &0.038072 &0.071572 \\ 
        275    &0.48917  &7.3078    &7.4682   &15.899 &17.500 &0.038988 &0.078164 \\ 
        280    &0.42135  &7.7841    &7.9715   &15.839 &17.401 &0.039971 &0.086856 \\ 
        285    &0.36189  &8.2804    &8.4993   &15.749 &17.262 &0.041046 &0.098934 \\ 
        290    &0.30912  &8.8039    &9.0600   &15.618 &17.071 &0.042252 &0.11700 \\ 
        295    &0.26142  &9.3670    &9.6680   &15.429 &16.806 &0.043653 &0.14723 \\ 
        300    &0.21691  &9.9960    &10.354   &15.145 &16.422 &0.045376 &0.20883 \\ 
        305    &0.17210  &10.768    &11.205   &14.661 &15.791 &0.047733 &0.40691 \\ 
        309.52 &0.097371 &12.745    &13.450   &12.745 &13.450 &  &  \\ 
	\end{tabular}
\end{table}

\begin{table}[H]
	\small
	\centering
	\caption{Derivatives of saturated properties of nitrous oxide.}
	\label{tab:lookup2}
	\begin{tabular}{cccccccc}
	$T$ & $\f{\d \nuv}{\d T}$ & $\f{\d \ul}{\d T}$ & $\f{\d \hl}{\d T}$  &  $\f{\d \uv}{\d T}$ & $\f{\d \hv}{\d T}$ & $\f{\d \cvv}{\d T}$ & $\f{\d \cpv}{\d T}$  \\ \hline
K & m\super{3}/(kmol-K) & J/(kmol-K) & J/(kmol-K) & J/(kmol-K) & J/(kmol-K) & J/(kmol-K\super{2}) & J/(kmol-K\super{2}) \\ \hline
182.33 & -0.86891 & 75535& 75751& 19850& 26217& 111.99& 167.04 \\
185& -0.6828& 75628& 75889& 19600& 25800& 116.6 & 179 \\ 
190& -0.58932& 75719& 76013& 19200& 25200& 119.4 & 187 \\ 
195& -0.4311& 75930& 76300& 18600& 24200& 124.7& 203.1 \\ 
200& -0.32068& 76215& 76672& 18000& 23200& 129.3& 219.7 \\ 
205& -0.24222& 76570& 77130& 17200& 21900& 133.2& 237.1 \\
210& -0.18556& 76980& 77660& 16500& 20500& 136.3& 255.6 \\ 
215& -0.14402& 77480& 78290& 15600& 19100& 138.9& 275.8 \\ 
220& -0.11312& 78050& 79030& 14500& 17600& 141.1& 298.6 \\ 
225& -0.08985& 78700& 79860& 13400& 15900& 142.9& 324.7 \\ 
230& -0.07211& 79450& 80800& 12300& 14100& 144.7& 355.2 \\ 
235& -0.05842& 80290& 81870& 11000& 12000& 146.5& 391.9 \\ 
240& -0.04777& 81250& 83090& 9400& 9700& 148.5& 436.7 \\ 
245& -0.03941& 82340& 84470& 7700& 7300& 1050.9& 492.5 \\ 
250& -0.032761& 83570& 86040& 5800& 4600& 154& 563.3 \\ 
255& -0.027442& 84990& 87830& 3600& 1500& 742.1& 655.1 \\ 
260& -0.023171& 86640& 89910& 1100& -2100& 162.8& 777 \\ 
265& -0.019708& 88570& 92340& -1800& -6200& 169.5& 943.6 \\ 
270& -0.016891& 90880& 95210& -5300& -10900& 178.3& 1179.4 \\ 
275& -0.014595& 93700& 98720& -9600& -16600& 189.9& 1528.4 \\ 
280& -0.012728& 97260& 103110& -15000& -23800& 205.8& 2077 \\ 
285& -0.011223& 101980& 108850& -22100& -33000& 228.1& 3014.4 \\
290& -0.010047& 108660& 116870& -32000& -45600& 260.7& 4829.6 \\ 
295& -0.009221& 119210& 129400& -47300& -64900& 312.4& 9183 \\ 
300& -0.008932& 140100& 153700& -76800& -101500& 408& 25968 \\
305& -0.008962& 154400& 170200& -96800& -126200& 471.4& 39616 \\
309.52& -0.016533& 437389& 496681& -423894& -517920&  &  \\ 
	\end{tabular}
\end{table}

\subsection{Derived Properties}

The saturated vapor compressibility factor $Z$ is,
\beqarrnn
Z(T) &=& \f{p \nuv}{\Ru T} \\
\Rightarrow \f{\d Z(T)}{\d T} &=& Z(T) \bigg(\f{1}{p} \f{\d p}{\d T} + \f{1}{\nuv} \f{\d \nuv}{\d T} - \f{1}{T} \bigg)
\eeqarrnn
where $Z$ is dimensional, $p$ and its derivative are obtained from the correlation as a function of temperature, and $\nuv$ and its derivative are interpolated from Table~\ref{tab:lookup1} and Table~\ref{tab:lookup2}, respectively.


\section{NASA Atmospheric Model}

The atmospheric model described in the current section is valid for the altitude range $h = 0-30,000$~m ($0-98,425$~ft), where $h$ is the absolute elevation above sea level--not above ground level. Note that in the rocket ascent formulation described in Sections~\ref{sec:steady_ascent} and \ref{sec:unsteady_ascent}, the variable $\zr$ represents the absolute elevation above sea level, hence it can be used directly in the NASA atmospheric model. However, the initial value of $\zr$ is not 0 and depends on the location of the rocket launch. Table~\ref{tab:steady_openrocket} shows the initial value of $\zr$ is 1402~m (4600~ft). The atmospheric model parameters are re-transcribed from \cite{hall2021}.

For $0 \leq h < 11,000$~m, the air temperature $T$ and pressure $p$ are,
\beqarrnn
T &=& c_1 h + c_2 \\
p &=& c_3 {\bigg(\f{T}{c_4}\bigg)}^{c_5}
\eeqarrnn
where $T$ is in K, $p$ is in Pa, and $c_1 = -0.00649$, $c_2 = 288.19$, $c_3 = 101,290$, $c_4 = 288.08$, $c_5 = 5.256$.

For $11,000 \leq h < 25,000$~m, the air temperature $T$ and pressure $p$ are,
\beqarrnn
T &=& c_1 \\
p &=& c_2 \exp \big[ c_3 - c_4 h \big]
\eeqarrnn
where $T$ is in K, $p$ is in Pa, and $c_1 = 216.69$, $c_2 = 22,650$, $c_3 = 1.73$, $c_4 = 0.000157$.

For $25,000 \leq h \leq 30,000$~m, the air temperature $T$ and pressure $p$ are,
\beqarrnn
T &=& c_1 h + c_2 \\
p &=& c_3 {\bigg(\f{T}{c_4}\bigg)}^{c_5}
\eeqarrnn
where $T$ is in K, $p$ is in Pa, and $c_1 = 0.00299$, $c_2 = 141.94$, $c_3 = 2488$, $c_4 = 216.6$, $c_5 = -11.388$.

In all altitude ranges, the air density $\rho$ is,
\beq
\rho = \f{p \Wa}{\Ru T}
\eeq
where $\Wa = 28.9647$~kg/kmol is the air molar weight, and $\Ru = 8314.5$ J/(kmol-K) is the universal gas constant.



\end{document}